\documentclass[final,10pt,a5paper]{phdimt}

\usepackage{verbatim}
\usepackage{xspace}
\usepackage{float}

\newfloat{algorithm}{tbp}{loa}
\floatname{algorithm}{Code Fragment}

{
\newcommand{\muskel}[0]{\xspace\mbox{\textbf{\texttt{muskel}}}\xspace}

\pdfinfo{
	/Title  (Tools and Model for High Level Parallel and Grid Programming)
	/Creator (MikTeX)
	/Producer (pdfTeX)
	/Author (Patrizio Dazzi)
	/CreationDate (T:20080327193000)
	/ModDate (D:20080214193000)
	/Subject (Computer Science)
	/Keywords (parallel computing, grid computing, programming models)
}
\pdfcatalog{
	/PageMode (/UseOutlines)
      /OpenAction (fitbh)
}

\title{Tools and Models for High Level Parallel and Grid Programming}

\author{Patrizio Dazzi}
\mail{p.dazzi@imtlucca.it}
\program{Computer Science and Engineering}
\coordinator{Prof. Ugo Montanari}
\coordinatorinst{University of Pisa, Italy}
\cycle{XX}
\year{2008}

\supervisor{Prof. Marco Danelutto}
\supervisorinst{University of Pisa, Italy}
\firstreviewer{Prof. Vladimir Getov}
\firstreviewerinst{University of Westminster, UK}
\secondreviewer{Dr. Christian Perez}
\secondreviewerinst{INRIA, France}
\begin{document}

% My vectors are bold, no arrows
\def\vec#1{{\bf #1}}

% Let's do not cheat on line spacing
\renewcommand\baselinestretch{1}
\baselineskip=14pt

\frontmatter

\newcommand{\code}[1]{{\tt #1}{}}
\newcommand{\SubType}[2]{\ensuremath{#1~\mbox{\code{<:}}~#2}}
\newcommand{\serif}[1]{\textsf{#1}\xspace}
\newcommand{\serifd}[2]{\textsf{#1}${}_{#2}$\xspace}
\newcommand{\lab}[2]{#1${}_{#2}$}

\pagestyle{empty} % following pages do not require page numbering
\maketitle
\makereviewerspage
% Alessio Botta 2008 - alebotta@gmail.com
%
% phdimt class file and demo
% A LaTex implementation of IMT Lucca PhD Thesis style
% Please send an email to let me know you used this style.

\begin{dedication} 
%Allo zio Giancarlo... \\ mi manchi... \\ ci manchi.
A tutta la mia famiglia, \\con un particolare pensiero per lo zio Giancarlo.
\end{dedication}
\pagestyle{plain} % numbering is shown from now on

\tableofcontents
\listoffigures
%\listoftables

\insertblankpage

% Alessio Botta 2008 - alebotta@gmail.com
%
% phdimt class file and demo
% A LaTex implementation of IMT Lucca PhD Thesis style
% Please send an email to let me know you used this style.

\begin{acknowledgements}
\addcontentsline{toc}{chapter}{Acknowledgements}
%The contribution of many people has made possible this thesis. Hence I want to thank everyone who contributed to this result. I apologize but this single page is not enough to mention all the people that deserve it.
%
First of all I wish to thank my supervisor \textbf{Marco Danelutto} who always supported and encouraged me, leaving me the freedom of experimenting with many different topics, far beyond I have expected. 
If I wouldn't dread being disrespectful I would say he is a real good friend.
I am very grateful to \textbf{Domenico Laforenza} the head of the High Performance Computing Laboratory, where I spent most of time of the last three years. I give my thanks also to my co-workers: \textbf{R. Perego}, \textbf{R. Baraglia},  \textbf{S. Orlando}, \textbf{F. Silvestri}, \textbf{N. Tonellotto}, \textbf{C. Lucchese}, \textbf{M. Coppola}, \textbf{D. Bocci}, \textbf{F. M. Nardini}, \textbf{G. Capannini} and \textbf{G. Tolomei}.
Many thanks go also to who co-authored my works: \textbf{M. Aldinucci}, \textbf{S. Campa}, \textbf{L. Presti}, \textbf{A. Panciatici}, \textbf{M. Pasin}, \textbf{M. Vanneschi} and \textbf{P. Kilpatrick}.
I thank \textbf{Marco Paquali}, he was both a valuable co-worker and a precious co-author  but, above all, he is a very good friend.
I am very grateful both to my reviewers \textbf{Vladimir Getov} and \textbf{Christian Perez}, and to the member of my internal thesis committee \textbf{Gianluigi Ferrari} and \textbf{Paolo Ciancarini} for giving me many helpful suggestions.
I express my gratitude to all my friends who shared with me the spare time. Among them \textbf{Francesco}, with whom I had many discussions about almost everything.
%
%I would like to express my gratitude also to \textbf{Gianluigi Ferrari}, and to \textbf{Paolo Ciancarini}, they have been part of my internal thesis committee and they gave me many helpful suggestions.

Now let me switch to Italian to... ringraziare tutta la mia pre\-zio\-sa famiglia, con particolare riferimento ai miei genitori che da anni mi supportano sia col loro affetto sia economicamente. Sono fermamente convito che possiate ritenervi i mi\-glio\-ri genitori che un figlio possa avere. L'ultimo ringraziamento va alla persona che da anni ha il monopolio del mio cuore, \textit{Chiara}, trovare le parole per ringraziarti \'e un compito impossibile, qualsiasi frase o parola non sarebbe mai abbastanza.
\end{acknowledgements}

% Alessio Botta 2008 - alebotta@gmail.com
%
% phdimt class file and demo
% A LaTex implementation of IMT Lucca PhD Thesis style
% Please send an email to let me know you used this style.

% Don't modify the following
% Note that if the table is too long, it will flow out of the page. If you have a very long CV, maybe you're smart enough to solve this problem by yourself! :)
\begin{center}
\vspace*{0.5cm}
{\Large \bf  Vita}
\addcontentsline{toc}{chapter}{Vita and Publications}
\end{center}
\begin{table}[h!]
\begin{center}
\renewcommand{\arraystretch}{1.25}
\begin{tabular*}{1\textwidth}{l p{8.5cm}}

% Edit content of this part (just the content!)
{\bf November 23, 1979} & Born, Carrara (Tuscany), Italy \\
& \\
{\bf July, 2004} & Master Degree in Computer Technologies\\  
& Final mark: 110/110\\
& University of Pisa, Pisa, Italy\\
& \\
{\bf March, 2005} & PhD Student in Computer Science and Engineering\\  
& IMT Lucca Institute for Advanced Studies\\
& Lucca, Italy\\
& \\
{\bf April, 2005 -- Present} & Graduate Fellow at HPC Lab\\  
& ISTI -- CNR, Pisa, Italy\\

% Don't modify the following
\end{tabular*}
\end{center}
\end{table}
% Comment the following if you do not prefer the publications to start on a different page
\clearpage
\begin{center}
\vspace*{0.5cm}
{\Large \bf  Publications}
\end{center}
\vspace*{0.5cm}
\sloppy
{\small
\begin{enumerate}
% Edit content of this part (just the content!)
\item Marco Aldinucci, Sonia Campa, Marco Danelutto, Patrizio Dazzi, Peter
  Kilpatrick, Domenico Laforenza, and Nicola Tonellotto.
\newblock ``Behavioural skeletons for component autonomic management on grids.''
\newblock In Marco Danelutto, Paraskevi Frangopoulou, and Vladimir Getov,
  editors, {\em Making Grids Work}, CoreGRID. Springer, June 2008.

%\item Marco Aldinucci, Sonia Campa, Marco Danelutto, Peter Kilpatrick, Patrizio
%  Dazzi, Domenico Laforenza, and Nicola Tonellotto.
%\newblock ``Behavioural skeletons for component autonomic management on grids.''
%\newblock In Marco Danelutto, Vladimir Getov, and Paraskevi Frangopoulou,
%  editors, {\em Proceedings of the CoreGRID Workshop on Grid Programming Model,
%  Grid and P2P Systems Architecture, Grid Systems, Tools and Environments},
%  pages 24--35, Heraklion, Greece, June 2007. CoreGRID, FORTH-ICS.

%\item Marco Aldinucci, Sonia Campa, Marco Danelutto, Peter Kilpatrick, Patrizio
%  Dazzi, Domenico Laforenza, and Nicola Tonellotto.
%\newblock ``Behavioural skeletons for component autonomic management on grids.''
%\newblock Technical Report TR-07-12, Dept. of Computer Science, University of
%  Pisa, May 2007.

\item Marco Aldinucci, Sonia Campa, Marco Danelutto, Marco Vanneschi, Peter
  Kilpatrick, Patrizio Dazzi, Domenico Laforenza, and Nicola Tonellotto.
\newblock ``Behavioral skeletons in gcm: automatic management of grid components.''
\newblock In Julien Bourgeois and Didier El~Baz, editors, {\em Proceedings of
  the 16th Euromicro Conference on Parallel, Distributed and Network-based
  Processing (PDP2008)}, pages 54--63, Toulose, France, February 2008. IEEE
  Computer Society Press.

\item Marco Aldinucci, Marco Danelutto, and Patrizio Dazzi.
\newblock ``Muskel: an expandable skeleton environment.''
\newblock {\em Scalable Computing: Practice and Experience}, 8(4):325--341,
  December 2007.

\item Marco Aldinucci, Marco Danelutto, Peter Kilpatrick, and Patrizio Dazzi.
\newblock ``From {Orc} models to distributed grid {Java} code.''
\newblock In {\em Proc. of the Integrated Research in Grid Computing Workshop},
  CoreGRID, April 2008.

\item Marco Danelutto and Patrizio Dazzi.
\newblock ``A java/jini framework supporting stream parallel computations.''
\newblock In {\em Proc. of Intl. PARCO 2005: Parallel Computing}, September
  2005.

\item Marco Danelutto and Patrizio Dazzi.
\newblock ``Joint structured/non structured parallelism exploitation through data
  flow.''
\newblock In Vassil~N. Alexandrov, G.~Dick van Albada, Peter M.~A. Sloot, and
  Jack Dongarra, editors, {\em Proc. of ICCS: International Conference on
  Computational Science, Workshop on Practical Aspects of High-level Parallel
  Programming}, LNCS, Reading, UK, May 2006. Springer Verlag.

%\item Marco Danelutto and Patrizio Dazzi.
%\newblock ``Workflows on top of a macro data flow interpreter exploiting aspects.''
%\newblock In {\em CoreGRID Workshop on Grid Programming Model, Grid and P2P
%  Systems Architecture, Grid Systems, Tools and Environments}, June 2007.

\item Marco Danelutto and Patrizio Dazzi.
\newblock ``Workflows on top of a macro data flow interpreter exploiting aspects.''
\newblock In Marco Danelutto, Paraskevi Frangopoulou, and Vladimir Getov
  editors, {\em Making Grids Work}, CoreGRID. Springer, June 2008.

%\item Marco Danelutto, Marcelo Pasin, Marco Vanneschi, Patrizio Dazzi, Luigi Presti,
%  and Domenico Laforenza.
%\newblock ``Pal: towards a new approach to high level parallel programming.''
%\newblock In Marian Bubak, Sergei Gorlatch, and Thierry Priol, editors, {\em
%  Proc. of the Integrated Research in Grid Computing Workshop}, CoreGRID,
%  Krak{\^U}w, Poland, October 2006. Academic Computing Centre CYFRONET AGH.

\item Marco Danelutto, Marcelo Pasin, Marco Vanneschi, Patrizio Dazzi, Luigi Presti,
  and Domenico Laforenza.
\newblock ``Pal: Exploiting java annotations for parallelism.''
\newblock In Marian Bubak, Sergei Gorlatch, and Thierry Priol, editors, {\em
  Achievements in European Research on Grid Systems}, CoreGRID Series, pages
  83--96. Springer, Krakow, Poland, 2007.

\item Patrizio Dazzi, Francesco Nidito, and Marco Pasquali.
\newblock ``New perspectives in autonomic design patterns for
  stream-classification-systems.''
\newblock In {\em Proceedings of the 2007 workshop on Automating service
  quality (WRASQ): Held at the International Conference on Automated Software
  Engineering (ASE)}, pages 34--37, New York, NY, USA, 2007. ACM.
\newblock ISBN 978-1-59593-878-7.

\item Ranieri Baraglia, Patrizio Dazzi, Antonio Panciatici, and Marco Pasquali.
\newblock ``Self-Optimizing Classifiers: Formalization and Design Pattern''
\newblock In {\em Proceedings of the CoreGRID Symposium: Held at Europar 08, August 25-26, 2008, Las Palmas de Gran Canaria, Canary Island, Spain}

\end{enumerate}
}
\fussy
% Comment the following if you do not prefer the publications to start on a different page
\clearpage
\begin{center}
\vspace*{0.5cm}
{\Large \bf  Presentations}
\end{center}
\vspace*{0.5cm}
\sloppy
{\small
\begin{enumerate}

% Edit content of this part (just the content!)
\item Patrizio Dazzi, ``PAL: towards a new approach to high level parallel programming'', at \emph{Academic Computing Centre CYFRONET AGH}, Krakow, Poland, 2006.
\item Patrizio Dazzi and Sonia Campa, ``Autonomic Features of Grid Component Model'', at \emph{GRIDSYSTEMS S.A.}, Palma de Mallorca, Spain, 2007.
\item Patrizio Dazzi and Nicola Tonellotto, ``Tutorial on Grid Component Model -- Non Functional framework: autonomic management'', at \emph{Tsinghua University}, Beijing, China, 2007.
\item Patrizio Dazzi, ``Programming Autonomic Applications with Grid Component Model'', at \emph{Univeristy of Pisa}, Pisa, Italy, 2008.

% Don't modify the following
\end{enumerate}
}
\fussy

\insertblankpage

% Alessio Botta 2008 - alebotta@gmail.com
%
% phdimt class file and demo
% A LaTex implementation of IMT Lucca PhD Thesis style
% Please send an email to let me know you used this style.

\begin{abstract} 
\addcontentsline{toc}{chapter}{Abstract}
%%% a brief statement of the problem;
%

When algorithmic skeletons were first introduced by Cole in late 1980 \cite{128874} the idea had an almost immediate success. The skeletal approach has been proved to be effective when application algorithms can be expressed in terms of skeletons composition. However, despite both their effectiveness and the progress made in skeletal systems design and implementation, algorithmic skeletons remain absent from mainstream practice. 
Cole and other researchers, respectively in \cite{cole:manifesto:02} and \cite{advske:pc:06}, focused the problem. They recognized the issues affecting skeletal systems and stated a set of principles that have to be tackled in order to make them more effective and to take skeletal programming into the parallel mainstream.
%
%%% a brief exposition of the method or procedures used;
%
In this thesis we propose tools and models for addressing some among the skeletal programming environments issues. We describe three novel approaches aimed at enhancing skeletons based systems from different angles. First, we present a model we conceived that allows algorithmic skeletons customization exploiting the macro data-flow abstraction. Then we present two results about the exploitation of metaprogramming techniques for the run-time generation and optimization of macro data-flow graphs. In particular, we show how to generate and how to optimize macro data-flow graphs accordingly both to programmers provided non-functional requirements and to execution platform features. The last result we present are the Behavioural Skeletons, an approach aimed at addressing the limitations of skeletal programming environments when used for the development of component-based Grid applications.
%
%%% a condensed summary of the findings of the study.
%
We validated all the approaches conducting several test, performed exploiting a set of tools we developed.
\insertblankpage
\end{abstract}

\mainmatter

\chapter{Introduction}

Computers are becoming tools of vital importance. They are used almost everywhere, they are used for work, for study, for fun and actually for solve problem.
Unfortunately, many problems require a huge amount of computational power to solve (as an example: genome mapping, portfolio risk-analysis, protein folding). Such a power cannot be obtained using a single processor. The only suitable solution is to distribute the application workload across many different computational resources. Resources those contemporaneously (``in parallel'') execute parts of the whole application. Programming applications that make use of several computational resources at the same time introduces some difficulties, as an example the communication and synchronization among the resources, or the application code and data decomposition and distribution. In order to ease this burden, since the early steps of computer science, researchers conceived and designed programming models and tools aiming at supporting the development of parallel applications.
Throughout the ages, a lot of models and tools have been proposed, presented in several different (sometime exotic) forms. Nevertheless, the main goal is always the same: find a good trade-off between simplicity and efficiency.
Indeed, a very abstract model simplifies the programming activity but can lead to a very inefficient exploitation of computing resources. Instead, a low-level model allows programmers to efficiently exploit the computational resources but requires to programmers a tremendous effort when the number of resources grows.
Since the nineties, several research groups have proposed the \emph{structured parallel programming environments} (SPPE). Since the structured parallel programming model was conceived, several works have been done about it.
Programming environments relying on this paradigm ask programmers to explicitly deal with the \emph{qualitative} aspects of parallelism exploitation, namely the application structure and problem decomposition strategies. All the low-level parallelism exploitation related aspects like communication, synchronization, mapping and scheduling are managed by compiler tools and run-time support.
In these environments parallelism is exploited by composing ``skeletons'', i.e. parallelism exploitation patterns. From language viewpoint, a skeleton is a higher-order function that behaves as a pure function (no side-effects). Several real world, complex applications have been developed using these environments. The skeletal approach has been proved to be quite effective, when application algorithms can be somehow expressed in terms of skeleton composition. Notwithstanding, skeletal programming has still to make a substantial impact on mainstream practice in parallel applications programming.

\section{Contribution of the thesis}

This thesis originates from the wish to address the issues that have limited the diffusion of structured parallel programming environments. These issues are well-known by the structured parallel programming models scientific community. They have been organically reported in two key papers \cite{cole:manifesto:02, advske:pc:06} where the authors describe both the issues and the features that the next generation of structured parallel programming environments have to support in order to address them. The features ``checklist'' includes, as an example, the ease of use, the integration of structured and unstructured form of parallelization, the support for code reuse, the heterogeneity and dynamicity handling.
Drawing a parallel with web programming model we can refer as ``Skeletons 2.0'' the next generation of structured parallel programming environments that address the issues that prevent the skeleton environment to became part of the mainstream practice in parallel applications programming.
Some groups of researchers involved in structured parallel programming developed skeleton systems that have partially addressed the ``Skeletons 2.0'' principles to different degrees in different combinations. Nevertheless, the research for addressing the presented issues has just started. Indeed, up to now, tools and models that are generally recognized as the best solutions for addressing the issues still do not exist.

The main goal of this thesis is to present an organic set of tools and models conceived, designed and developed to address most of these issues, therefore form the base of a next generation skeleton system. The scientific contribution of the thesis is organized in three main parts. They reports four results we obtained in the last three years. These research results as has been already presented in published papers.
Some results have been achieved with actual experiments conducted using software tools and packages designed and developed to the purpose. Some of them are simple, proof-of-concept tools, like JJPF \cite{DaDa05parco} or PAL \cite{pal}. Some others are custom version of existing framework, like \muskel with the support for developing unstructured form of parallelism \cite{muskelJournal} or \muskel with an aspect oriented programming support \cite{muskaspects:cg_book:08}. Others are part of complex international research project focused on Grid computing, like the Behavioural Skeletons \cite{pdp08:beske}.

Our first contribution copes with the lack of models supporting the integration of unstructured form of parallelization in skeleton systems.  In fact, if on the one hand structured parallel programming environments raise the level of abstraction perceived by programmers and guarantee good performance, on the other hand they restrict the freedom of programmers to implement arbitrary parallelism exploitation patterns. In order to address this issue we propose a \textit{macro data-flow}   based approach that can be used to implement mixed parallel programming environments providing the programmer with both structured and unstructured ways of expressing parallelism. Structured parallel exploitation patterns are implemented translating them into data-flow graphs executed by a distributed macro data-flow interpreter. Unstructured parallelism exploitation can be achieved by explicitly programming data-flow (sub)graphs.
To validate the approach, we modified a skeleton system that in its original form does not deal with unstructured parallelism: \muskel. We extended \muskel, in collaboration with the research staff that developed it.
Our customized \muskel is implemented exploiting (macro) data-flow technology, rather than more usual skeleton technology relying on the usage of implementation templates. Using data-flow, the extended \muskel supports the development of both classical, predefined skeletons, and programmer-defined parallelism exploitation patters.
Our extended version provides two mechanisms to the \muskel programmers for unstructured parallelism exploitation. First, we provide primitives that allow to access the fundamental features of the data-flow graph generated out of the compilation of a skeleton program. Namely, methods to deliver data to and retrieve data from data-flow graph. We provide to programmers the ability to instantiate a new graph in the task pool by providing the input task token and to redirect the output token of the graph to an arbitrary data-flow instruction in the pool. Second, we provide the programmer with direct access to the definition of data-flow graphs, in such a way he can describe his particular parallelism exploitation patterns that cannot be efficiently implemented with the available skeletons.
The two mechanisms can be jointly used to program all those parts of the application that cannot be easily and efficiently implementing using the traditional skeletons subsystem.
Unfortunately, this approach is not free from shortcomings. In fact, exploiting unstructured parallelism interacting directly with data-flow graph requires to programmers to reason in terms of program-blocks instead of a monolithic program.

In order to ease the generation of macro data-flow blocks and in general to provide mechanism easing the use of structured parallel programming environment, we exploited some \textit{metaprogramming} techniques. 
Exploiting these techniques the programmers are no longer requested to deal with complex application structuring but simply to give hints to the metaprogramming support using high-level directives. The directives drive the automatic application transformation. 
In this thesis we present two results we obtained regarding the exploitation of metaprogramming techniques for parallel programming. The first result is ``Parallel Abstraction Layer'' (PAL). A java annotation based metaprogramming framework that restructures applications at bytecode-level at run-time in order to make them parallel. The parallelization is obtained asynchronously executing the annotated methods. Each method call is transformed in a macro data-flow block that can be dispatched and executed on the available computing resources.
PAL transformations depend on both on the resources available at run-time and the hints provided by programmers. 
The other result concerns the integration of the Aspect Oriented Programming mechanisms with our modified \muskel skeleton framework. 
We make this integration in two distinct phases, in the first phase we integrated the AOP mechanisms in order to achieve very simple code transformation. In the second phase we implemented a more complex integration to obtain a support enabling the development of \textit{workflows} which structure and processing are optimized at run-time depending on the available computational resources.

In this thesis we present also a model to address two other issues: the lack of support for code reuse, and the lack of support for handling of dynamicity.
The \muskel framework, addresses this last point through the definition of the \emph{Application Manager}, namely an entity able to observe, at run-time, the behavior of the parallel application and in case of faults or application non-functional requirement violations it reacts aiming to fix the problem.
The dynamicity handling is a very important feature for next generation parallel programming systems, especially for the ones designed for computational Grids. Indeed, Grid are often composed by heterogeneous computer  
and managed by different administration policies. To address these additional difficulties most of the models and tools conceived and developed for parallel programming have to be re-thought and adapted.
Actually, the \muskel framework, at least in its original form, is designed to be exploited in cluster and network of workstations rather than in Grids. Indeed, some of the implementation choices done when it was developed limit its exploitation on Grids, in particular the ones related with communication protocol and with the mechanisms for recruiting computational resource.
On the other hand, several studies recognized that component technology could be leveraged to ease the development of Grid Application  \cite{armstrong99toward, 383872}. Indeed, a few component based model have been proposed by parallel computing scientific community for programming Grids, as CCA \cite{cca}, CCM \cite{DenPerPriRib} and GCM \cite{gcm:coregrid:07}.
The GCM represents one of the main European scientific community efforts for designing and developing \cite{gridcomp} a grid component model. We contributed to the design of GCM and its reference implementation together with the research group that developed \muskel and with several European research groups. In particular, we focused our contribution on the GCM autonomic features. We referred to the \muskel \textit{Application Manager} approach, generalizing it and extending the approach to make it suitable for components based models. Indeed, each GCM component with a complete support of autonomic features has an \textit{Autonomic Manager} that observes the component behavior. In case the behavior turns out to be different from the expected one the manager trigger a component reconfiguration.
In other words, GCM autonomic features provide programmers with a configurable and straightforward way to implement autonomic grid applications. Hence, they ease the development of application for the Grids. Nevertheless, they rely fully on the application programmer's expertise for the setup of the management code, which can be quite difficult to write since it may involve the management of black-box components, and, notably, is tailored for the particular component or assembly of them. As a result, the introduction of dynamic adaptivity and self-management might enable the management of grid dynamism, and uncertainty aspects but, at the same time, decreases the component reuse potential since it further specializes components with application specific management code. In order to address this problem, we propose the \emph{Behavioural Skeletons} as a novel way to describe autonomic components in the GCM framework. Behavioural Skeletons aim to describe recurring patterns of component assemblies that can be equipped with correct and effective management strategies with respect to a given management goal. Behavioural Skeletons help the application designer to i) design component assemblies that can be effectively reused, and ii) cope with management complexity. The Behavioural Skeletons model is an effective solution for handling dynamicity, supporting reuse both of functional and non-functional code. We want to point out that we have not the ``sole rights'' concerning the Behavioural Skeletons model. Indeed, it has been developed in conjunction with the other authors of the two papers about Behavioural Skeletons we published \cite{pdp08:beske, heraklion-beske}.

This thesis is not our first attempt of design programming model for parallel programming. In a previous work we developed JJPF, a Java and Jini based Parallel Framework, and investigated the possibilities offered by structured parallel programming. In \cite{DaDa05parco} we described the architecture of JJPF. JJPF was specifically designed to efficiently exploit affordable parallel architectures, such as a network of workstations. Its reactive fault-tolerance support and its dynamic support for task distribution as well as for resources recruiting were designed to enable an efficient exploitation of resources in highly dynamic environment. In particular, JJPF exploits the Jini technologies to dynamically find and recruit the available computational resources.
JJPF provide to programmers an API enabling the development of task-parallel application following the master-slave paradigm. It also provides an high-level support for data sharing among slaves. JJPF ease the parallel programming task hiding most of low-level error prone issues to programmers.
As we stated above, JJPF is implemented in Java. It simplifies the code portability among heterogeneous architectures. For the communications among master and slaves JJPF exploits the JERI. It is a variant of RMI allowing the protocol customization and as a consequence an optimization of its performance in several situations. For the performance purpose JJPF also provides an alternative to the java distributed class-loader that reduces the class-loading latency in some situations.
Some problems encountered during the design of JJPF still remain open. Moreover, during the realization of JJPF we faced directly with the development complexity of this kind of software so we think that some kind of software engineering is needed to facilitate reuse and maintenance of source code.

\section{Thesis Outline}
As we already stated, in this thesis we report our contribution to address the issues that are typical of traditional structured parallel programming environments. The contribution is organized in three main parts. Each part is presented in a dedicated chapter. Moreover, there are three more chapters: an Introduction chapter (this one, actually), a Conclusion chapter and another one that introduces the problems we face in this thesis and outlines the state-of-the-art of existing solutions.
In the remain of this section we describe the content of each chapter.

\paragraph{Chapter \ref{parallel_issues}}
In this chapter we take into account the problems related to programming parallel applications, the existing solutions and their main limitations. In particular, after a general introduction to the different parallel programming models, the topic is focused on the limitations that prevent the structured parallel programming models from spreading and to become part of the mainstream practice.
Section \ref{sec:fromseq} gives a bird's-eye view both on the parallel architectures and on the fields in which parallelism has traditionally been employed. Section \ref{sec:stateofart} reports a selection of the main parallel programming models distinguishing between the implicit (Section \ref{functional-logic-model}) and explicit (Section \ref{sec:explicit}) approaches. The explicit approaches are further discussed subdividing them, with respect to the abstraction presented to programmers, in high-level (Section \ref{dataflow-model}) and low-level (Section \ref{low-level-model}) ones. For each of them are presented the Pros and Cons. The chapter reports also some other notable approaches (Section \ref{other-appr}). Then the Chapter present the structured approach, an approach conceived in order to overcome the limitations of traditional approaches (Section \ref{structured-model}). Some tools based on the structured parallel programming models are presented (Section \ref{old-fashion}) and others are reported as well as references to the literature. The models are presented highlighting their features and main limitations. Section \ref{sec:openissues} reports the issues that next generation skeleton system should own to address the existing limitations. Finally, the chapter introduces (Section \ref{sec:ourefforts}) our contributions to the field placing them in the proper context, showing how such contributions can be exploited for addressing the issues related to structured parallel programming environments.

\paragraph{Chapter \ref{skeleton_customization}}
In this Chapter we discuss a methodology that can be exploited in order to provide to programmers the possibility to mix structured and unstructured ways of expressing parallelism while preserving most of the benefits typical of structured parallel programming models. The methodology is based on the data-flow model. Unstructured parallelism exploitation is achieved by explicitly programming data-flow graphs.
Section \ref{sec:introDFmuskel} briefly recalls the structured programming models outlining their main advantages and limitations. In particular, the section focuses on the skeleton customization issue. Namely the lack of flexibility of skeletal systems in expressing parallel form different from the ones that are ``bundled'' with the skeleton framework and their compositions. Then the section introduces the macro data-flow based approach we conceived in order to address of this limitation and reports the related work: alternative approaches addressing the structured parallel programming limitations.
Section \ref{sec:templ-dataflow} introduces both the classical template-based implementation of skeleton systems and the more recent data-flow technologies based one used in \muskel.
Section \ref{sec:unstruc} describes the details of our contribution, i.e. how we exploited the methodology presented to extend the \muskel framework. %, discussing how skeletons customization is supported exploiting data-flow implementation.
Finally, Section \ref{sec:results} reports the experimental results we obtained conducting some test using our customized \muskel framework.

\paragraph{Chapter \ref{muskelWorkflow}} In this Chapter we introduce some novel metaprogramming techniques for the generation and optimization of macro data-flow blocks. This Chapter presents our efforts aimed at providing metaprogramming tools and models for optimizing at run-time the execution of structured parallel applications. The approaches are based on the run-time generation of macro data-flow blocks from the application code. The Chapter discusses how we exploited these techniques both in our modified \muskel framework as well as in other frameworks we developed.
Section \ref{motivations} presents the motivations behind our contributions. Section \ref{PAL} presents PAL, our first result in the field. The core of PAL framework is its metaprogramming engine that transforms at run-time an annotated sequential java code in a parallel program exploiting both programmer hints and information about executing platforms.
Section \ref{PALimpl} describes the details of our PAL prototype implementation. Section \ref{PALtests} reports the experimental results we obtained testing PAL framework. Section \ref{PALmotivations} discusses the motivations that convinced us to integrate the PAL approach to our modified \muskel framework. Section (\ref{metamuskel} describes the preliminary attempts we made integrating metaprogramming techniques in \muskel showing how Aspect Oriented Programming can be exploited to do some simple code transformations. %in order to normalize a skeleton code.
Section \ref{muskworkflows} describes how we further enhanced \muskel making it able to exploit metaprogramming for run-time code optimizations. In particular, how it can be exploited to optimize the parallel execution of computations expressed as workflows. Section \ref{sec:imple} describes the implementation details of workflows transformations and Section \ref{sec:perfresults} presents the results of some experiments we conducted. Finally Section \ref{sec:gendifferences} presents a comparison of the two approaches.

\paragraph{Chapter \ref{mdf_as_components}}
In this Chapter we present some results about the customization of skeletons applied to the Grid Component Model.
In this chapter we present the Behavioural Skeletons model, an approach, we contribute to conceive and validate, aimed at provide programmers with the ability to implement autonomic grid component-based applications completely taking care of the parallelism exploitation details by simply instantiating existing skeletons and by providing suitable, functional parameters. The model has been specifically conceived to enable code reuse and dynamicity handling.
Section \ref{sec:introComponents} describes how component-based applications can ease the task of developing grid applications. Section \ref{sec:GCMintro} outlines the grid component model focusing on its autonomic features. After, Section \ref{sec:BeSke} presents the Behavioural Skeletons model, Section \ref{sec:BeSkeSet} reports a set of noteworthy Behavioural Skeletons and Section \ref{sec:BeSkeImpl} describe their GCM implementation. Section \ref{sec:BeSkeExp} describes a set of experiment we conducted to validate the Behavioural Skeletons model.

\paragraph{Chapter \ref{thesis_concl}}
This Chapter summarizes the materials contained in the previous chapters and discusses the conclusions of
the thesis. Finally, the future work related to the thesis is introduced.

\insertblankpage
\chapter{High-Level Parallel Programming} \label{parallel_issues}

As we already stated in the Introduction, using several processors (or computational resources) at the same time (in parallel), however, introduces some difficulties. The conceptual barrier encountered by the programmers in efficiently coordinating many concurrent activities towards a single goal is an example of such barriers. To address these difficulties software developers need high-level programming models for sensibly raising the abstraction of computational resources. This is a fundamental requirement to avoid programmers having to deal with low-level coordination mechanisms. In fact, low-level parallel programming is an error prone approach that distracts programmers from qualitative aspects of parallelization.
Throughout the ages, researchers conceived and developed several models for high-level parallel programming.
However, most of current implementations of very high-level programming models often suffer from low performance. This because of the abstraction penalty, which actually has historically limited the usage of high-level programming techniques in high performance computing. For this reason, nowadays most of parallel programs are developed exploiting lower-level language, even if a higher-level language would make the coding easier.
%Several researchers have tried to address the limitation of these approaches. 
% 
\emph{Structured parallel programming models} were conceived to be an alternative both to very high-level models and to low-level models.
Structured parallel programming models ask programmers to explicitly deal with the \emph{qualitative} aspects of parallelism exploitation, namely the application structure and problem decomposition strategies. 
Compilers and run-time supports manage all the low-level parallelism exploitation related aspects like communication, synchronization, scheduling and mapping.
%
%All the low-level parallelism exploitation related aspects like communication, synchronization, mapping and scheduling are managed by compiler tools and run-time support. 
The \emph{Structured Way} is driven by those two observations: there are some things programmers do better than compilers, and there are some things that compilers do better than programmers.
Nevertheless, also the structured models are not perfect and free from limitations. In fact, for some years researchers very expert in structured parallel programming models have outlined the features that the next generation of structured models have to provide in order to address these limitations \cite{cole:manifesto:02, advske:pc:06}.
In next three chapters of this thesis we present some results we obtained as an attempt of address some of these limitations.

\paragraph{Chapter road-map} %perspective
\emph{The chapter starts with a bird's-eye view both on the parallel architectures and on the fields in which parallelism has traditionally been employed (Section \ref{sec:fromseq}). Then, it reports the main parallel programming models (Section \ref{sec:stateofart}) distinguishing between the implicit (Section \ref{functional-logic-model}) and explicit (Section \ref{sec:explicit}) approaches. The explicit approaches are further subdivided in high-level (Section \ref{dataflow-model}), and low-level (Section \ref{low-level-model}) ones. The chapter reports also some other notable approaches (Section \ref{other-appr}). Then the Chapter present the structured approach, an approach conceived in order to overcome the limitations of traditional approaches (Section \ref{structured-model}). Some tools based on the structured parallel programming models are presented (Section \ref{old-fashion}) highlighting their features and main limitations. %(Section \ref{old-fashion-limits}).
Then Section \ref{sec:openissues} reports the issues that next generation skeleton system should own to address the existing limitations. Finally, the chapter introduces (Section \ref{sec:ourefforts}) our contributions to the field placing them in the proper context, showing how they can be exploited for addressing some of the issues related to structured parallel programming environments.
}

\section{From sequential to parallel architectures}\label{sec:fromseq}
The Von Neumann architecture is a very common and well-known computer design model. It has a very simple formulation and can be described as a sequential process running in a linear address space. It consists in a processing unit and a single separate storage structure to hold both instructions and data. The Von Neumann model ``implements'' a universal Turing machine. It represents the common ``referential model'' of specifying \textit{sequential architectures}, in contrast with \textit{parallel architectures}.
In a parallel architecture many instructions are carried out simultaneously. Parallel computers operate on the principle that large problems can almost always be divided into smaller ones, which may be carried out at the same time.
Parallel architectures exist in several forms and levels. They range from superscalar processors to computational Grids.%: bit-level parallelism, instruction level parallelism, data parallelism, and task parallelism.
In this section we briefly mention some of the most common forms of parallelism, without claiming to be exhaustive but only to give an idea of the variety of the existing forms of parallelism.

\paragraph{Bit-level parallelism} is a form of parallelization based on increasing processor word size. It leads to a reduction of the number of instructions the processor must execute in order to perform an operation on variables whose sizes are greater than the length of the word. (For instance, consider a case where a  16-bit processor must add two 32-bit numbers. The processor must first add the 16 lower-order bits from each number, and then add the 16 higher-order bits, and the carry from the previous add requiring two instructions to complete a single operation. A 32-bit processor would be able to complete the operation using a single instruction).
Historically, 4-bit microprocessors were replaced with 8-bit, then 16-bit, then 32-bit microprocessors. This trend generally came to an end with the introduction of 32-bit processors, which has been a standard in general purpose computing for two decades. Only recently, with the proliferation of processors based both on the IBM PowerPC G5 processor and on the x86-64 architectures, the 64-bit processors have become commonplace.

\paragraph{Instruction-level parallelism} is a form of parallelization based on the simultaneous execution of  instructions part of a computer program. Even if ordinary programs are typically written according to a sequential execution model where instructions execute one after the other and in the order specified by the programmer, in some significant cases there is no need to follow this order. ILP allows the compiler and the processor to overlap the execution of multiple instructions or even to change the order in which instructions are executed. Due to its nature, ILP requires an hardware support; micro-architectural techniques that are used to exploit ILP include (for a better description see \cite{ILParallelism}):
\begin{itemize}
\item Instruction pipelining, where the execution of multiple instructions can be partially overlapped.
\item Superscalar execution, in which multiple execution units are used to execute multiple instructions in parallel. In typical superscalar processors, the instructions executing simultaneously are adjacent in the original program order.
\item Out-of-order execution, where instructions execute in any order that does not violate data dependencies. Note that this technique is orthogonal w.r.t. both pipelining and superscalar.
\item Register renaming, which refers to a technique used to avoid unnecessary serialization of program operations imposed by the reuse of registers by those operations, used to enable out-of-order execution.
\item Speculative execution, which allows the execution of complete instructions or parts of instructions before being certain whether this execution should take place or not. A commonly used form of speculative execution is control flow speculation where instructions following a control flow instruction (e.g., a branch) are executed before the target of the control flow instruction is determined. Several other forms of speculative execution have been proposed and are in use including speculative execution driven by value prediction, memory dependence prediction and cache latency prediction.
\item Branch prediction, which is used to avoid stalling for control dependencies to be resolved. Branch prediction is used with speculative execution.
\end{itemize}

\paragraph{Data parallelism} is a form of parallelization of computer code across multiple processors in parallel computing environments.
%One of the most successful paradigm for parallelism exploitation is the data-parallel programming paradigm [103].
This paradigm is useful for taking advantage of the large amounts of data parallelism that is available in many scientific/numeric applications.
The data parallelism is exploited by performing the same operation on a large amount of data, distributed across the processors of the machine. From the programmer viewpoint, languages based on data-parallel paradigm (such as HPF, sketched in Section \ref{other-appr}) are pretty similar to sequential languages. The main difference is that certain data types are defined to be parallel. Parallel data values consist of a collection of standard, scalar data values.
%
%These languages contain predefined operations on parallel variables that either operate on the parallel variable element-wise (e.g. multiplying every element by a scalar value), or operate on the parallel value as a whole (e.g. summing all elements of the parallel variable).

The data-parallel paradigm has some main virtues that have led to its success. %The first virtue of this model is that data-parallel codes are fairly easy to write and debug. Just as in a serial program, the programmer sees a sequential flow of control The values making up a parallel value are automatically spread across the machine, although typically the programmer does have the option of influencing how data is placed.
Parallel data types are typically static in size (e.g. arrays); their distribution across the machine is usually done at compile-time. Any synchronization or communication that is needed to perform an operation on a parallel value is automatically added by the compiler/run-time system.
The processors collectively compute operations on parallel data values; 
%Operations on parallel data values are collectively computed by the processors; 
computation load usually distributed directly linking data values and computations through the owner computes rule. As data values, computation load is statically distributed across the processors of the system.
%
%The model is easy for a programmer to understand the performance of a program. Given the size of a parallel value to be operated on, the execution time for an operation is likely to be predictable.
%
%Since the execution of each operation is independent of the others, and there are is overhead due to the dynamic management of data values and computations, the execution time for the program as a whole is predictable as well. Faithful performance models can therefore been developed for this kind of languages.
The data parallelism approach typically offers very good scalability. Because operations may be applied identically to many data items in parallel, the amount of parallelism is dictated by the problem size. Higher amounts of parallelism may be exploited by simply solving larger problems with greater amounts of computation.
Data parallelism is also simple and easy to exploit. Because data parallelism is highly uniform, it can usually be automatically detected by an advanced compiler, without forcing the programmer to manage explicitly processes, communication, or synchronization.
Many scientific applications may be naturally specified in a data-parallel manner. In these settings, programs data layout is often fixed; the most used data structures are large arrays. Operations on whole data structures, such as adding two arrays or taking the inner product of two vectors, are common, as are grid-based methods for solving partial differential equations (PDEs).
In spite of this, data parallelism has a significant drawback: the limited range of applications for which data-parallel is well suited. Applications with data parallelism tend to be static in nature; the control flow of a data-parallel program is mostly data independent. Many applications are more dynamic in nature and do not have these characteristics. To run in parallel, these dynamic applications need to %exploit control parallelism by
perform independent operations at the same time. These applications, which may be as simple as recursively computing Fibonacci numbers or as complex as computer chess and n-body simulations, are nearly impossible parallelize using data parallelism.

%Data parallelism focuses on distributing parts of the source data item across different parallel computing nodes. In a multiprocessor system executing a single set of instructions (SIMD), data parallelism is achieved when each processor performs the same task on different pieces of distributed data. In some situations, a single execution thread controls operations on all pieces of data. In others, different threads control the operation, but they execute the same code.

\paragraph{Task parallelism} is a form of parallelization of computer code across multiple processors in parallel computing environments. %Task parallelism focuses on distributing execution processes (threads) across different parallel computing nodes. In a multiprocessor system, task parallelism is achieved when each processor executes a different thread (or process) on the same or different data. The threads may execute the same or different code. In the general case, different execution threads communicate with one another as they work.
Task parallelism focuses on distributing execution processes  across different parallel computing nodes. In the task-parallel paradigm the program consists of a set of (potentially distinct) parallel tasks that interact through explicit communication and synchronization. Task parallelism may be both synchronous and asynchronous. A major advantage of task parallelism is its flexibility. %Because of its emphasis on explicit coordination of individual tasks (or processes, as they are often called), task parallelism can be used to exploit both structured and unstructured forms of parallelism.
Many scientific applications contain task parallelism. For example, in a climate model application the atmospheric and ocean circulation may be computed in parallel. A task-parallel language can express this relationship easily, even if different methods are used for the two circulation models. Another natural application of task-parallel languages is reactive systems in which tasks must produce output in response to changing inputs, in a time-dependent manner. %Tasks may also be organized as a pipeline to exploit pipeline parallelism [54, 55, 53].
Another common structured paradigm exploits parallelism on different data items through task replication. For example, the elaboration of a video stream may involve the filtering on each single frame. In a task-parallel language the filter may be farmed out by spreading different frames on different worker processes, each of them computing the same function.
In the task parallelism approach the interactions between tasks are explicit, thus the programmer can write programs that exploit parallelism not detectable automatically by compiler techniques. %The programmer may also carefully tune the application so that it includes only the communication and synchronization that is actually necessary or efficient, hence reducing reliance on compiler optimization.
In general, task parallelism is less dependent on advanced compiler technology than the data parallelism; in many cases, all that is strictly necessary is the translation of task interactions into appropriate low-level primitives on the target architecture.
%However, compiler technology is still important as a means of guaranteeing correct execution and permitting representations of communication and synchronization that are convenient for the programmer.
A disadvantage of the task-parallel programming model is that it requires extra effort from the programmer to create explicit parallel tasks and
manage their communication and synchronization. % It is also often convenient to consider data owned by different tasks as being part of a single data structure; many task-parallel languages do not support this view directly.
Because communication and synchronization are explicit, changing the manner a program is parallelized may require extensive modifications to the program text.

\bigskip

Due to their nature data and task parallelism (unlike the bit level and instruction level parallelism) cannot be fruitfully exploited using a single CPU system but they are well-tailored for multi-processors or cluster computers, typically referred as parallel computers.

For many years parallel computers has been mainly used in high performance computing, but they have spread in recent years as convenient and effective way to increase the computational power of personal computers and workstations due to physical constraints preventing frequency scaling of CPUs. Hence, parallel architectures are becoming the dominant paradigm in computer architecture, mainly in the form of multicore processors \cite{Asanovic:EECS-2006-183}. Indeed, if a problem requires a huge computational capacity to be rapidly solved and such a power cannot be obtained using a single \textit{processing element} (PE) the only suitable solution is to use many processors simultaneously. Traditionally, parallel architectures have been motivated by numerical simulations of complex systems and ``Grand Challenge Problems'' such as:

\begin{itemize}
\item weather and climate forecasting
\item chemical and nuclear reactions simulations
\item biological, human genome analysis
\item geological, seismic activity analysis
\item mechanical devices and electronic circuits' behavior simulations
\end{itemize}
Today, also commercial applications need the development of faster and faster computers. These applications require to process large amounts of data in sophisticated ways. Example applications include:
\begin{itemize}
\item parallel databases, data mining
\item web search engines, web based business services
\item computer-aided medical diagnosis
\item management of national and multi-national corporations
\item advanced graphics and virtual reality, particularly in the entertainment industry
\item networked video and multi-media technologies
\item collaborative working environments
\end{itemize}
Unfortunately, as we already stated before, using several PEs at the same time introduces some difficulties. Among the others: (i) code and data have to be decomposed and distributed among the computational resources, (ii) work and communications of resources have to be simultaneously coordinated and (iii) fault-tolerance has to be managed.
Thus, the design and implementation of software systems that can ease this burden is very important. Indeed, since the early steps of computer science, researchers conceived and designed programming models, systems and tools aiming at supporting the development of parallel applications. Such systems must find a good balance between the simplicity of the interface presented to the programmers and their implementation efficiency. Finding a good trade-off is a grand challenge. Indeed, a very abstract model simplifies the programming activity but can lead to a very inefficient exploitation of computing resources. Instead, a low-level model allows programmers to use efficiently the computational resources but requires tremendous efforts from the programmers when the number of resources grows.

\section{Parallel programming models: \\State-of-the-art}\label{sec:stateofart}
A good way to organize the state of art of parallel programming models for reporting purpose is to divide them with respect to their level of abstraction. Therefore, in this section we report a selection of the main parallel programming models, proposed by computer scientist over the years, classifying them with respect to the level of abstraction provided to programmers. With respect to this aspect, the parallel programming models can be roughly partitioned in two main classes: the implicit parallel models and the explicit ones.
The former completely cover up parallelism to programmers. Typically, they are exploited by functional and logic languages. The latter ask programmers to deal directly with parallelism. These models can be further partitioned, w.r.t. the abstraction perspective, in three categories: high, medium and low-level programming models.

In the remaining of this section we describe for each category, by way of examples, some programming models and tools belonging to it showing the models Pros \& Cons. In particular, Section \ref{functional-logic-model} describes the functional and logic models as an example of implicit models for parallel programming, Section \ref{dataflow-model} shows the data-flow model as a representative of high-level explicit models. In Section \ref{low-level-model} we outline the low-level approaches describing the OpenMP and MPI frameworks. Then, in Section \ref{other-appr} we report some other notable approaches. Finally, we describe the structured approach  in Section \ref{structured-model}, it is one of the main medium-level models. Here we describe also some our past contributions in the field (Section \ref{JJPF}).

% Section \ref{dataflow-model} describes dataflow programming model with its limitation. Section \ref{macrodataflow-model} reports macro dataflow model, an extension of dataflow created to addresses the problems of traditional approach. Section \ref{structured-MDF-Environments} describes two examples of SPPE that exploit parallelism using macro dataflow model. Section \ref{unstructuredIssue} reports the limitation of traditional SPPE approach. Section \ref{workPerformed} reports the work we done in this area. Section \ref{related-work} illustrate some works related to our work.
%\subsection{Functional and logic approaches to parallel programming}
\subsection{Implicit approaches}
\label{functional-logic-model}
These systems present to programmers a programming model entirely devoid of parallelism and completely isolated from the underlying implementation mechanism. Such systems typically present functional or logical models of computation. They are often referred to as being ``declarative'' systems, since the programmer makes a series of declarations defining the properties of a solution to some problem, rather than specifying a precise series of operations which will lead to the solution. Thus, languages of this type are neither parallel nor sequential, having no notion at all of a flow of control.

Functional languages are based on the lambda calculus. It is a very simple, but powerful language to define expressions and their transformation rules. The only objects present are identifiers, single argument function definitions (``abstractions'') and applications of functions to arguments. A ``program'' consists of a collection of such objects. The program execution is performed applying a top-level function to an argument.
This type of function application is the only operation present and involves the replacement of a function-argument pair with a copy of the function body (from its definition) in which occurrences of the ``free'' variable have been replaced by copies of the actual argument.
This simple system can be shown to provide as much computational power as any other fundamental computing mechanism (e.g. the Turing machine). A particularly powerful aspect of the model is the ability to define ``higher order functions'', namely, functions taking functions as input parameter. Other convenient features such as multiple argument functions, localized definitions and data structures may all be defined as lambda expressions.

In the same way, a high-level functional program is simply a function definition that refers to other functions in its body. A ``call'' of the program involves supplying arguments to this function and ``execution'' consists of using the function definitions (conceptually using the application by substitution technique from the lambda calculus) to obtain an alternative, but equivalent representation of the function and arguments pair, namely a more useful representation of the original program and the ``input''.

The key point of this approach is that execution may progress from the initial to the final representation in any fashion that preserves the equivalence. In particular, it will often be possible to execute many transformation steps concurrently since the conventional problems associated with changes of state have been discarded along with the notions of state and store themselves. A quite common way to represent the program as it evolves is as a graph, in which nodes represent function applications and the children of a node are the (``input'') arguments of the corresponding application. The process of expanding and contracting the graph is referred to as ``graph reduction''.

Exploiting this approach, the parallelization via decomposition is simple. The abstract execution model allows candidate nodes to be expanded at any time, while function applications may be evaluated as soon as arguments are available.%, like in data-flow model (see below).
Thus, a potentially parallel process is generated every time a node reaches one of these states.

It is important to realize that this approach does not imply that every functional program is a highly parallel one. As a trivial, well-known, example, consider defining a function to compute factorials.

The obvious definition will look something like this:

{\small
\begin{eqnarray*}
{\tt factorial}~{\tt 0}~{\tt =}&~{\tt 1}&\\
{\tt factorial}~{\tt n}~{\tt =}&~{\tt n}&\times~~{\tt factorial}~{\tt (n - 1)}
\end{eqnarray*}
}

Such a function would execute in a sequential way on a typical graph reduction machine, irrespective of the number of available processors. A more complex definition notes that

{\small
\begin{eqnarray*}
{\tt factorial}~~{\tt 0}&~{\tt =}&~{\tt 1}\\
{\tt factorial}~~{\tt n}&~{\tt =}&~{\tt product}~{\tt 1}~{\tt n}\\
{\tt product}~~{\tt a}~~{\tt a}&~{\tt =}&~{\tt a}\\
{\tt product}~~{\tt a}~~{\tt b}&~=&~{\tt (product}~{\tt a}~\lfloor~\frac{{\tt a}~{\tt +}~{\tt b}}{{\tt 2}}~\rfloor)~\times~({\tt product}~(\lfloor~\frac{{\tt a}~{\tt +}~{\tt b}}{{\tt 2}}~\rfloor~{\tt +}~{\tt 1})~{\tt b})
\end{eqnarray*}
}

This definition produces significant potential parallelism. Although declarative systems involve no explicit notion of execution sequence, it is unfortunately clear that, in order to optimize the parallel execution programmers must be aware of the execution mechanisms.

%The main problem for the programmer of a functional system comes with the realistic distribution of the available parallelism. Indeed the structure of the graphs produced is specific to each problem instance. Furthermore, this structure only becomes apparent during execution and evolves dynamically. Thus any mapping scheme which tries to distribute the graph and the associated workload effectively must be both dynamic and general purpose.

%This problem can be tackled in two ways. A possibility consists in balancing the work dynamically in a localized manner by allowing idle processors to steal work (effectively portions of the expanding graph) from busy neighbors \cite{workStealing1,workStealing2,workStealing3}. The other way take a more global view: the graph is stored as a globally accessible ``pool of packets'' which in practice is distributed across the local memories of executing elements. An interconnection network deals with accesses to non-local packets. There is a difficult trade-off here between the locality of access and lack of global scheduling of the former method, and the more complicated global access and distribution of the latter.

An alternative approach recognizes the difficulty of automating distribution process and introduces program annotations that programmers exploit to drive the execution mechanism in order to %. These are guaranteed to preserve the semantics of the computation, but may
improve its efficiency. Such additions may be argued to move the model out of this category, in that the programmer is now partially responsible (and aware) for the task of parallel decomposition. Similarly, \cite{PaulKelly} discusses a language which allows program partitioning and interconnection structure to be described in a declarative style.

\medskip

%A functional program contains no explicit notions of communication or synchronization. However, in a realistic implementation these are introduced as a by-product of decomposition, distribution and sharing, and must be handled by the system itself, possibly driven by programmers annotations.

Another category of implicit systems consists in parallel logic languages. They are based on Horn clauses, a restricted form of first order logic. The computational model focuses on the definition and investigation of relationships described as predicates, among data objects described as input arguments to these predicates. As in functional programming, the specification of a computation consists of a collection of predicates and clauses. In the logic model the role of the outermost function application, is played by the outermost predicate together with its arguments.
The arguments interpretation is similar: ``execution'' consists of deciding whether the predicate is true given the arguments and the associated definitions. Furthermore, it is possible to specify the outermost predicate with unbound arguments to find bindings to the arguments that allow the predicate to be satisfied, or to determine that no such bindings exist.

At an abstract level, the process of evaluation may be seen as expanding and searching a tree of possibilities presented by consideration of the various dependencies between appropriate predicates and clauses. As with graph reduction, the semantics of pure logic languages often allow this process to proceed at many points in parallel.
Four principal kinds of (implicitly exploitable) parallelism can be identified in logic programs:
\begin{itemize}
\item \textit{Unification parallelism} arises when arguments of a goal are unified with those of a clause head with the same name and arity. The different argument terms can be unified in parallel as can the different subterms in a term \cite{parUnifi}. Unification parallelism is very fine-grained and has been exploited by building specialized processors with multiple unification units.
\item \textit{Or-parallelism} arises when more than one rule defines some relation and a procedure call unifies with more than one rule head; the corresponding bodies can then be executed in parallel with each other. Or-parallelism is a way of efficiently searching for solutions to the query, by exploring alternative solutions in parallel.
\item \textit{Independent and-parallelism} arises when more than one goal is present in the query or in the body of a procedure, and the run-time bindings for the variables in these goals are such that two or more goals are independent of one another, i.e., their resulting argument terms after applying the bindings of the variables are either variable-free or have non-intersecting sets of variables. Parallel execution of such goals result in and-parallelism.
\item \textit{Dependent and-parallelism} arises when two or more goals in the body of a procedure have a common variable and are executed in parallel. Dependent and-parallelism can be exploited in two ways: (i) the two goals can be executed independently until one of them accesses/binds the common variable. %Note that it is also possible to continue executing the two goals independently in parallel (i.e., executing each without regard to the other goal) even after the common variable has been accessed; in such a case, after the two goals nish, the bindings produced by each will have to be checked for compatibility (this compatibility check at the end is called back uni cation).
(ii) Once the common variable is accessed by one of the goals, if it is bound to a structure, or stream (the goal generating this binding is called the producer), and this structure is read as an input argument of the other goal (called the consumer) then parallelism can be further exploited by having the consumer goal compute with one element of the stream while the producer goal is computing the next element. Case (i) is very similar to independent and-parallelism. Case (ii) is sometimes also referred to as stream-parallelism and is useful for speeding up producer-consumer interactions.
\end{itemize}

\begin{figure}
{\scriptsize
\begin{verbatim}
fib(0, 1).
fib(1, 1).
fib(M, N) :-  [ M1 is M - 1, fib(M1, N1) ],
              [ M2 is M - 2, fib(M2, N2) ],
              N is N1 + N2.
\end{verbatim}
}
\caption{Fibonacci program parallelizable with independent and-parallelism}
\label{logic-example}
\end{figure}

Figure \ref{logic-example} show a simple program for computing the Fibonacci number. The two lists of goals, each enclosed within square brackets above, have no data-dependencies among themselves and hence can be executed independently in parallel with each other. However, the last subgoal N is N1 + N2 depends on the outcomes of the two and-parallel subgoals, and should start execution only after N1 and N2 get bound. Consider that, as in case of functional languages, the programmers in order to exploit the potential application parallelism should give a proper structure to the program.

It should be pointed out that exist some extensions for logic programming language with explicit constructs for concurrency. They can be largely put into three categories:
\begin{itemize}
\item those that add explicit message passing primitives to Prolog, e.g., Delta Prolog \cite{12074} and CS-prolog \cite{163131}. Multiple Prolog processes are run in parallel that communicate with each other via messages.
\item those that add blackboard primitives to Prolog, e.g., Shared Prolog \cite{92417}. These primitives are used by multiple Prolog processes running in parallel to communicate with each other via the blackboard.
\item those based on guards, committed choice, and data-flow synchronization, e.g., Parlog \cite{5390}, GHC \cite{Ued86}, and Concurrent Prolog \cite{39085}.
\end{itemize}
As for the functional languages, the extensions of parallel logic languages move the approach outside the category of implicit parallel programming models.

Similarities between functional and logic styles are emphasized in \cite{alice}.

\paragraph{Summarizing Pros and Cons}
\emph{Implicit parallel programming models provide programmers a very expressive programming metaphor: programmers can implement parallel application without actually deal with parallelism. Unfortunately, this ease is paid in terms of efficiency. In order to address such performance issues researchers introduced some annotation mechanisms and communication primitives, through which programmers can drive the code parallelization. Nevertheless, such additions place the model out of highly abstract systems category because the programmer exploiting annotations is partly responsible and aware for the task of decomposition.}

\subsection{Explicit models} \label{sec:explicit}
The inefficient exploitation of available parallelism caused by the absence of parallel structure in implicit parallel programs is the main reason why explicit parallel programming models exist. These models are based on the assumption that programmers are often the best judges of how parallelism can be exploited for a particular application. Actually, in nearly every case the use of explicit parallelism will obtain a better efficiency than implicit parallelism models.

\subsection{High-level explicit models: data-flow}
\label{dataflow-model}
The models belonging to this category still not require programmers to deal with the several issues related with parallel programming. For instance communications, fault-tolerance, heterogeneity, data decomposition and task granularity. Programmers are only required to write their applications as a set of independent instructions that interact each other through well-known interfaces, so that automatic tools can execute it in parallel. The data-flow model of computation is the main representative of this class of models.

In the data-flow model (for a deep description see \cite{CullerArvind, 1013209, silc98asynchrony, 612574}) the computations are represented by a graph of ``operator'' or ``instruction'' nodes connected by edges along which data items flow. Each node receives by its input edges the data ``tokens'', it performs some simple, stateless, calculation and distributes resultant data tokens on its output edges. A node may only perform its operation once it has received all the data tokens required, from all of its inputs. Thus, each node may compute in parallel, subject only to the availability of data. The processes of associating output tokens with appropriate operator nodes and of deciding which are ready for execution is known as ``matching'' process. %Ready operators must then be selected for actual execution. %These processes are usually separated (at least in principle) from the ``execution units''. Indeed, in most cases, for any realistic problem there will be more operator nodes in the graph than available processors.

Under this paradigm there is no current operation, and each operator is free to execute when all its input tokens are available. The model is naturally concurrent, and the concurrency grain depends on the operations grain.

The data-flow model has the single-assignment property. Values are data tokens that are carried from their producing node to the node that consumes them; there is no concept of a variable with a state that can be arbitrarily updated later. In data-flow, identifiers may be used to name these data tokens. Such identifiers are thus either undefined (not yet produced) or carry a single unique value; they cannot be updated. A node with all input data available is called ``fireable''. When a node is ``fireable'' is ready to be run on a data-flow interpreter. Each data-flow interpreter is called ``actor''. The features of a data-flow model were listed by Ackerman in its 1982 milestone paper \cite{ackerman}. They are:
\begin{itemize}
\item side effects free;
\item locality of effect;
\item equivalence of instruction scheduling with data dependencies;
\item single-assignment semantics;
\item an unusual notation for iterations;
\item lack of history sensitivity in procedures.
\end{itemize}
Synchronization is automatically provided by the token transport mechanism. Parallelism is exploited in data-flow architectures by allowing any actor to execute on any processor and by allowing as many enabled actors to fire as there are processors to execute them. When there are a sufficiently large number of processors, only actors that do not have the input data available are not enabled.

A key feature of the model is that the order of actor execution does not affect the result. Thus, the data-flow model naturally achieves high degrees of parallelism. Nevertheless, traditional data-flow presents three major problems when considered for large distributed (grid) environments.
\begin{itemize}
\item The granularity of traditional data-flow is too small for many distributed architectures, for instance related to distributed memory access time (where latencies are measured in hundreds to thousands of microseconds). The overhead of token transport and actor scheduling and instantiation requires that the granularity of computation be at least hundreds of thousands, and perhaps million of instructions.
%\item Second, traditional data-flow systems have program graphs with a topology that is fixed at compile-time. A static topology implies that the types of objects must be definable at compile-time in order to include in the program graph the subgraphs that implement their function can
\item  The programming abstraction provided to programmers is quite different with respect to the traditional sequential one.
\end{itemize}
The main difference between this approach and those discussed above is that whereas a graph reducer manipulates the graph by modifying both data and the ``instruction code'' itself, a data-flow graph is statically defined by the program and only data is manipulated.

Data-flow based languages %are ``functional'' but
may be dressed up to resemble sequential imperative languages \cite{50455}, particularly in case of ``scientific'' applications. The compilation process from high-level language to the underlying data-flow graph is quite similar to the process of expansion in graph reduction. It is equivalent to the decomposition phase of parallel implementation.

All the problems of distribution, communication and synchronization are associated with the data-flow graph and the interactions between its node operators. Although the structure of the graph is static, it will only be apparent during (or even after) execution that some sections of the graph were more active than others. Thus, a good distribution scheme is difficult to obtain without any additional information, for instance in the form of programmer annotations.

\subsubsection{Macro-Dataflow approaches}
\label{macrodataflow-model}
The macro data-flow model extends the traditional data-flow model addressing its main problems. There are two principal differences with traditional data-flow. First, the granularity of the actors is considerably larger (indeed in this case they are named ``macro'' actors). This allows to achieve a good scalability when the degree of parallelism, namely the number of recruited PEs, increases. Second, some actors \cite{38811} can maintain state information between firings, providing an effective way to model side-effects and non-determinism, these actors are called ``persistent'' actors. %Third, the structure of macro data-flow program graphs is not fixed. Graphs can grow at runtime depending on the elaboration.
Some examples of existing and widely used macro-actors implement high-level functions such as: matrix multiplication, Gaussian elimination or image convolution instead of individual machine instructions. Macro actors can be described as follows.

\paragraph{Regular actors} are similar to actors in the data-flow model. Specifically, all regular actors of a given type are functionally equivalent. A regular actor is enabled and may execute when all of its input tokens are available. It performs some computation, generating output tokens that depend only on its input tokens. It may maintain internal state information during the course of a single execution, but no state information is preserved from one execution to another; regular actors, therefore, represent pure functions.

\begin{figure}
  \center
  \includegraphics[width=250pt]{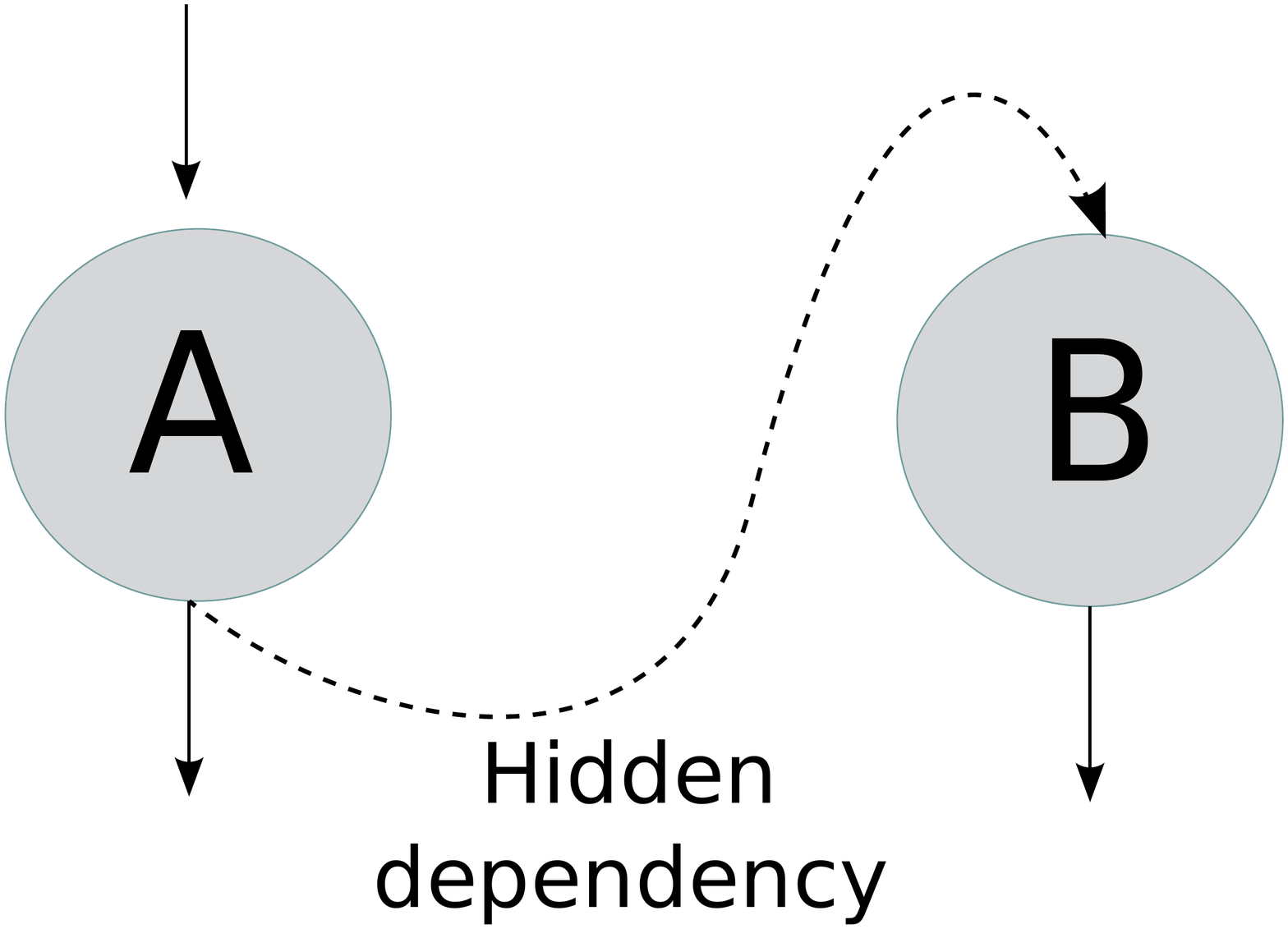}
  \caption{Hidden Dependency}\label{hiddenDep}
\end{figure}

\paragraph{Persistent actors} maintain state information that is preserved from one execution to the next. Output tokens generated by a persistent actor during different executions are not necessarily the same for the same input tokens. The state corresponds to member variables (instance variables) in the object-oriented paradigm. This correspondence implies that several different actors may share the same state, (as an example with the enqueue and dequeue operations on a queue). The model guarantees that the actors that share state will be executed in mutual exclusion, that is, no two actors that share the same state will ever be executing simultaneously. (This can be modeled in stateless data-flow using a single ``state'' token and a non-deterministic merge operator \cite{547755}).
The introduction of state means that the arcs of the program graph no longer model all dependencies in the program; there are implicit dependencies via the shared state. For example, consider the program graph fragment in Figure \ref{hiddenDep}. Suppose that actors A and B share state. If the execution of A occurs first, there is a hidden dependency, based on the state, between A and B. Because of this hidden dependency, the results of the A and B operations depend not only on their arguments and the object history, but also on the order of execution.

\bigskip

If on the one hand the persistent macro actors approach addresses the one limitation of the traditional data-flow model, on the other hand it makes the programming model more complicated and requires to programmers to pay more attention when programming parallel applications. In particular, the introduction of state has one very important consequence: some programs will be deterministic, and others not. Non-determinism is not necessarily bad. There are in fact many ``correct'' non-deterministic applications. Thus, it is the responsibility of the programmer to guarantee higher-level notions of correctness.
Due to the additional complexity they introduce, several existing macro data-flow systems do not support persistent actors.

\subsubsection{A notable MDF approach: the Mentat framework}\label{related-work}

Mentat is one of the most known and used macro data-flow system \cite{grimshaw93mentat}. It is an object-oriented parallel processing system for MIMD architectures developed at the University of Virginia. The computation model used in Mentat is a data-driven macro data-flow computation model based on the object-oriented paradigm.
%The computational grain of the macro data-flow blocks is fixed, it is adjusted at compile time depending on the target architecture. It is  medium sized tradeoff to guarantee both a sufficient computational complexity to amortize overhead costs and a sufficient number of MDF blocks to exploit the available parallelism degree. The mentat macro-actors may maintain state between executions to model the concept of persistent actors.
%
There are two primary components of Mentat: the Mentat Programming Language (MPL) and the Mentat run-time system. MPL is an object-oriented programming language based on C++.
The computational grain of the macro data-flow block is the Mentat class instance, which consists of contained objects (local and member variables), their procedures, and a thread of control. Programmers are responsible for identifying those object classes that are of sufficient computational complexity to allow efficient parallel execution. Instances of Mentat classes are used just like ordinary C++ classes. The data and control dependencies between Mentat class instances involved in invocation, communication, and synchronization are automatically detected and managed by the compiler and run-time system without programmer intervention.

\paragraph{MPL} is an extended C++ designed for developing parallel applications by providing parallelism encapsulation. Parallelism encapsulation takes two forms, intra-object encapsulation and inter-object encapsulation. In intra-object encapsulation of parallelism, callers of a Mentat object member function are unaware of whether the implementation of the member function is sequential or parallel, i.e., whether its
program graph is a single node or a parallel graph. In inter-object encapsulation of parallelism, programmers of code fragments (e.g., a Mentat object member function) need not concern themselves with the parallel execution opportunities between the different Mentat object member functions they invoke.
The basic idea in the MPL is to allow the programmer to specify those C++ classes that are of sufficient computational complexity to warrant parallel execution. Programmers can select which classes should be executed in parallel using a \textbf{mentat} keyword in the class definition. Instances of Mentat classes are called Mentat objects. Mentat classes are very similar to C++ class instance but with some minor differences (described below).
The compiler generates code to construct and execute data dependency graphs in which the nodes are Mentat object member function invocations, and the arcs are the data dependencies found in the program. Thus, it transparently generates inter-object parallelism encapsulation. All the communications and synchronizations are managed by the compiler.
MPL is built around four main extensions to the C++ language. The extensions are Mentat classes, Mentat object instantiation, the return-to-future mechanism, and guarded select/accept statements.

%Intra-Object and Inter-Object Parallelism Encapsulation
A key feature of Mentat is the transparent encapsulation of parallelism within and between Mentat object member function invocations. The hiding of whether a member function implementation is sequential or parallel is called intra-object parallelism encapsulation.
Similarly, the inter-object parallelism encapsulation consists in the exploitation of parallelism opportunities between Mentat object member function invocations in a transparent way to the programmer.
Intra-object parallelism encapsulation and inter-object parallelism encapsulation can be combined. Indeed, inter-object parallelism encapsulation within a member function implementation is intra-object parallelism encapsulation as far as the caller of that member function is concerned. Thus, multiple levels of parallelism encapsulation are possible, each level hidden from the level above.

Not all class objects should be Mentat objects. In particular, objects that do not have a sufficiently high communication ratio, i.e., whose object operations are not sufficiently computationally complex, should not be Mentat objects.
The programmer defines a Mentat class by using the keyword \textbf{mentat} in the class definition. The programmer may further specify whether the class is persistent, sequential, or regular.
Persistent and sequential objects maintain state information between member function invocations, while regular objects do not. Thus, regular object member functions are pure functions. Because they are pure functions, the system is free to instantiate new instances of regular classes at will. Regular classes may have local variables much as procedures do, and may maintain state information for the duration of a function invocation.
The programmer binds Mentat variables to persistent Mentat objects using two reserved member functions for all Mentat class objects: create() and bind().
The create() call tells the system to instantiate a new instance of the appropriate class whereas the bind() function binds Mentat variables to an already existing instance. The member function destroy() destroys the named persistent Mentat object.
The return-to-future function (\textbf{rtf()}) is the Mentat analog to the return of C. Its purpose is to allow Mentat member functions to return a value to the successor nodes in the macro data-flow graph in which the member function appears.
The select/accept statements of Mentat is a guarded statement that derives directly from the ADA \cite{ada-language} one.  Guarded statements permit the programmer to specify a set of entry points to a monitor-like construct. The guards are boolean expressions based on local variables and constants. A guard is assigned to each possible entry point. If the guard evaluates to true, its corresponding entry point is a candidate for execution. The rules vary for determining which of the candidates is chosen to execute. It is common to specify in the language that it is chosen at random. This can result in some entry points never being chosen.
There are two types of guard-actions supported by Mentat: accepts, tests, and non-entries. Accept is similar to the accept of ADA. Tests are used to test whether a particular member function has any outstanding calls that satisfy the guard. When a test guard-action is selected, no parameters are consumed. In Mentat there is no ``else''
clause as in ADA. However, using the priority options, the programmer can simulate one by specifying that the clause is a non-entry statement and giving the guard- statement a lower priority than all other guard-statements. Then, if none of the other guards evaluates to true, it will be chosen.
The priority of the guard-statement determines the order of evaluation of the guards. It can be set either implicitly or explicitly. The token priority determines which call within a single guard-statement priority level will be accepted next. The token priority is the maximum of the priorities of the incoming tokens. Within a single token priority level, tokens are ordered by arrival time.

To give an idea of the programming model in Figure \ref{Mentat-example} we report a simple Mentat program. The program computes recursively the Fibonacci number. It is composed by two classes, the first one recursively computes the Fibonacci number exploiting the second one for computing the sum of partial results. Clearly, in this case the efficiency is low because the amount of computation done by the macro actors computing the mentat object adder\_class is very small.

\begin{figure}
{\scriptsize
\begin{verbatim}
mentat class fibonacci_class {
public:
   int fibonacci_class::fibonacci(int n) {
      fibonacci_class fib;
      adder_class adder;

      // if the index is 0 or 1 it returns 1 to return-to-future function
      if (n == 0 || n == 1)
         rtf(1);
      else  { // otherwise it call the add method and itself recursively
         rtf(adder.add(fib.fibonacci(n - 1), fib.fibonacci(n - 2)));
      }
      return(1);
   }
};

mentat class adder_class {
public:
   int adder_class::add(int arg1, int arg2) {
      // rtf function pass the result to the successor in data-flow graph
      rtf(arg1 + arg2);
      return(arg1 + arg2);
   }
};
\end{verbatim}
}
\caption{Fibonacci computation with Mentat}
\label{Mentat-example}
\end{figure}

Unfortunately, there are a number of issues and limitation that MPL programmers must be aware of that can lead to unpredictable program behavior, related both to Mentat implementation and model. Among the others:
\begin{itemize}
\item The use of static member variables for Mentat classes is not allowed. Since static members are global to all instances of a class, they would require some form of shared memory between the instances of the object.
\item Mentat classes cannot have any member variables in their public definition. If data members were allowed in the public section, users of that object would need to be able to access that data as if it were local. If the programmer wants the effect of public member variables, appropriate member functions can be defined.
\item Programmers cannot assume that pointers to instances of Mentat classes point to the member data for the instance.
\item Mentat classes cannot have any friend classes or functions. This restriction is necessary because of the independent address space of Mentat classes.
\item It must be possible to determine the length of all actual parameters of Mentat member functions, either at compile-time or at run-time. This restriction follows from the need to know how many bytes of the argument to send. Furthermore, each actual parameter of a Mentat member function must occupy a contiguous region of memory
in order to facilitate the marshaling of arguments.
\item Mentat object member function parameter passing is call-by-value. All parameters are physically copied to the destination object. Similarly, return values are by-value.
\item if a Mentat member function returns a pointer, the programmer must explicitly delete the reference when the function is finished using the value.
\item semantic equivalence to the sequential program is not guaranteed when persistent objects are used. This is trivially true for programs that have select/accept statements; there are no serial equivalents.
\end{itemize}

\paragraph{Summarizing Pros and Cons}
\emph{Data-flow model is inherently parallel, it represents each computation as a graph made by operators and instructions where each node can be potentially executed in parallel. This model permit to programmers to express parallel applications in a very abstract way, indeed programmers are not required to deal with low-level issues related to the running architecture. The main problem of Data-flow model is the fine-granularity of instruction that prevent its exploitation in most distributed architectures and in large grid environments. This limitation led to the development of the macro data-flow model (MDF). The MDF model allows programmers to define code fragment in place of instruction as nodes in DF graph. Unfortunately, such additions impair the high-level abstraction, like in case of the implicit models. Hence, programmers have both to deal with data/application decomposition and to assure semantic equivalence with respect to the sequential program, especially when exploiting persistent actors.}

\subsection{Low-level explicit models: MPI and OpenMP}\label{low-level-model}

The low-level approaches provide to the programmers a programming metaphor where parallelism is represented by means of primitives in the form of special-purpose directives or function calls. Most parallel primitives are related to process synchronization, communication or task partitioning.  The total amount of computational cost for executing these primitive is considered as parallelization overhead.
The advantage of explicit parallel programming is the absolute programmer control over the parallel execution. A very skilled parallel programmer takes advantage of explicit parallelism to produce very efficient code. However, programming with explicit parallelism is often difficult and error prone, because of the extra work involved in planning the task division and synchronization of concurrent processes.
In this section we report two of the main approaches to low-level parallel computing: MPI and OpenMP. The former is suitable for distributed architectures whereas the latter is appropriate for multicore and multiprocessor architectures.

\subsubsection{MPI}
MPI is a message-passing library, proposed as a standard by a broadly based committee of vendors, implementors, and programmers. MPI was designed for high performance on both massively parallel machines and on workstation clusters. The Message Passing Interface is meant to provide essential synchronization and communication functionality between a set of processes, mapped into different computer instances, in a language independent way, plus a few features that are language specific. The programming metaphor of MPI is based on the ``process'' concept. An MPI program consists of autonomous processes, executing their own code, in a \textit{Multiple Instructions, Multiple Data stream} (MIMD) style, i.e. Multiple autonomous processors simultaneously executing different instructions on different data. Distributed systems are generally recognized to be MIMD architectures. The processes communicate exploiting MPI communication primitives. Typically, each process executes in its own address space, although shared-memory implementations of MPI are possible. MPI does not specify the execution model for each process. A process can be sequential, or can be multi-threaded, with threads possibly executing concurrently. The intended interaction of MPI with threads is that concurrent threads be all allowed to execute MPI calls, and calls be reentrant; a blocking MPI call blocks only the invoking thread, allowing the scheduling
of another thread. MPI does not provide mechanisms to specify the initial allocation of processes to an MPI computation and their binding to physical processors.
MPI mapping of processes on PEs happens at run-time, through the agent that starts the MPI program, normally called \textbf{mpirun} or \textbf{mpiexec}.

\begin{figure}
  \center
  \includegraphics[width=300pt]{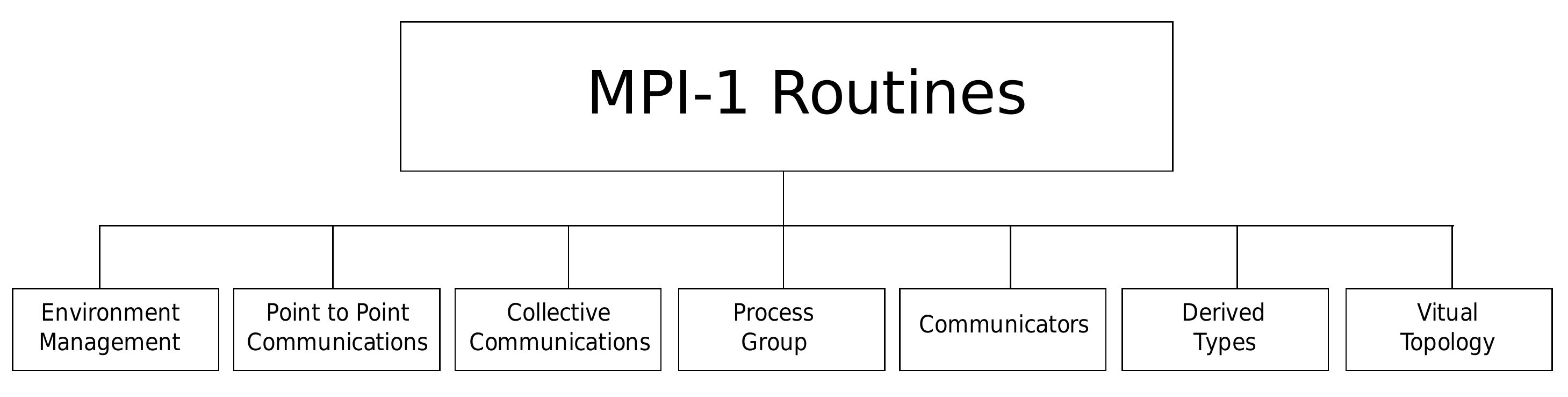}
  \caption{MPI-1 Routines}\label{mpi-1}
\end{figure}

MPI primitives include, but are not limited to, point-to-point rendez-vous type send/receive operations, combining partial results of computations (gathering and reduction operations), choosing between a Cartesian or graph-like logical process topology, exchanging data between process pairs (send and receive operations),  synchronizing nodes (barrier operation) as well as obtaining network-related information such as the number of processes in the computing session, current processor identity that a process is mapped to, neighboring processes accessible in a logical topology, and so on. Point-to-point operations come in synchronous, asynchronous, buffered, and ready forms in order to allow both relatively stronger and weaker semantics for the synchronization aspects of a rendezvous-send. Many outstanding operations are possible in asynchronous mode, in most implementations. Figure \ref{mpi-1} reports the main classes of MPI primitives.
There are two versions of the MPI standard that are currently popular: version 1.2 (also called MPI-1), which emphasizes message passing and has a static run-time environment, and MPI-2.1 (MPI-2), which includes features such as parallel I/O, dynamic process management and remote memory operations.
Figure \ref{MPI-example} show a simple Hello World MPI program. It defines two roles: master and slave. The master ask slaves to process the ``Hello word'' string and then return it. The master eventually print on screen the string received by slaves. The roles are specified by means of the MPI process id. The process number 0 is the master whereas the others are slaves.

As shown in Figure \ref{MPI-example} MPI Hello World programmer is in charge of:
\begin{itemize}
\item initialize MPI
\item find the available resources and manage them
\item implement by hands a way to differentiate the master and the slaves
\item prepare the data the master sends
\item send the data to slaves
\item make the slaves receive the data
\item implement the slave data processing
\item prepare the data the slaves send
\item make the master receive the data, collecting it and processing it
\item finalize MPI
\end{itemize}
furthermore, he must allocate memory buffers, manage fault(s) and distribute data by hands. It is easy to understand that implement a complex application with MPI is a very difficult and error prone task because MPI programmers must manage all the aspects of the application parallelization. On one hand, it guarantees maximum programming flexibility, but on the other hand such a freedom is paid in terms of programming complexity.

\begin{figure}
{\scriptsize
\begin{verbatim}
#include <mpi.h>
#include <stdio.h>
#include <string.h>
#define BUFSIZE 128
#define TAG 0

int main(int argc, char *argv[])
{
  char idstr[32], buff[BUFSIZE];
  int numprocs, myid, i;
  MPI_Status stat;

  /* MPI programs start with MPI_Init; all 'N' processes exist thereafter */
  MPI_Init(&argc,&argv);

  /* find out the number of available PEs */
  MPI_Comm_size(MPI_COMM_WORLD,&numprocs);

  /* and this processes' rank is */
  MPI_Comm_rank(MPI_COMM_WORLD,&myid);

  /* At this point, all the programs are running equivalently, the rank is
     used to distinguish the roles of the programs in the SPMD model  */
  if(myid == 0)
  {
    /* rank 0 process sent a string to all the other processes */
    for(i=1;i<numprocs;i++)
    {
      sprintf(buff, "Hello %d! ", i);
      MPI_Send(buff, BUFSIZE, MPI_CHAR, i, TAG, MPI_COMM_WORLD);
    }

    /* rank 0 process sent a string to all the other processes */
    for(i=1;i<numprocs;i++)
    {
      MPI_Recv(buff, BUFSIZE, MPI_CHAR, i, TAG, MPI_COMM_WORLD, &stat);
      printf("%d: %s\n", myid, buff);
    }
  }
  else
  {
    /* receive from rank 0: */
    MPI_Recv(buff, BUFSIZE, MPI_CHAR, 0, TAG, MPI_COMM_WORLD, &stat);
    sprintf(idstr, "Processor %d ", myid);
    strcat(buff, idstr);
    strcat(buff, "reporting for duty\n");

    /* send to rank 0: */
    MPI_Send(buff, BUFSIZE, MPI_CHAR, 0, TAG, MPI_COMM_WORLD);
  }

  /* MPI Programs end with MPI Finalize */
  MPI_Finalize();
  return 0;
}
\end{verbatim}
}

\caption{Hello Word example implemented using MPI}
\label{MPI-example}
\end{figure}

\subsubsection{OpenMP}
Like MPI, OpenMP (Open Multi-Processing) is a specification defined by a group of major computer hardware and software vendors for multi-platform multiprocessing  programming. It consists of a set of compiler directives, library routines, and environment variables that influence run-time behavior. Unlike MPI, it is mainly targeted to shared memory multiprocessing. Indeed, it is used in conjunction with MPI on distributed architectures made of multicore/multiprocessor machines.
OpenMP uses multiple, parallel threads to accomplish parallelism. A thread is a single sequential flow of control within a program. OpenMP uses a directive-based method to tell explicitly to the compiler how to distribute programs across parallel threads.

\begin{figure}
  \center
  \includegraphics[width=300pt]{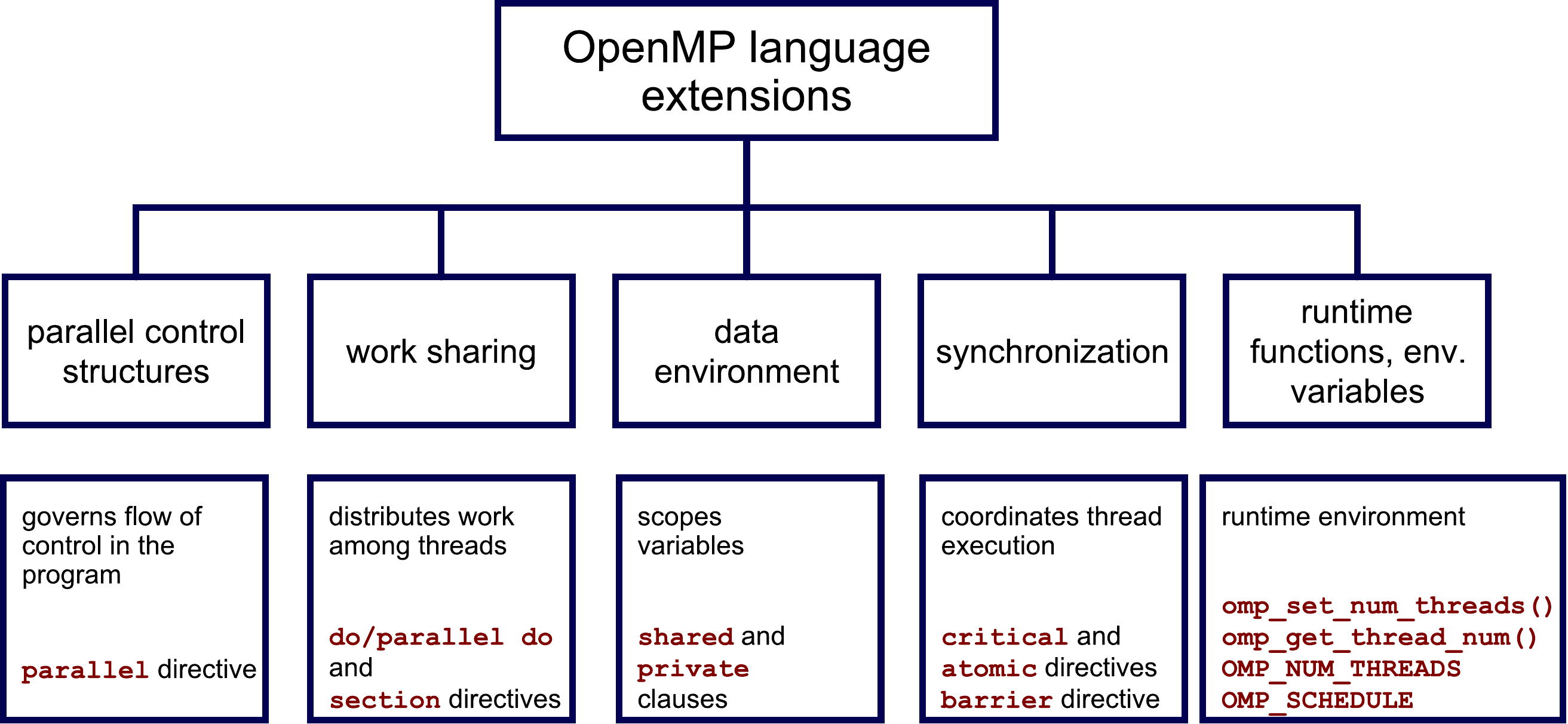}
  \caption{OpenMP language extensions}\label{openMP}
\end{figure}
The core elements of OpenMP are the constructs for thread creation, workload distribution (work sharing), data environment management, thread synchronization, user level run-time routines and environment variables. OpenMP programmers exploit such constructs to manage all the aspects of application parallelization. Figure \ref{openMP} shows the classes of existing OpenMP language extensions.

Even if the OpenMP approach to parallel programming has to be considered as a low-level one, OpenMP code is more straightforward than MPI code. This is mainly due to the memory model indeed, relying on a shared memory model. The OpenMP application does not need to deal with message passing hence data are not directly split and divided among PEs but handled through compiler directives.

An OpenMP program is a C++ or Fortran program with OpenMP pragma statements/directives placed at appropriate points. The pragma statement directs the compiler how to process the block of code that follows the pragma. An OpenMP-enabled compiler recognizes the pragma directives and produces a parallelized executable suitable for running on a shared-memory machine. In C/C++, an OpenMP directive has the general form:

\[ {\tt \#\ pragma\ \ omp\ \ directive-name\ [clause,...]\ newline} \]

The \#pragma omp directive tags a block for parallel or various types of work sharing execution, variable scoping and synchronization considerations. One or more clauses are optional and may be in any order. The clauses are used to explicitly define the scoping of enclosed variables.
In OpenMP there are two main constructs:
\begin{itemize}
\item A parallel region is a block of code that will be executed by multiple threads. This is the fundamental parallel construct.
\item A work-sharing construct divides the execution of the enclosed code region among the members of the team that encounter it. Work-sharing constructs do not launch new threads. These constructs are identified by DO/FOR, SECTIONS and WORKSHARE (Fortran only) directives.
\end{itemize}

Since OpenMP is a shared memory programming model, most variables in OpenMP code are visible to all threads by default. However, sometimes private variables are necessary to avoid a race condition and there is a need to pass values between the sequential part and the parallel region. Another important issue is the synchronization and scheduling of the threads. These are managed through clauses appended to the OpenMP directive. Thus, the different types of clauses are Data Scoping, Synchronization and Scheduling clauses.

\begin{figure}
{\scriptsize
\begin{verbatim}
int main (int argc, char *argv[]) {
        int nthreads, tid, i, chunk;
        float a[N], b[N], c[N];
	
        /* Some initializations */
        for (i=0; i < N; i++) a[i] = b[i] = i * 1.0;
        chunk = CHUNKSIZE;
	
        #pragma omp parallel shared(a,b,c,nthreads,chunk) private(i,tid)
        {
                tid = omp_get_thread_num();
                if (tid == 0) {
                        nthreads = omp_get_num_threads();
                        printf("Number of threads = %d\n", nthreads);
                }
                printf("Thread %d starting...\n",tid);
                #pragma omp for schedule(dynamic,chunk)
                for (i=0; i<N; i++) {
                        c[i] = a[i] + b[i];
                        printf("Thread %d: c[%d]= %f\n",tid,i,c[i]);
                }
        } /* end of parallel section */
}
\end{verbatim}
}
\caption{Factorial example in OpenMP}
\label{openMP-example}
\end{figure}

In Figure \ref{openMP-example} we report an OpenMP example program. The example uses two pragma directives. The outer \#pragma omp parallel tags a block for parallel execution. The \textbf{shared()} clause specifies common variables, and \textbf{private()} specifies the variables restricted to exclusive use by a process. The inner \#pragma omp for schedule directive specifies distribution across threads. The threads share the variables a, b, c and chunk; the iteration variable i is private in each thread. The expression tells the compiler to perform parallel execution of the for-loop and to split the iteration space into blocks of size chunk.

The current version of OpenMP presents some issues, some related to the implementation and others related to the model. For instance a reliable error handling,  fine-grained mechanisms for controlling thread-processor mapping or synchronization among a subset of threads. The model related issues, clearly more difficult to overcome include inefficient parallelism exploitation in distributed-memory platforms and a limited scalability that actually depends by memory architecture.

\paragraph{Summarizing Pros and Cons}
\emph{Low-level approaches allow programmers to control all the aspects of parallel applications and their execution. Exploiting low-level approaches skilled programmers can implement very efficient parallel applications. The freedom and efficiency allowed by the model are paid in terms of expressiveness and ease of use. Indeed, programmers have to manage ``by hand'' all the issues related to data and program decomposition, fault tolerance, load balancing and communications.}

\subsection{Other notable approaches}\label{other-appr}
Other two noteworthy explicit parallel approaches are Cilk and High Performance Fortran. 

The first one is quite similar to OpenMP, indeed it consists in an enriched version of C language, it requires that the computing resources share the main memory hence can be used for programming parallel applications running in multiprocessor machines but not in distributed architecture like clusters. It enriches GNU C with a few Cilk-specific keywords. Using them  programmers expose the parallelism identifying elements that can safely be executed in parallel. Using such information the run-time environment, in particular the scheduler,  decides during execution how to distribute the work among processors.
The first Cilk keyword is \textbf{cilk}, which identifies a function written in Cilk. Since Cilk procedures can call C procedures directly, but C procedures cannot directly call or spawn Cilk procedures, this keyword is needed to distinguish Cilk code from C code. Other keywords are: spawn, sync, inlet and abort.
The first two keywords are all Cilk programmers have to use to start using the parallel features of Cilk: \textbf{spawn} indicates that the procedure call it modifies can safely operate in parallel with other executing code. Note that from the point of view of the scheduler it is not mandatory to run this procedure in parallel; the keyword only inform the scheduler that it can run the procedure in parallel. sync indicates that execution of the current procedure cannot proceed until all previously spawned procedures have completed and returned their results to the parent frame. The two remaining Cilk keywords are slightly more advanced, and concern the use of inlets. Typically, when a Cilk procedure is spawned, it can only return its results to the parent procedure by putting those results in a variable in the parent's frame, as we assigned the results of our spawned procedure calls in the example to x and y. The alternative is to use an \textbf{inlet}. An inlet is a function internal to a Cilk procedure that handles the results of a spawned procedure call as they return. One major reason to use inlets is that all the inlets of a procedure are guaranteed to operate atomically with regards to each other and to the parent procedure, thus avoiding the bugs that could occur if the multiple returning procedures tried to update the same variables in the parent frame at the same time. 
The \textbf{abort} keyword can only be used inside an inlet; it tells the scheduler that any other procedures that have been spawned off by the parent procedure can safely be aborted. 

High Performance Fortran is an extension of Fortran 90 defined by the high performance fortran forum with constructs that support data-parallel computations. It consists in a portable language for data-parallel computations. HPF uses a data parallel model of computation to support spreading the work of a single array computation over multiple processors. This allows efficient implementation on both SIMD and MIMD style architectures. It provides a number of basic data parallel functions as built-in array operators and intrinsic functions. It also provides constructs, such as the \textbf{where} and the \textbf{forall}, which assist in programming more complex data parallel functions.
The simplest data parallel operations are the elementwise operations. For any base operation on a data type, programmers can extend that operation to an array operation. For binary (and higher degree) operations, the arrays must have the same shape. The result of the operation is another array of that shape, in which the elements are defined by the elementwise extension of the base operation.
A more advanced set of operations operate on an entire array to produce a single answer, they implement a behavior generally known as reduction. Reduction can be defined for any associative, binary operation that produces a result of the same element type by successively accumulating the results of applying that operation to elements of the array. Commonly used operations include arithmetic operators like addition, multiplication, maximum, and minimum and boolean operators. As an example, HPF programmers  can define reduction with addition, usually called sum reduction, over any array whose element type can be added.

\subsection{Structured approach}\label{structured-model}
Highly abstract approaches and low-level approaches represent the two extremes in parallel programming models. The formers completely automate the aspects of parallelization, namely do not ask programmers (at least in their ``pure'' version) to give any information about application, like data distribution and synchronization, communication mechanisms, executing environment or code sequences to run in parallel. The latter, opposite, approaches do not automate anything and ask programmers to deal, almost entirely, with the application parallelization aspects.

As we outlined in previous sections, several researchers have tried to address the limitation of these approaches enriching them with additional features.
Some other work was done trying to conceive alternative models. In particular, since the nineties, several research groups have proposed the \emph{structured parallel programming environments}(SPPE). Since the structured parallel programming model was conceived, several works have been done about it, also from a foundational point of view  \cite{lithium:sem:CLSS}, \cite{aldinuc:sem:parco2003}, \cite{128874}. Programming environments relying on this paradigm (i.e. \cite{772854}) ask programmers to explicitly deal with the \emph{qualitative} aspects of parallelism exploitation, namely the application structure and problem decomposition strategies. All the low-level parallelism exploitation related aspects like communication, synchronization, mapping and scheduling are managed by compiler tools and run-time support.

The \emph{structured way} is driven by two observations: that there are some things people do better than compilers, and that there are some things that compilers do better than people. Rather than have either do the complete job, it exploits the comparative advantages of each. Indeed the management of tens to thousands of asynchronous tasks, where timing-dependent errors are quite common, is beyond the capacity of most programmers whereas compilers are very good at ensuring that events happen in the right order and can more readily and correctly manage communication and synchronization than programmers. On the other hand, data decomposition strategies and \textit{computational grain} can be successful managed by programmers but not efficiently by compilers.

The environments following this way are those based on the algorithmic skeleton concept. A skeleton, is a known and widely used pattern of parallelism exploitation originally conceived by Cole \cite{128874} and later on by different research groups to design high-performance structured parallel programming environments.

Basically, structured parallel programming systems allow a parallel application to be coded by properly composing a set of basic parallel skeletons. These basic skeletons usually include skeletons modeling embarrassingly parallel computations (farms), computations structured in stages (pipelines) as well as common data parallel computation patterns (map/forall, reduce, scan). Each skeleton is parametric; in particular, it accepts as a parameter the kind of computation to be performed according to parallelism exploitation pattern it models.
As an example, a farm skeleton takes as a parameter the worker, i.e. the computation to be performed on the single input task (data item). As a further example, a pipeline takes as parameters the pipeline stages. Such parameters may be either parameters modeling sequential portions of code (sequential skeletons) or even other skeletons, in turn. Therefore, a farm skeleton may take as a worker a two stage pipeline. The composition of the two expresses embarrassingly parallel computations where each input task (data item) is processed by two stages. Parallelism is exploited both by using different resources to compute independent input tasks and by using different resources to compute the first and the second stage onto a single input task.

A skeleton (in its original formulation) is formally an higher order function taking one or more other skeletons or portions of sequential code as parameters, and modeling a parallel computation out of them. Cole's skeletons represent parallelism exploitation patterns that can be used (instanced) to model common parallel applications. Later, different authors figure out that skeletons can be used as constructs of an explicitly parallel programming language, actually as the only way to express parallel computations in these languages \cite{darlington:parle:93, orlando-grosso}. Recently, the skeleton concept evolved, and became the coordination layer of structured parallel programming environments (\cite{van:assist:02, skie:PC:1999, Darlington1996}).
In any case, a skeleton can be considered as an abstraction modeling a common, reusable parallelism exploitation pattern.
Skeletons can be provided to the programmer either as language constructs \cite{orlando-grosso, Darlington1996, skie:PC:1999} or as libraries \cite{teti-fgcs, mlws, stigliani:europar:00, kuchen:europar:2002}. Usually, the set of skeletons includes both data-parallel and task parallel patterns.

\subsubsection{Traditional skeleton approaches}\label{old-fashion}
From the nineties, several research groups proposed or currently propose programming environments supporting parallel computations based on the algorithmic skeleton concept.  They are implemented as frameworks, languages or libraries. Among the others, we mention Kuchen's C++ MPI skeleton library \cite{kuchen:europar:2002}, Serot's SKiPPER environment, $P^3L$, Lithium, a first version of \muskel and JJPF. In particular, the last one, JJPF, represents our approach to traditional SPPE.
In the rest of this section we present a more detailed description about the programming model of $P^3L$, \muskel and JJPF to describe the ``concept behind'' SPPE models. We developed this last one, whereas all the other skeleton environments presented in this section have been developed by the Parallel and Distributed Architecture Group, part of the Department of Computer Science at University of Pisa. This group has a deep background on skeleton environment, indeed the group began to work in this field from the very beginning the skeleton model were conceived. We collaborated with several researchers belonging to this group, also for the conception and the design of the results presented in this thesis.

\paragraph{\textbf{$P^3L$}}
is a high-level structured explicitly parallel language developed in the nineties \cite{p3l}. Using $P^3L$ parallelism can be expressed only by means of a restricted set of parallel constructs each corresponding to a specific parallel form. Sequential parts are expressed by using an existing language also called the host sequential language of $P^3L$. Being a SPPE its constructs can be hierarchically composed to express more complex parallel forms. This compositional property relies on the semantics associated with the various $P^3L$ constructs and their compositions. In fact, each of them can be thought of as a data-flow module. In $P^3L$ each module computes in parallel or sequentially a function on a given stream of input data and produces an output stream of results. The lengths of both the streams are identical and the ordering is preserved, i.e.
\begin{center}
$[in_1,...,in_n] \rightarrow M \rightarrow [out_1,...,out_n]$
\end{center}
\noindent where $M$ is the data-flow module corresponding to a generic $P^3L$  construct  $[in_1,...,in_n]$ is the input stream, $[out_1,...,out_n]$ is the output stream, $n$ is the length of both the streams and every output data item $out_i$ is obtained by applying the function computed by $M$ on the input data item $in_i$. The types of the input and the output interface of each $P^3L$ construct i.e. the types of every $in_i$ and every $out_i$ have to be declared statically. Actually the compiler performs type checking on these interfaces when the $P^3L$ constructs are to be composed. Another feature of $P^3L$ is its interface with the host sequential language. The interface has been designed to make easier portability between different host languages. In fact, sequential parts are completely encapsulated into the constructs of $P^3L$.
Parameter passing between $P^3L$ constructs are handled by linguistic constructs that are external to the specific host sequential language while the data types that can be used to define the interface of the $P^3L$ constructs are a fixed subset of those usually available in the most common languages.
The first $P^3L$ compiler adopted as host sequential language C and C++. The constructs included since the first $P^3L$ compiler were
\begin{itemize}
\item{The \textbf{farm} construct} which models processor farm parallelism. In this form of parallelism a set of identical workers execute in parallel the independent tasks that come from an input stream and produce an output stream of results.
\item{The \textbf{map} construct} which models data parallel computations. In this form of parallelism each input data item from an input stream is decomposed into a set of partitions and assigned to identical and parallel workers. The workers do not need to exchange data to perform their data parallel computations. The results produced by the workers are recomposed to make up a new data item of an output stream of results.
\item{The \textbf{pipe} construct} which models pipeline parallelism. In this form of parallelism a set of stages execute serially over a stream of input data producing an output stream of results.
\item {The \textbf{loop} construct} which models computations where for each input data item a loop body has to be iteratively executed until a given condition is reached and an output data item is produced.
\item{The \textbf{sequential} construct} which corresponds to a sequential process that for each data item coming from an input stream produces a new data item of an output stream
\end{itemize}
The sequential constructs constitute the leaves of the hierarchical composition because the computations performed by them have to be expressed in terms of the host sequential language. %Hence the structure of a $P^3L$ program can be expressed using a macro data-flow graph.

\paragraph{\muskel}

\cite{muskel:qos:pdp:05} is a full Java framework, providing programmers with structured ways of expressing parallel programs. The muskel environment represents a sensible evolution of the Lithium one \cite{teti-fgcs}. It inherits from Lithium the \textit{normal form} \cite{653465} and macro data-flow \cite{Da01PPL, 772854} implementation techniques as well as the general structure of the run-time support.

\textit{Normalization} consists in transforming the original skeleton tree (or composition) into a program that is basically a task farm with sequential workers \cite{pdcs:nf:99}. Such optimization basically substitute skeleton subtrees by skeleton subtrees providing a better performance and efficiency in the target machine resource usage than the original skeleton tree. Previous results demonstrated that full stream parallel skeleton subtrees can be collapsed to a single farm skeleton with a (possibly huge) sequential worker leading to a service time which is equal or even better that the service time of the uncollapsed skeleton tree \cite{128874}.

The \muskel macro data-flow run-time support consists in deriving a graph of macro data-flow blocks from skeleton trees and dispatching them to computational resources running macro-actors.

\muskel adds to Lithium a limited form of resource discovery and fault tolerance features as well as the whole \textit{Application Manager} concept.

The \textit{Application Manager}(AM) is an entity that takes care of assuring that the application non-functional requirement were satisfied. The requirements are specified by programmers in a performance contract. The AM actively observes the application behavior and in case of faults or performance contract violations it reacts aiming to fix the problem, as an example, in case of a computational resource fault it recruits a new resource in the computation.

Using \muskel a programmer can implement parallel programs that match the task farm or the pipeline parallelism exploitation patterns as well as arbitrary composition of the two. Despite the limited amount of patterns supported, however, a large range of applications can be programmed, for instance all embarrassingly parallel applications, parameter sweeping applications and multistage applications.

A task farm computation can be defined just using a Farm object. The Farm constructor takes a parameter representing the computation performed by the farm workers. This computation can be either a sequential computation or another parallelism exploitation pattern (another Farm or a Pipeline one). A pipeline computation can be defined using a Pipeline object. The Pipeline constructor takes two parameters that can either be sequential computation objects or in turn parallel exploitation patterns. Pipelines with more stages can be obtained composing several Pipeline objects.
Then the programmer has to add an \textit{Application Manager} to the application code, and he must also specify the performance contract he pretends to be respected on the target architecture. This is done instantiating an application manager and specifying a performance contract.
\muskel supports two different kinds of contracts. The first one requires a constant parallelism degree, that is, it requires that a constant number of processing elements are dedicated to the parallel execution of our parallel program. The second one requires that a given throughput is maintained in terms of task processed per unit time.
Both of these kinds of contracts can be specified before the computation of the parallel \muskel program actually starts and can be changed during the program execution. The management of the parallel
computation in such a way that the contracts are satisfied is completely handled by an independent execution flow. Therefore, the submission of a new performance contract to the application manager
immediately triggers all those (possibly additional) activities needed to satisfy the contract. The possibility to change the performance contracts during the execution of the parallel applications allows the programmer to implement some kind of application dependent dynamic execution strategy.
Once the program has been specified along with its performance contract the programmer must supply the list/stream of tasks to be computed. When all the elements belonging to the list/stream have been processed, the parallel execution of the program is terminated and the relative results can be fetched.

During the computation of the parallel program the \muskel run-time automatically discovers available processing elements. In case there are no enough resources to satisfy the contract, an error is signaled to the programmer.

As we stated before, in case of faults the Application Manager recruits new resources among the available ones to substitute the faulty one. In case the application manager recognizes that the performance contract specified by the programmer cannot be satisfied, it raises an Exception.
Being any task to be computed a fireable macro data flow instruction, it is completely independent of any other task needed to compute the parallel application. Therefore, it can be scheduled on any one of the available resources. However, the normal form concept implemented in \muskel, only generates fully independent macro data flow instructions. That is, no result of an instruction is needed to compute another instruction. In this case, most of the scheduling problems we just mentioned disappear.

\begin{figure}[t]
\begin{center}
\includegraphics[width=0.95\linewidth]{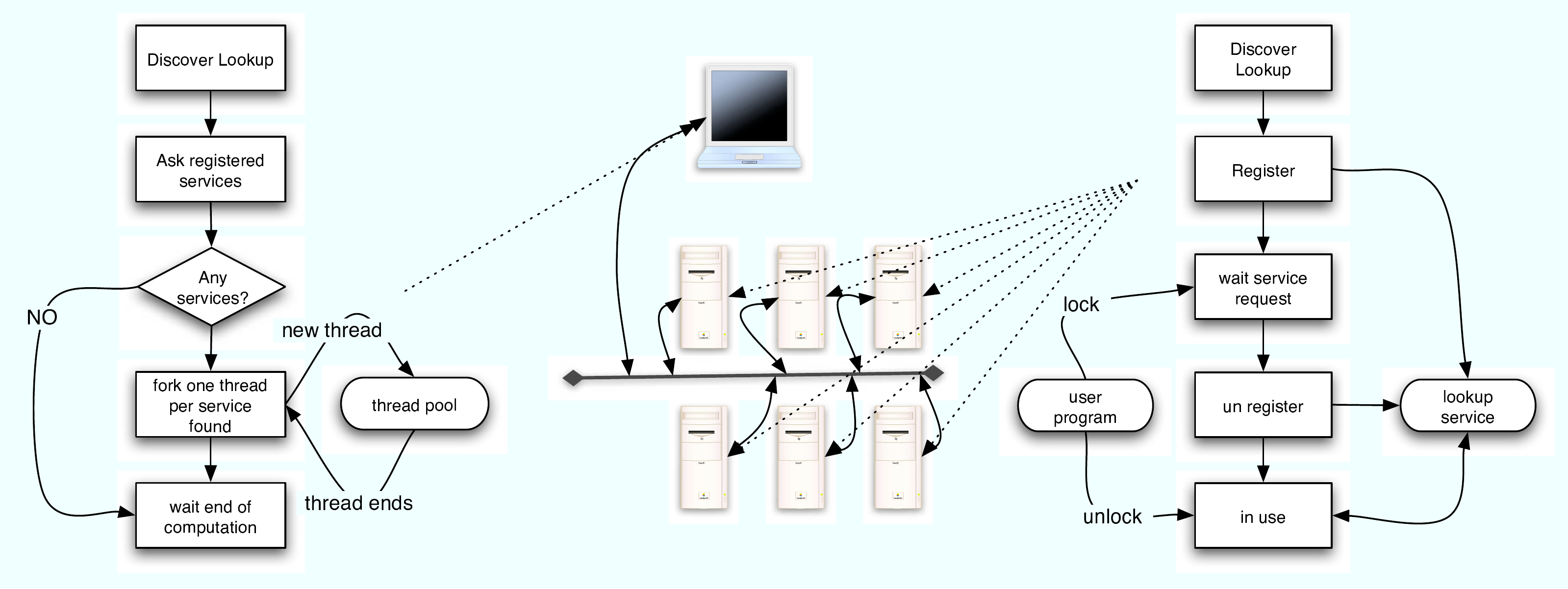}
\end{center}
\caption{Simplified state diagram for the generic {\sf JJPF}\ \textit{client} (left) and \textit{service} (right)}
\label{fig:servizioEcliente}
\end{figure}

\paragraph{JJPF}\label{JJPF} is a parallel programming framework built on top of plain Java that can run stream parallel applications on several parallel/distributed architectures ranging from tightly coupled workstation clusters to generic workstation networks and grids. In a sense, JJPF represents our approach to old-fashioned structured parallel programming environments.
It directly inherits from the early versions of Lithium and \muskel \cite{teti-fgcs}. Both Lithium and \muskel exploit plain RMI Java technology to distribute computations across nodes, and rely on NFS (the network file system) to distribute the application code to the remote processing elements.  JJPF, instead, is fully implemented on top of JINI and Java and relies on the Jini Extensible Remote Invocation (JERI) mechanism to distribute code across the remote processing nodes involved in stream parallel application computation.
JJPF exploits the stream parallel structure of the application in such a way that several distinct goals can be achieved:
\begin{itemize}
\item \textit{load balancing is achieved} across the computing elements participating in the computation
\item processing elements available to participate to the computation of stream parallel application are \textit{automatically discovered and recruited} exploiting standard Jini mechanisms
\item \textit{faulty processing elements are automatically substituted} by fresh ones (if any) in a seamless and automatic way. Therefore, the stream parallel applications computations resist to both node and network faults. Programmers do not need to add a single line of code in his application to deal with faulty nodes/network, nor it has to take any other kind of action to get advantage of this feature.
\end{itemize}
JJPF has been tested using both synthetic and real applications, on both production workstation networks and on clusters, with very nice and encouraging results.
{\sf JJPF} has been designed to provide programmers with an environment supporting the execution of stream parallel applications on a network of workstations, exploiting plain Java technology. Overall JJPF provides a distributed server providing a stream parallel application computation service. Programmers must write their applications in such a way they just exploit an arbitrary composition of task farm and pipeline patterns. Task farm only applications are directly executed by the distributed server, while applications exploiting composition of task farm and pipeline patterns are first processed to get their \textit{normal form}. % \cite{pdcs:nf:99},
A distributed environment that exploits task parallel computations, permits to implement different applications in really different applicative and hardware contexts.
{\sf JJPF} is based on a master-worker architecture. JJPF defines two entities: ``client'', that is the application code (the master), and ``service'', that consists in distributed server instances (the workers) that actually compute results out of input task data to execute client program.
Figure \ref{fig:servizioEcliente} sketches the structure of the two components. The client component basically recruits available services and forks a control thread for each one of them. The control thread, in turn, fetches uncomputed task items from the task vector, delivers them to the remote service and retrieves the computed results, storing them to the result repository. Low-level activities, like resource recruiting, program deployment and data transfer are performed directly by the framework exploiting the JINI technology \cite{jini}.
The key concept in {\sf JJPF}\ is that service discovery is automatically performed in the client run time support. Not a single line of code dealing with service discovery or recruiting is to be provided by application programmers.
{\sf JJPF}\ achieves automatic load balancing among the recruited services, due to the scheduling adopted in the control threads managing the remote services. Furthermore, it handles faults in service nodes automatically taking care of the tasks assigned to a service node in such a way that in case the node does not respond any more they can be rescheduled to other service nodes.
This is only possible because of the kind of parallel applications that are supported in {\sf JJPF}, that is stream parallel computations. In this case, there are natural \textit{descheduling points} that can be chosen to restart the computation of one of the input tasks, in case of failure of a service node.
JJPF has demonstrated good scalability both in embarrassingly parallel application and in more ``problematic'' applications.

\section{Open issues in structured approaches}\label{sec:openissues}
Despite being around since long time and despite the progress made in skeletal system design and implementation, the skeleton systems did not take off as expected. Nowadays, the skeleton system usage is actually restricted to small communities grown around the teams that develop the skeleton systems. Cole focused very well the problem in his manifesto \cite{cole:manifesto:02}. Here he stated four principles that have to be tackled in skeletal systems to make them effective and successful:

\paragraph{I) Propagate the concept with minimal conceptual disruption}
It means that skeletons must be provided within existing programming environments without actually requiring the programmers to learn entirely new programming languages. In order to make them widely used by practitioners they should not require further conceptual baggage.

\paragraph{II) Integrate ad-hoc parallelism}
Many parallel applications are not obviously expressible as instances of skeletons. Some have phases that require the use of less structured interaction primitives. For example, Cannon's well-known matrix multiplication algorithm \cite{156619} invokes an initial step in which matrices are skewed across processes in a manner which is not efficiently expressible in many skeletal systems. It is unrealistic to assume that skeletons can provide all the parallelism we need. We must construct our systems to allow the integration of skeletal and ad-hoc parallelism in a well-defined way.

\paragraph{III) Accommodate diversity}
All the existing skeleton systems have a common core of simple skeletons and a variety of more exotic forms. When described informally, the core operations are straightforward. Instead, precise specification reveals variations in semantics that reflect the ways skeletons are applied in real algorithms. The result is that some algorithms, which intuitively seem to represent an instance of a skeleton, cannot be expressed in certain systems because of constraints imposed by the specification. Hence, skeletal systems should provide mechanisms to specialize skeletons, in all those cases where specialization does not radically change the nature of the skeleton, and consequently the nature of the implementation.

\paragraph{IV) Show the pay-back}
A new technology will only gain acceptance if it can be demonstrated that adoption offers some improvement over the status quo. The structural knowledge embedded in skeletons should allow optimization within and across uses that would not be realistically achievable by hand, i.e. demonstrate that the effort required to adopt a skeletal system is immediately rewarded by some kind of concrete results: shorter design and implementation time of applications, increased efficiency, increased machine independence of the application code, etc.

\bigskip

The second and the third points are specifically technical whereas the first and the last one are actually a kind of ``advertising'' ones, in a sense. All these points, however, have impacts on both the way the skeleton systems are designed and on the way they are implemented. The Cole's analysis is not the only one, \cite{advske:pc:06} extends it adding some other features a skeleton environment have to address to be suitable for the computational grids.
In particular, the authors present three more requirements for Skeletal systems:

\paragraph{V) Support code reuse}
that is allow programmers to reuse with minimal effort existing sequential code;

\paragraph{VI) Handle heterogeneity} i.e. implement skeletons in such a way skeleton programs can be run on clusters/networks/grids hosting heterogeneous computing resources (different processors, different operating systems, different memory/disk configurations, etc.);

\paragraph{VII) Handle dynamicity} i.e. implement in the skeleton support mechanisms and policies suitable to handle typical dynamic situations, such as those arising when non-dedicated processing elements are used (e.g. peaks of load that impair load balancing strategies) or from sudden unavailability of processing elements (e.g. network faults, node reboot).

Summarizing, the next generation of Skeletal Systems, that drawing a parallel with web programming model we can refer as ``Skeletons 2.0'', have to integrate ad-hoc parallelism and provide mechanisms to specialize skeletons in order to express customized form of parallel exploitation. They have to support code reuse, handle heterogeneity and dynamicity in order to be exploited in grid environments. Moreover, such features must be provided with minimal conceptual disruption, hence without requiring the programmers to learn entirely new programming languages or environments but integrating ``Skeletons 2.0'' principles inside the existing programming tools, possibly without changing their programming abstraction.

Some Skeletal systems have addressed the ``Skeletons 2.0'' principles to different degrees in different combinations. Next section reports some of the most notable among these systems.

\subsection{Attempts to address issues}

In its ``manifesto'' paper Murray Cole, together with the check-list of issues that next generation of skeleton system should address, sketches the eSkel library \cite{cole:manifesto:02}. eSkel consists in Cole's attempt to address the issues he present in his ``manifesto'' paper.
More in detail, eSkel is a library of C functions and type definitions that extends the standard C binding to MPI with skeletal operations. Its underlying conceptual model is the SPMD distributed memory model, inherited from MPI, and its operations must be invoked from within a program that has already initialized an MPI environment. eSkel provides programmers with some language primitives performing complex operations that can be integrated with the traditional MPI functions. eSkel implements skeletons as collective MPI operations. In \cite{cole:manifesto:02,BenoitCGH05} authors describe how the manifesto issues are addressed in eSkel. eSkel also provides some code reuse facilities (check-list point V) as most C and C++ code can simply be adapted in eSkel programs. In eSkel heterogeneous architectures are supported (VI) through the usage of MPI, much in the sense heterogeneous architectures are supported through the usage of Java in muskel. However, current implementation of eSkel does not support custom, programmer defined, MPI data types in the communication primitives, that actually use MPI\_INT data buffers, and therefore heterogeneous architectures can be targeted using proper MPI implementations just when all the nodes have the same type of processors. No support for dynamicity handling (VII) is provided in eSkel, however.

Some other groups involved in structured parallel programming research, developed programming systems that partially address the issues above presented.
Schaeffer and his group at the University of Alberta that implemented a system were programmers can insert new parallelism exploitation patterns in the system \cite{bromling:parco:2001}. Kuchen Muesli \cite{muesli-home} is basically a C++ library built on top of MPI providing stream parallel skeletons, data parallel objects and data parallel operations as C++ template classes. The programming interface is definitely very good, as the full power of object oriented paradigm along with templates is exploited to provide Muesli programmers with user-friendly skeletons, and consequently C++ programmers can develop parallel applications very rapidly. In particular, Muesli does not require any MPI specific knowledge/action to write a skeleton program. Therefore, point (I) is very well addressed here. Points (II) and (III) are addressed providing the programmer with a full set of (data parallel) operations that can be freely combined. The payback (IV) is mainly related to the OO techniques exploited to provide skeletons. Code reuse (V) is supported as it is supported in eSkel, as programmers can use C++/C code to build their own skeletons as well as sequential code to be used in the skeletons. Even in this case there is limited support to heterogeneity (VI): the MPI code in the Skeleton library directly uses MPI\_BYTE buffers to implement Muesli communications, and therefore MPI libraries supporting heterogeneous architectures may be used just in case the nodes sport the same kind of processor and the same C/C++ compiler tool-set. Dynamicity handling (VII) is not supported at all in Muesli.

Gorlatch's and its research group presented a grid programming environment HOC \cite{gorlatch:hoc:dagstuhl:05}, which provides suitable ways of developing component based grid applications exploiting classical skeleton components. The implementation exploits Web Services technology. Overall, the HOC programming environment addressed principles (I) and (IV). Points (II) and (III) rely on the possibility given to programmers to insert/create new HOCs in the repository. Point (VI) is handled via Web Services. This technology is inherently multiplatform, and therefore heterogeneous target architectures can be easily used to run HOC programs. Point (V) is guaranteed as sequential code can easily (modulus the fact some XML code is needed, actually) be wrapped in Web Services. However, no support to (VII) is included in the current HOC version.

%Below we outline two of them: eSkel and our customized version of \muskel. The former has been developed by the Cole's group, the latter is a modified version of the original ``old-fashioned'' \muskel environment. It represent our approach to next generation skeletal environments.

\section{Our efforts in designing ``Skeletons 2.0''\\ systems}\label{sec:ourefforts}
Even though Cole and other research groups, focused on skeleton system, designed and developed skeleton systems that own some of the features required to be a next generation skeleton system, the research for addressing the presented issues is just started. In fact, up to now tools and model that are generally recognized as the best solutions for addressing the issues presented in \cite{cole:manifesto:02} and in \cite{advske:pc:06} simply do not exist. In the Chapters \ref{skeleton_customization}, \ref{muskelWorkflow} and \ref{mdf_as_components} we present some models and the concerning tools that we designed and developed in order to contribute to research for next generation skeleton systems.

More in detail, in Chapter \ref{skeleton_customization} we propose a \textit{macro data-flow based approach} designed supporting the integration of unstructured form of parallelization in skeleton systems, hence addressing the issue number II. To validate the approach we modified a skeleton system that in its original form does not deal with unstructured parallelism: \muskel. We extended \muskel, in collaboration with the research staff that develop it, to integrate it with a methodology that can be used to implement mixed parallel programming environments providing the programmer with both structured and unstructured ways of expressing parallelism.
The methodology is based on data-flow. Structured parallel exploitation patterns are implemented translating them into data-flow graphs executed by a distributed macro data-flow interpreter. Unstructured parallelism exploitation can be achieved by explicitly programming data-flow (sub)graphs. The modified \muskel provides suitable ways to interact with the data-flow graphs derived from structured pattern compilation in such a way that mixed structured and unstructured parallelism exploitation patterns can be used within the same application.
%
%The new \muskel builds a sort of data-flow abstraction, defined using the \textit{normal form} model, starting from a structured parallel program. Then it use a master that send data-flow code and data to a set of macro data-flow actors (each one running on a different Processing Element) that run the code using data received as input. The master is used also to assure that the whole computation follow the specifications defined in the application performance contract.
%
Two mechanisms provided to the \muskel programmers for unstructured parallelism exploitation. First, we provide primitives that allow accessing the fundamental features of the data-flow graph generated out of the compilation of a skeleton program. Namely, methods to deliver data to and retrieve data from data-flow graph. We provide to programmers the ability to instantiate a new graph in the task pool by providing the input task token and to redirect the output token of the graph to an arbitrary data-flow instruction in the pool. Second, we provide the programmer with direct access to the definition of data-flow graphs, in such a way he can describe his particular parallelism exploitation patterns that cannot be efficiently implemented with the available skeletons.
The two mechanisms can be jointly used to program all those parts of the application that cannot be easily and efficiently implementing using the skeletons subsystem.
%
%
%\paragraph{Metaprogramming Run-time Optimizations}
%
Unfortunately, this approach is not free from shortcomings In fact exploiting unstructured parallelism interacting directly with data-flow graph requires to programmers to reason in terms of program-blocks instead of a monolithic program. Hence, at a first sight this approach may look like  the ones present in the other early macro data-flow models. Nevertheless, we want to point out that the effort required to customize an application made by a composition of existing skeleton is not comparable with the complexity of developing it from scratch as a set of macro data-flow blocks. % Moreover, for such a customization, graphical tools can be exploited to aim the task of linking the blocks. Actually, several tools exist that are tailored to a very similar purpose, namely the workflow graphical designer. Indeed, they are intended to provide mechanisms to link/unlink code blocks (or components). We tested the suitability of graphical tools as an operative way to customizing skeletal application developing such a tool and integrating it in \muskel.

In order to ease the generation of macro data-flow blocks, and therefore provide programmers with a easier way to express program-blocks, we exploited some \textit{metaprogramming techniques} that are successfully used for code transformation in fields like web development and component based programming \cite{fraclet, Spring, ejb3}. Exploiting these techniques the programmers are no longer requested to deal with complex application structuring but simply give hints to the metaprogramming support using high-level directives. The directives are used by the support to drive the application transformation. Chapter \ref{muskelWorkflow} presents our efforts aimed at providing metaprogramming tools and models for ease the generation of macro data-flow blocks and their run-time optimization.
In particular, two results are presented. The first is ``Parallel Abstraction Layer'' (PAL). A java annotation \cite{javaAnnotation} based metaprogramming framework that restructures applications at bytecode-level at run-time in order to make them parallel. The parallelization is obtained asynchronously executing the annotated methods. Each method call is transformed in a macro data-flow block that can be dispatched and executed on the available computing resources.
PAL transformations depend on the resources available at run-time, the programmers hints and the available \textit{adapters}. An adapter is a specialized entity that instructs the PAL transformation engine to drive the code transformation depending on the available parallel tools and frameworks. %Experimental results shows that the PAL approach is effective and efficient in order to handle resource heterogeneity and dynamicity.  Actually, runtime code transformation brings to a very good exploitation of computational resources.
The other result presented in the chapter concerns the integration of the Aspect Oriented Programming \cite{aop1, AOP} mechanisms (more in detail the AspectJ framework \cite{aspectj}) with our modified \muskel skeleton framework. The first step in this direction was exploiting AspectJ to implement \textit{aspect driven program normalization} (see \cite{pdcs:nf:99}) in \muskel. The second step consisted in testing the integration of \muskel with AspectJ to in a more complex scenario. Hence, we exploited the aspect oriented programming support integrated in \muskel in order to develop \textit{workflows} which structure and processing are optimized at run-time depending on the available computational resources. Let us point out that we introduced metaprogramming techniques for easing the generation of macro data-flow blocks (in particular to address the issue number I) but as a corollary we obtained the possibility to optimize the application and adapt it at run-time with respect to the executing environment (addressing the issues number III and VI).

The other two main issues to address are the support for code reuse (V) and the handling of dynamicity (VII).
As we already discussed when we introduced \muskel, it addresses this last point through the definition of the \emph{Application Manager}. The dynamicity handling is a very important feature for next generation parallel programming systems, especially for the ones designed for computational Grids.
Actually, \muskel framework, at least in its original form, is designed to be exploited in cluster and network of workstations rather than in Grids. Indeed, some of its features limit its exploitation on Grids, in particular:
\begin{itemize}
\item \muskel communicates with the resources it recruits exploiting the RMI protocol, that (at least in its original version) uses TCP ports that are typically blocked by firewall;
\item the computational resources are found by \muskel exploiting multicast communications that are often blocked by firewall;
\item the recruitment of a computational resource requires to \muskel programmers to run a proper application on the resource, hence to have an account on it;
\item the Application Manager is a centralized entity. This represents a twofold limitation in Grid environment: it is a  single point of failure and a bottle-neck that curb the scalability of the approach.
\end{itemize}
We addressed most of these limitations exploiting ProActive Parallel Suite \cite{proactive} to implement the macro data-flow distributed interpreters (see the experimental results presented in Chapter \ref{skeleton_customization}). ProActive
provides mechanisms to tunnel RMI communications and ease the deployment of Grid applications. Indeed, it has been successfully used for developing applications in the Grid5000 \cite{grid5000} platform. ProActive support for Grids has became more complete since it began to support the component based development, in particular the support for the CoreGrid Grid Component Model \cite{gcm:coregrid:07}. Indeed, several studies recognized that component technology could be leveraged to ease the development of Grid Application  \cite{armstrong99toward, 383872} and a few component based model have been proposed by parallel computing scientific community for programming Grids \cite{cca, DenPerPriRib, gcm:coregrid:07}.
Component-based software development can be considered an evolutionary step beyond object-oriented design. Object-oriented techniques have been very successful in managing the complexity of modern software, but they have not resulted in significant amounts of cross-project code reuse. Furthermore, sharing object-oriented code is difficult because of language incompatibilities, the lack of standardization for inter-object communication, and the need for compile-time coupling of interfaces. Component-based software development addresses issues of language independence (seamlessly combining components written in different programming languages) and component frameworks define standards for communication among components. Finally, the composition compatibility is evaluated providing a meta-language specification for their interfaces.
The GCM represents one of the main European scientific community efforts for designing and developing \cite{gridcomp} a grid component model. We contributed to the design of GCM and its reference implementation together with the research group that developed \muskel and with several European research groups. In particular, we focused our contribution, in the context of the CoreGrid Programming model virtual institute, on GCM autonomic features. Therefore, by designing the autonomic features of GCM components, each component is able to react dynamically to changes in the executing environment. We referred to the \muskel application manager approach, generalizing and extending the approach to make it suitable for components based models. Indeed, each GCM component with a complete support of autonomic features has an \textit{Autonomic Manager} that observes the component behavior. In case the behavior turns out to be different from the one expected the manager trigger a component reconfiguration.
%che tra l'altro abbiamo contribuito a sviluppare ben separando la parte funzionale da quella non-funzionale, anche qui portando l'esperienza dello sviluppo nella separation of concerns.
%Quello che ci vuole è che le diverse parti eseguite in modo distribuito sulla griglia siano capaci di autogestirsi. Parti che in una architettura tipo griglia
%
In other words, GCM autonomic features provide programmers with a configurable and straightforward way to implement autonomic grid applications. Hence, they ease the development of application for the Grids. Nevertheless, they rely fully on the application programmer's expertise for the set-up of the management code, which can be quite difficult to write since it may involve the management of black-box components, and, notably, is tailored for the particular component or assembly of them. As a result, the introduction of dynamic adaptivity and self-management might enable the management of grid dynamism, and uncertainty aspects but, at the same time, decreases the component reuse potential since it further specializes components with application specific management code.
In Chapter \ref{mdf_as_components}, we propose \emph{Behavioural Skeletons} as a novel way to describe autonomic components in the GCM framework. Behavioural Skeletons aim to describe recurring patterns of component assemblies that can be (either statically or dynamically) equipped with correct and effective management strategies with respect to a given management goal. Behavioural Skeletons help the application designer to i) design component assemblies that can be effectively reused, and ii) cope with management complexity by providing a component with an explicit context with respect to top-down design (i.e. component nesting). We consider the Behavioural Skeletons,  coupled with the CoreGRID Grid Component, a good structured parallel programming model for handling dynamicity (VII), supporting reuse both of functional and non-functional code (V). The model defines characters as the \textit{Skeleton designers} and the \textit{Expert users} that can design new skeletons and customize the existing ones (II and III), whereas, standard \textit{users} can easily (I) exploit the existing ones.

\chapter{Mixing Structured and Macro-Dataflow approaches}\label{skeleton_customization}
%\sloppy
% %\begin{chapterabstract}
% Programming models based on algorithmic skeletons promise to raise the level of abstraction perceived by programmers when implementing parallel applications, while guaranteeing good performance figures. In the meanwhile, however, they restrict the freedom of programmers to implement arbitrary parallelism exploitation patterns. In fact, efficiency is achieved by restricting the parallelism exploitation patterns provides to the programmers to the useful ones for which efficient implementations, as well as useful and efficient compositions, are known.
% In this chapter we present a modified version of the  \muskel framework, a Java framework targeting clusters and networks of workstations providing to programmers a skeleton based parallel programming environment.
% \muskel is implemented exploiting (macro) data flow technology, rather than more usual skeleton technology relying on the usage of implementation templates. Using data-flow, \muskel efficiently implements both classical, predefined skeletons, and programmer-defined parallelism exploitation patters. This allows to overcome part of the problems that Cole identified in his skeleton ``manifesto'' as the problems impairing skeleton success in the parallel programming arena.

\paragraph{Chapter road-map} %perspective
\emph{
In this chapter we describe our contribution to skeleton customization. We start with an introduction on structured programming model outlining its main advantages and recalling its main limitations. In particular, we focus on the skeleton customization issue. Namely the lack of flexibility of skeletal systems in expressing parallel form different from the ones ``bundled'' with the skeleton framework. Then we briefly introduce the data-flow approach we conceived to address of this limitation and we report related work: alternative approaches addressing the structured parallel programming limitations (Section \ref{sec:introDFmuskel}).% (Section \ref{sec:relatedMuskelMDF}).
Besides, we introduce classical implementation template and more recent data-flow technologies as used to design and implement skeleton systems (Section \ref{sec:templ-dataflow}).
Then, we describe the details of our contribution, i.e. our extended version of \muskel framework, discussing how skeletons customization is supported exploiting data-flow implementation (Section \ref{sec:unstruc}). Finally, we report the experimental results we obtained exploiting our customized \muskel (Section \ref{sec:results}).
}
%\end{chapterabstract}

\section{Data-flow enables skeleton customization}\label{sec:introDFmuskel}
We already introduced structured parallel programming models in the previous chapter, where we described their Pros and Cons. Let us to briefly recall here their main features and limitations.

Structured parallel programming models provide the programmers with native high-level parallelism exploitation patterns that can be instantiated, possibly in a nested way, to implement a wide range of applications
\cite{cole:manifesto:02,kuchen:europar:2002,kuchen-optim,skie:PC:1999,teti-fgcs}.
In particular, those programming models hide to programmers ``assembly level'' of parallel programming, i.e. by avoiding a direct interaction with the distributed execution environment via communication or shared memory access primitives and/or via explicit scheduling and code mapping. Rather, the high-level native, parametric parallelism exploitation patterns provided encapsulate and abstract from all these parallelism exploitation related details.
%
%As an example, to implement an embarrassingly parallel application processing all the data items in an input stream or file, the programmers simply instantiates a ``task farm'' skeleton by providing the code necessary to (sequentially) process each input task item. The system, either a compiler and run time tool based implementation or the library based one, takes care of devising the proper distributed resources to be used, to schedule proper tasks on the resources and to distribute input tasks and gather output results according to the process mapping used.
%
In contrast, when using a traditional parallel programming system, the programmers have usually to explicitly program code for distributing and scheduling the processes on the available resources and for moving input and output data among the involved processing elements.
The cost of this appealing high-level way of dealing with parallel programs is paid in terms of programming freedom. The programmer (or skeleton system user) is normally not allowed to use arbitrary parallelism exploitation patterns, but he must only use the ones provided by the system. They usually include all those reusable patterns that have efficient distributed implementations available.
This is mainly aimed at avoiding the possibly for the programmers to write code that can potentially impairs the efficiency of the implementation provided for the available, native parallel patterns.
This is a well-known problem (See chapter \ref{parallel_issues}). %Cole recognized its importance in his ``manifesto'' paper \cite{cole:manifesto:02}.

In this Chapter we discuss the methodology we conceived, designed and used to modify the \muskel parallel programming environment in order to provide to programmers the possibility to mix structured and unstructured ways of expressing parallelism while preserving most of the benefits typical of structured parallel programming models.
The methodology is based on the macro data-flow model. Structured parallel exploitation patterns are implemented translating them into macro data-flow graphs executed by the %scalable, efficient, 
distributed macro data-flow interpreters. Unstructured, user-defined parallelism exploitation patterns are achieved by explicitly programming data-flow graphs. These (macro) data-flow graphs can be used in the skeleton systems in any place where predefined skeletons can be used, thus providing the possibility to seamlessly integrate both kind of parallelism exploitation within the same program.
% %
%The system provides suitable ways to interact with the data flows graphs derived from structured pattern compilation in such a way that mixed structured and unstructured parallelism exploitation patterns can be used within the same application.
%
The mechanisms enabling data-flow graphs customization provide programmers the possibility to program new parallelism exploitation patterns.

The methodology has been developed together with the other authors of \cite{muskelJournal}, we all contributed in a substantially equal way to the conception, design and implementation of the approach.

%
%We are currently extending our skeleton framework with other tools that allow to expand the non-functional features of the skeleton set, thus providing the programmers new possibilities to customize the skeleton set used. In particular we are considering the possibility to use Java 1.5 annotations and AOP (Aspect-Oriented Programming) techniques to associate to the skeletons different non-functional properties such as security or parallelism exploitation related properties.

%\ref{sec:templ-dataflow}\ introduces classical implementation template and more recent data flow technologies as used to design and implement skeleton systems.
%Section \ref{sec:struct}\ introduces \muskel, our extensible skeleton system and \S \ref{sec:unstruc}\ describes how expandability of the skeleton set is implemented %as well as the currently undergoing \muskel\ extensions related to non-functional features introduced via annotations and/or AOP techniques.
%Section \ref{sec:results}\ presents experimental results. %and eventually \S \ref{sec:relatedMuskelMDF}\ discussed related work.

\bigskip

%\section{Related work}
%\label{sec:relatedMuskelMDF}
Macro data-flow implementation for algorithmical skeleton programming environment was introduced in late '90 \cite{MDF:parco:99} and then has been used in other contexts related to skeleton programming environments \cite{772854}.
%Cole suggested in \cite{cole:manifesto:02} that ``we must construct our systems to allow the integration of skeletal and ad-hoc parallelism in a well defined way'', and that structured parallel programming environments should ``accommodate diversity'', that is ``we must be careful to draw a balance between our desire for abstract simplicity and the pragmatic need for flexibility''. Actually, his eSkel \cite{eskel:europar:05,eskel-site} MPI skeleton library addresses these problems by allowing programmers to program their own peculiar MPI code within each process in the skeleton tree. Programmers can ask to have a stage of a pipeline or a worker in a farm running on $k$ processors. Then, the programmer may use the $k$ processes communicator returned by the library for the stage/worker to implement its own parallel pipeline stage/worker process. As far as we know, this is the only attempt to integrate ad hoc, unstructured parallelism exploitation in a structured parallel programming environment. The implementation of eSkel, however, is based on process templates, rather than on data flow.

Cole eSkel, we already presented in the previous chapter, addresses these problems by allowing programmers to program their own peculiar MPI code within each process in the skeleton tree. Programmers can ask to have a stage of a pipeline or a worker in a farm running on $k$ processors. Then, the programmer may use the $k$ processes communicator returned by the library for the stage/worker to implement its own parallel pipeline stage/worker process. As far as we know, this is the only attempt to integrate ad hoc, unstructured parallelism exploitation in a structured parallel programming environment. The implementation of eSkel, however, is based on process templates, rather than on data flow.

Other skeleton libraries, such as Muesli \cite{kuchen:europar:2002,kuchen-optim,muesli-home}, provide programmers with a quite large flexibility in skeleton programming following a different approach. They provide a number of data parallel data structures along with elementary, collective data parallel operations that can be arbitrary nested to get more and more complex data parallel skeletons. However, this flexibility is restricted to the data parallel part, and it is anyway limited by the available collective operations.

CO2P3S \cite{shaeffer-europar00} is a design pattern based parallel programming environment written in Java and targeting symmetric multiprocessors. In CO2P3S, programmers are allowed to program their own parallel design patterns (skeletons) by interacting with the intermediate implementation level \cite{bromling:parco:2001}. Again, this environment does not use data flow technology but implements design patterns using proper process network templates.

JaSkel \cite{DBLP:conf/ccgrid/FerreiraSP06} provides a skeleton
library implementing the same skeleton set than \muskel. In JaSkel,
however, skeletons look much more implementation templates, according
to the terminology used in Section \ref{sec:templ-dataflow}. However, it looks like the programmer can exploit the full OO programming methodology to specialize the skeletons to his own needs. As the programmer is involved in the management of support code too (e.g. he has to specify the master process/thread of a task farm skeletons) JaSkel can be classified as a kind of ``low-level, extensible'' skeleton system, although it is not clear from the paper whether entirely new skeletons can be easily added to the system (actually, it looks like it is not possible at all).

%There are several works proposing aspect-oriented techniques for parallel programming. \cite{bruno04}%DBLP:conf/aosd/HarbulotG04} discusses an approach using AOP to separate concerns in scientific code. In \cite{Sobral:aosd-acp4is06,SobralIpdps2006} a usage of AOP is proposed aimed at separating the concerns of partitioning and distributing data and performing concurrent computations. This is far from the usage we think to make of AOP techniques in this work, however, in that it requires a much more ``template oriented'' approach w.r.t. the one followed in \muskel.

\section{Template based vs. data-flow based skeleton systems}
\label{sec:templ-dataflow}
A skeleton based parallel programming environment provides programmers with a set of predefined and parametric parallelism exploitation patterns. The patterns are parametric in the kind of basic computation executed in parallel and, possibly, in the execution parallelism degree or in some other execution related parameters. As an example, a pipeline skeleton takes as parameters the computations to be computed at the pipeline stages. In some skeleton systems these computations can be either sequential computations or parallel ones (i.e. other skeletons) while in other systems (mainly the ones developed at the very beginning of the skeleton related research activity) these computations may only be sequential ones.

\begin{figure}
\centerline{\includegraphics[scale=0.48]{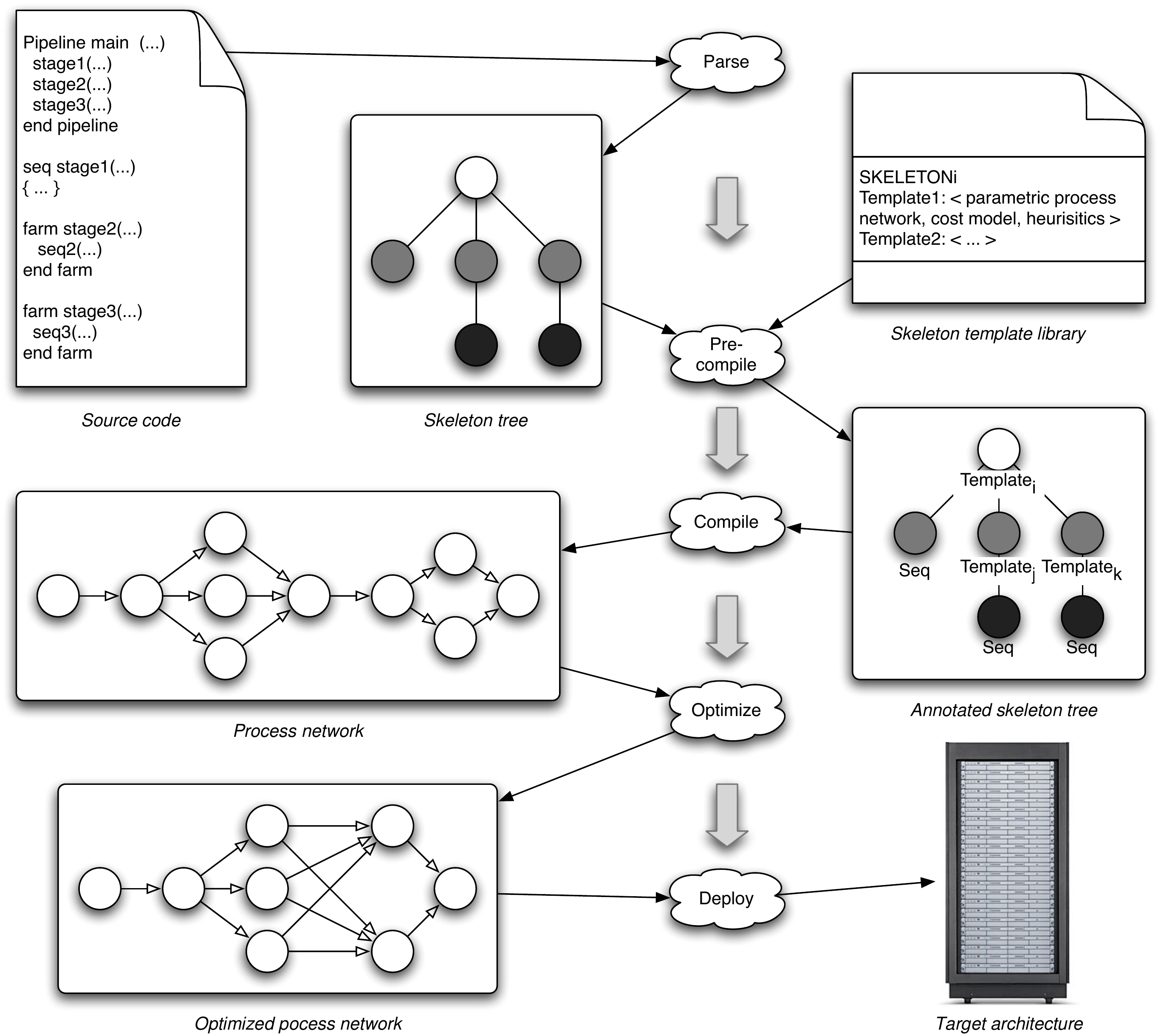}}
\caption{Skeleton program execution according to the implementation template approach.}
\label{fig:template}
\end{figure}

The first attempts to implement skeleton programming environments all relied on the implementation template technology. %Original Cole skeletons \cite{cole-th}, Darlington's group skeleton systems \cite{darlington:parle:93,darli-to-2,darli-to-3}, Kuchen's Muesli \cite{kuchen:europar:2002,muesli-home}\ and our group P3L \cite{orlando-grosso}\ and ASSIST \cite{van:assist:02}\ all use this implementation schema.
As discussed in \cite{libro-susanna}, in a implementation template based skeleton system each skeletons is implemented using a parametric process network picked up among the ones available for that particular skeleton and for the kind of target architecture at hand in a template library (see \cite{kuchen-farm}, discussing several implementation templates, already appeared in bibliography, all suitable to implement task farms, that is embarrassingly parallel computations implemented according to a master-worker paradigm). The template library is designed once and for all by the skeleton system designer and summarizes his knowledge concerning implementation of the parallelism exploitation patterns modeled by skeletons.
Therefore, the compilation process of a skeleton program, according to the implementation template model, can be summarized as follows:
\begin{enumerate}
\item the skeleton program is parsed, a skeleton tree is derived, hosting the precise skeleton structure of the application. The skeleton tree has nodes marked with one of the available skeleton, and leaves marked with sequential code (sequential skeletons).
\item the skeleton tree is traversed, in some order, and templates from the library are assigned to each one of the skeleton nodes, but the sequential ones, that always correspond to the execution of a sequential process on the target machine. During this phase, parameters of the templates (e.g. the parallelism degree or the kind of communication mechanisms used) are fixed, possibly exploiting proper heuristics associated to the library entries
\item the enriched skeleton tree is used to generate the actual
  parallel code. Depending on the system that may involve a
  traditional compilation step (e.g. in P3L when using the Anacleto
  compiler \cite{anacleto-australia} or in ASSIST when using the
  \textbf{astcc} compiler tools
  \cite{assist:imp:europar:03,assist:parco:03}) or exploiting proper
  parallel libraries (e.g. in Muesli \cite{muesli-home} and eSkel
  \cite{eskel-site} exploiting MPI within a proper skeleton library hosting templates
\item the parallel code is eventually run on the target architecture, possibly exploiting some kind of loader/deploy tool.
\end{enumerate}
Figure \ref{fig:template} summarizes the process leading from a skeleton source code to the running code exploiting template technology.

\begin{figure}
\centerline{\includegraphics[scale=0.48]{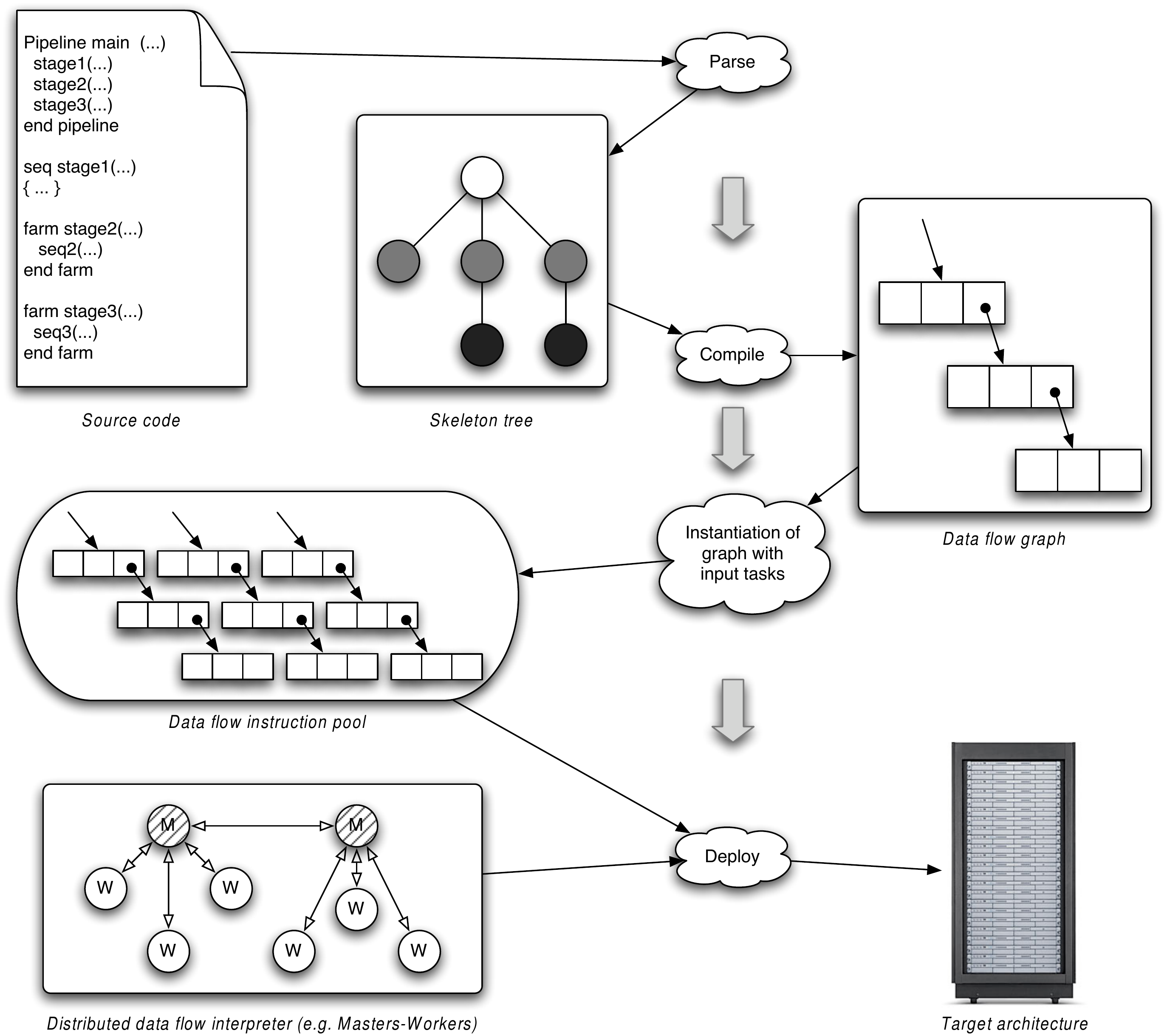}}
\caption{Skeleton program execution according to the data-flow approach.}
\label{fig:dataflow}
\end{figure}

More recently, an implementation methodology based on data-flow has been proposed \cite{MDF:parco:99}. In this case the skeleton source code is used to compile a data-flow graph and the data-flow graph is then executed on the target architecture exploiting a suitable distributed data-flow interpreter engine. The approach has been used both in the implementation of Lithium \cite{tesiteti,teti-fgcs} and in Serot's SKIPPER skeleton environment \cite{serot02}. In both cases, the data-flow approach was used to support fixed skeleton set programming environments. We adopted the very same implementation approach to develop our version of the  \muskel framework, modifying it in collaboration with the original developers, enriching it with a data-flow implementation to support extensible skeleton sets.

When data-flow technology is exploited to implement skeletons, the compilation process of a skeleton program can be summarized as follows:
\begin{enumerate}
\item the skeleton program is parsed, a data-flow graph is derived. The data-flow graph represents the pure data-flow behavior of the skeleton tree in the program
\item for each one of the input tasks, a copy of the data-flow graph is instantiated, with the task appearing as an input token to the graph. The new graph is delivered to the distributed data-flow interpreter ``instruction pool''
\item the distributed macro data-flow interpreter fetches fireable instructions from the instruction pool and the instructions are executed exploiting the nodes in the target architecture. Possibly, optimizations are taken into account (based on proper heuristics) that try to avoid unnecessary communications (e.g. caching tokens that will eventually be reused) or to adapt the computation grain of the program to the target architecture features (e.g. delivering more than a single fireable instruction to remote nodes to decrease the impact of communication set up latency, or multiprocessing the remote nodes to achieve communication and computation overlap).
\end{enumerate}
Figure \ref{fig:dataflow} summarizes the process leading from skeleton source code to the running code exploiting this data-flow approach.

The two approaches just outlined appear very different, but they have been successfully used to implement different skeleton systems. %Just for supporting what it will be presented in Section \ref{sec:aop}, we want
Let us to point out a quite subtle difference in the two approaches.

On the one side, when using implementation templates, the process network eventually run on the target architecture is very close to the one the programmer has in mind when instantiating skeletons in the source code. In some systems the ``optimization'' phase of Figure \ref{fig:template} is actually empty and the program eventually run on the target architecture is build out of plain juxtaposition of the process networks making up the templates of the skeletons using in the program. Even in case the optimization phase do actually modify the process network structure (in Figure \ref{fig:template} the master/slave service process of the two consecutive farms are optimized/collapsed, for instance), the overall structure of the process network does not change too much.

On the other side, when a data-flow approach is used the process network run on the target architecture is completely different from %has almost nothing to do with
the skeleton tree exposed by programmer in the source code. Rather, the skeleton tree is used to implement the parallel computation in a correct and efficient way, exploiting a set of techniques and mechanisms that are much more close to the techniques and mechanisms used in operating systems rather than to those used in the execution of parallel programs, both structured and unstructured.
Under a slightly different perspective, this can be interpreted as follows:
\begin{itemize}
\item skeletons in the program ``annotate'' sequential code by providing the meta information required to efficiently implement the program in parallel;
\item the support tools of the skeleton programming environment (the macro data-flow graph compiler and the distributed macro data-flow interpreter, in this case) ``interpret'' the meta information to accurately and efficiently implement the skeleton program, exploiting (possibly at run-time, when the target architecture features are known) the whole set of known mechanisms supporting implementation optimization (e.g. caches, pre-fetching, node multiprocessing, etc.).
\end{itemize}
Under this perspective, the macro data-flow implementation for parallel skeleton programs opens new perspectives in the design of parallel programming systems where parallelism is dealt with as a ``non-functional'' feature, specified by programmers
%via annotations or exploiting Aspect-Oriented Programming (AOP) techniques,
and handled by the compiling/run-time support tools in the more convenient and efficient way w.r.t. to the target architecture at hand. In the following Chapters of this thesis will be presented some techniques we exploited to provide programmers methodologies aiming the expression of non-functional requirements and their run-time enforcement.

\section{\muskel}
\label{sec:struct}
We already introduced \muskel and its programming model in the Chapter \ref{parallel_issues}. There we also outlined how we modified \muskel, collaborating with its original developers, in order to provide programmers with mechanisms enabling skeleton customizations.
In this section we give a more detailed explanation both of the original \muskel and of the enhanced version we proposed.

\begin{figure}
\centerline{\includegraphics[scale=0.8]{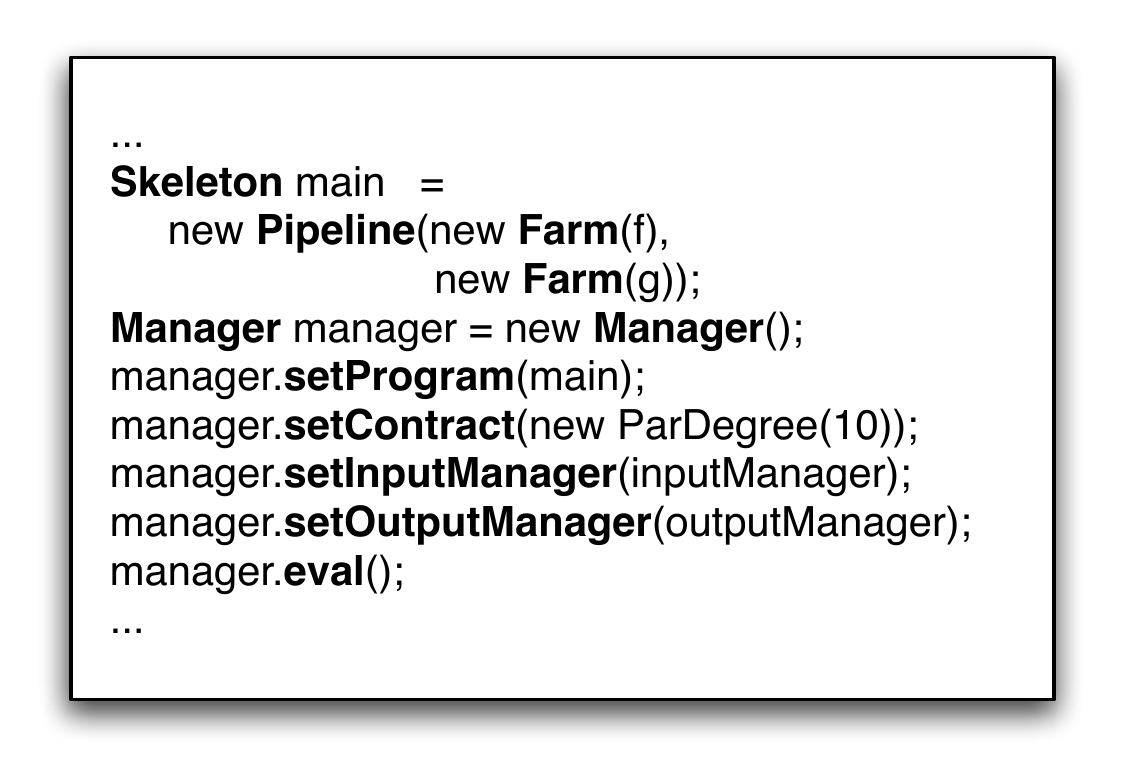}}
\caption{Sample \muskel code: sketch of \textit{all} (but the sequential portions of code) the coded needed to set up and execute a two-stage pipeline with parallel stages (farms).}
\label{fig:code}
\end{figure}

\muskel is skeleton programming environment derived from Lithium \cite{teti-fgcs}, it provides the stream parallel skeletons of Lithium, namely stateless task farm and pipeline. These skeletons can be arbitrary nested, to program pipelines with farm stages, as an example, and they process a single stream of input tasks to produce a single stream of output tasks. \muskel implements skeletons exploiting data-flow technology and Java RMI facilities.
\muskel programmers can express parallel computations simply using the provided \textbf{Pipeline} and \textbf{Farm} classes. For instance, to express a parallel computation structured as a two-stage pipeline where each stage is a farm, \muskel programmers should write a code such as the one of Figure \ref{fig:code}.
The two classes \textbf{f} and \textbf{g} implement the \textbf{Skeleton} interface, i.e. supplying a \textbf{compute} method with the signature
\[ \texttt{Object compute(Object t)} \]
computing \textbf{f} and \textbf{g} respectively. The \textbf{Skeleton} interface represents the ``sequential'' skeleton, that is the skeleton always executed sequentially and only aimed at wrapping sequential code in such a way such code can be used in other, non-sequential skeletons.

In order to execute the program, a \muskel programmer first sets up a \textbf{Manager} object. Then, using proper methods, he specifies the program to execute, the performance contract required (in this case, the parallelism degree required for the execution), the input data source (the input stream manager, which is basically an iterator providing the classical \textbf{boolean hasNext()} and \textbf{Object next()} methods) and who is in charge of processing the output data (the output stream manager, just providing a \textbf{void deliver(Object)} method processing a single result of the program).
Eventually he can ask parallel program execution simply issuing
an \textbf{eval} call to the manager.
When the call terminates, an output file is produced.

Actually, the \textbf{eval} method execution happens in steps.
First, the manager looks for available processing elements using a simplified, multicast based peer-to-peer discovery protocol, and recruits the required remote processing elements. Each remote processing element runs a data-flow interpreter.
Then the skeleton program (the \textbf{main} of the example depicted in Figure \ref{fig:code}) is compiled into a macro data-flow graph (actually capitalizing on  normal form results shown in \cite{pdcs:nf:99,teti-fgcs})
and a thread is forked for each one of the remote processing elements recruited.
Then the input stream is read. For each task item, an instance of the macro data-flow graph is created and the task item token is stored in the proper place (initial data-flow instruction(s)). The graph is placed in the task pool, the repository for data-flow instructions to be executed.
Each thread looks for a fireable instruction in the task pool and delivers it for execution to the associated remote data-flow interpreter. The remote interpreter instance associated to the thread is initialized by being sent the serialized code of the data-flow instructions, once and for all before the computation actually starts. Once the remote interpreter terminates the execution of the data-flow instruction, the thread either stores the result token in the proper ``next'' data-flow instruction(s) in the task pool, or it directly writes the result to the output stream, invoking the \textbf{deliver} method of the output stream manager.
%
%Currently, the task pool is a centralized one, associated with the centralized manager. We are currently investigating the possibility to distribute both task pool and manager, in such a way this bottleneck will eventually disappear.
%
%The \texttt{manager}\ takes care of ensuring that the performance contract is satisfied.
If a remote node ``fails'' (e.g. due to a network failure, or to the node failure/shutdown), the manager looks for another node and starts dispatching data flow instructions to the new node instead \cite{muskel:qos:pdp:05}. As the manager is a centralized entity, if it fails, the whole computation fails. However, the manager is usually run on the machine of the \muskel user, which is assumed to be safer than the remote nodes recruited as remote interpreter instances.

The policies implemented by the \muskel managers are \textit{best effort}. The \muskel framework tries to do its best to accomplish user requests. In case it is not possible to completely satisfy the user requests, the framework accomplishes to establish the closest configuration to the one implicitly specified by the user with the performance contract. In the example above, the framework tries to recruit 10 remote interpreters. In case only $n<10$ remote interpreters are found, the parallelism degree is set exactly to $n$. In the worst case, that is if no remote interpreter is found, the computation is performed sequentially, on the local processing element.

In the current version of \muskel, the only performance contract actually implemented is the \textbf{ParDegree} one, asking for the usage of a constant number of remote interpreters in the execution of the program. %The prototype has been thought to support at least another kind of contract: the \texttt{ServiceTime}\ one. This contract can be used to specify the maximum amount of time expected between the delivery of two program result tokens. Thus with a line code such as \texttt{manager.setContract(new ServiceTime(500))}, the user may ask to deliver one result every half a second (time is in ms, as usual in Java).
We do not enter in more detail in the implementation of the distributed data-flow interpreter here. The interested reader can refer to \cite{MDF:parco:99,muskel:qos:pdp:05}. Instead, we will try to give a better insight into the compilation of skeleton code into data-flow graphs.

A \muskel parallel skeleton code is described by the grammar:
\[ {\sf P}\ ::= {\sf seq}(\mathit{className}) \mid {\sf pipe}({\sf P}, {\sf P}) \mid
{\sf farm}({\sf P}) \]
where the \textbf{className}s refer to classes implementing the \textbf{Skeleton} interface,
and a macro data-flow instruction is a tuple:
$\langle \mathit{id}, \mathit{gid}, \mathit{opcode}, {\cal I}^n, {\cal O}^k \rangle $
where \textit{id} is the instruction identifier, \textit{gid} is the graph identifier (both are either integers or the special \textit{NoId} identifier), \textit{opcode} is the name of the \textbf{Skeleton} class providing the code to compute the instruction (i.e. computing the output tokens out of the input ones) and ${\cal I}$ and ${\cal O}$ are the input tokens and the output token destinations, respectively.
An input token is a pair $\langle \mathit{value}, \mathit{presenceBit} \rangle$ and an output token destination is a pair $\langle  \textit{destInstructionId},\textit{destTokenNumber} \rangle$.
With these assumptions, a data-flow instruction such as:
\[  \langle
a,b,\texttt{f},\langle \langle 123, \texttt{true} \rangle , \langle \texttt{null}, \texttt{false} \rangle \rangle,
\langle \langle i,j \rangle
\rangle \rangle
 \] \\
is the instruction with identifier \textit{a} belonging to the graph with identifier \textit{b}. It has two input tokens, one present (the integer 123) and one not present yet. It is not fireable, as one token is missing. When the missing token will be delivered to this instruction, coming either from the input stream or from another instruction, the instruction becomes fireable. To be computed, the two tokens must be given to the \textbf{compute} method of the \textbf{f} class. The method computes a single result that will be delivered to the instruction with identifier \textit{i} in the same graph, in the position corresponding to input token number \textit{j}.
The process compiling the skeleton program into the data-flow graph
can therefore be more formally described as follows. We define a
pre-compile function $PC[\ ]$ as:\smallskip\\
{\small \hspace*{2em}\ $ PC[ {\sf seq}\ ( {\texttt{f}})]_{gid}\ = \lambda i. \{
\langle \mathit{newId}(), gid, \texttt{f},
\langle \langle \texttt{null}, \texttt{false} \rangle \rangle,
\langle \langle i, \mathit{NoId} \rangle \rangle
\rangle \}\  $\smallskip\\
\hspace*{2em}\ $ PC[ {\sf farm}( {\sf P}\ )]_{gid}\ = C[{\sf P}]_{gid}\ $\smallskip\\
\hspace*{2em}\ $PC[ {\sf pipe}\ ( {\sf P}_1, {\sf P}_2) ] _ {gid}\ =
\lambda i. \{
   C[{\sf P}_1]_{gid}\ (getId(C[{\sf P}_2]_{gid}\ )),
 C[{\sf P}_2]_{gid}(i)
\}\ $ }
\smallskip\\
\noindent where $\lambda x.T$ is the usual function representation
($(\lambda x.T)(y) = T\hspace{-0.75ex}\mid_{x=y})$ and  $getID()$ is
the function returning the $id$ of the first instruction in its
argument graph, that is, the one assuming to receive the input token
from outside the graph, and a compile function $C[]$ such
as:

%\smallskip\\  
{\small
\hspace*{2em}\ \[ C[ {\sf P}\ ] = PC[ {\sf P}\ ] _ {newGid()}\ (\texttt{NoId}) \] } 
%\smallskip\\

\noindent where \textbf{newId()} and \textbf{newGid()} are stateful functions returning a fresh (i.e. unused) instruction and graph identifier, respectively. The compile function returns therefore a graph, with a fresh graph identifier, hosting all the data-flow instructions relative to the skeleton program. The result tokens are identified as those whose destination is \textbf{NoId}.
As an example, the compilation of the \textbf{main} program
{\sf pipe}({\sf farm}({\sf seq}(\texttt{f})), {\sf farm}({\sf seq}(\texttt{g})))
produces the data flow graph:\smallskip\\{\small
$\{  \langle 1, 1, \texttt{f}, \langle \langle \texttt{null}, \texttt{false}
\rangle \rangle , \langle \langle 2, 1 \rangle \rangle \rangle\
,
\langle 2, 1, \texttt{g}, \langle \langle \texttt{null}, \texttt{false} \rangle \rangle , \langle \langle \texttt{NoId}, \texttt{NoId} \rangle \rangle \rangle
\}$
} \smallskip\\
\noindent (assuming that identifiers and token positions start from 1).

When the application manager is told to actually compute the program, via an \textbf{eval()} method call, the input file stream is read looking for tasks to be computed. Each task found is used to replace the data field of the lower \emph{id} data-flow instruction in a new $C[ {\sf P}\ ]$ graph.
In the example above, this results in the generation of a set of independent graphs such as:\smallskip\\
{\small
$\{  \langle 1, i, \texttt{f}, \langle \langle \texttt{task$_i$}, \texttt{true}
\rangle \rangle , \langle \langle 2, 1 \rangle \rangle \rangle\
,
\langle 2, i, \texttt{g}, \langle \langle \texttt{null}, \texttt{false} \rangle \rangle , \langle \langle \texttt{NoId}, \texttt{NoId} \rangle \rangle \rangle
\}$
}
\smallskip\\

\noindent for all the tasks ranging from $task_1$ to $task_n$.

All the resulting instructions are put in the task pool of the distributed interpreter in such a way that the control threads taking care of ``feeding'' the remote data-flow interpreter instances can start fetching the fireable instructions.
The output tokens generated by instructions with destination tag equal to \textbf{NoId} are directly delivered to the output file stream by the threads receiving them from the remote interpreter instances. Those with a non-\textbf{NoId} flag are delivered to the proper instructions in the task pool that will eventually become fireable.

%\subsection{Expanding \muskel\ skeleton facilities}
%\label{sec:unstruc}
%In this section, we will discuss how the skeleton facilities provided by \muskel\ can be extended to accomplish user peculiar needs.
%%
%Two aspects are taken into account.
%%
%First, the mechanisms used to allow programmers to define their own skeletons are discussed, along with their \muskel\ implementation. Using these mechanisms, the programmers may declare and use arbitrary, possibly ``unstructured''\footnote{w.r.t. classical skeletons frameworks}\ new skeletons.
%%
%Then, we will discuss how other, alternative mechanisms based on Java annotations and/or AOP techniques that are currently being used to provide further expandability of the \muskel\ skeleton set, in particular characterizing existing skeletons with new, non-functional features.

\begin{figure}
\centerline{\includegraphics[scale=0.7]{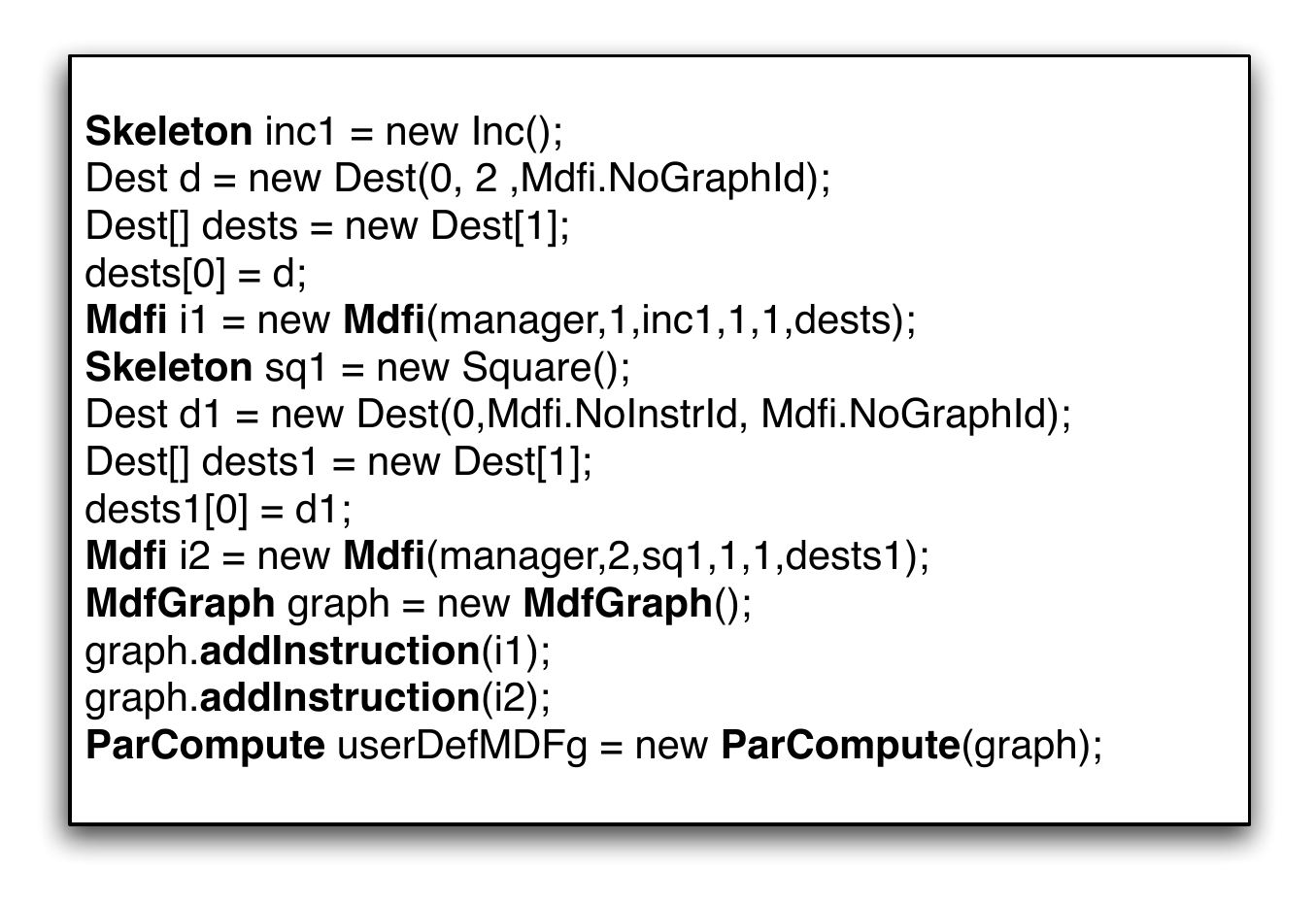}}
\caption{Custom/user-defined skeleton declaration.}
\label{fig:ucode}
\end{figure}

\subsection{Programmer-defined skeletons}\label{sec:unstruc}
In order to introduce completely new parallelism exploitation patterns, our version of the \muskel framework provides programmers with mechanisms that can be used to design plain, arbitrary macro data-flow graphs. A macro data-flow graph can be defined creating some \textbf{Mdfi} (macro data-flow instruction) objects and connecting them in a \textbf{MdfGraph}\ object.
As an example, the code in Figure \ref{fig:ucode} is the code needed to program a data-flow graph with two instructions. The first one computes the \textbf{compute} method \textbf{inc1}  on its input token and delivers the result to the second instruction. The second one, computes the \textbf{sq1}  \textbf{compute} method on its input token and delivers the result to a generic ``next'' instruction (this is modeled giving the destination token tag a \textbf{Mdfi.NoInstrId} tag). The \textbf{Dest} stuff in the code is meant to represent destination of output tokens as triples hosting the graph identifier, the instruction identifier and the destination input token targeted in this instruction. Macro data-flow instructions are build stating the manager they refer to, their identifier, the code executed (must be a \textbf{Skeleton} object) the number of input and output tokens and a vector with a destination for each one of the output tokens.
%
%We do not enter all the details relative to arbitrary macro data-flow graphs building here (complete description is provided with the \muskel documentation). The example is just to give the flavor of the tools provided in the \muskel environment.
Take into account that the simple macro data-flow graph of Figure \ref{fig:ucode} is actually the very same macro data-flow graph derived compiling a primitive \muskel skeleton code such as:
\begin{center}
\verb9Skeleton main = new Pipeline(new Inc(), new Sq()))9
\end{center}
\noindent More complex, programmer-defined macro data-flow graph may comprehend instructions delivering tokens to an arbitrary number of other instructions, as well as instructions gathering input tokens from several  distinct other instructions.

\textbf{MdfGraph} objects are used to create new \textbf{ParCompute} objects. The \textbf{ParCompute} objects can be used in any place were a \textbf{Skeleton} object is used. Therefore programmer-defined parallelism exploitation patterns can be used as pipeline stages or as farm workers, for instance. The only limitation on the graphs that can be used in a \textbf{ParCompute} object consists in requiring that the graph has a unique input token and a unique output token.

When executing programs with programmer-defined parallelism exploitation patterns the process of compiling skeleton code to macro data-flow graphs is slightly modified.
When an original \muskel skeleton is compiled, the process described above is applied.
When a programmer-defined skeleton is compiled, the associated macro data-flow graph is directly taken from the \textbf{ParCompute} instance variables where the graph supplied by the programmer is maintained. Such graph is linked to the rest of the graph according to the rules relative to the skeleton where the programmer-defined skeleton appears.
\begin{figure}[!ht]
\centerline{\includegraphics[scale=0.6]{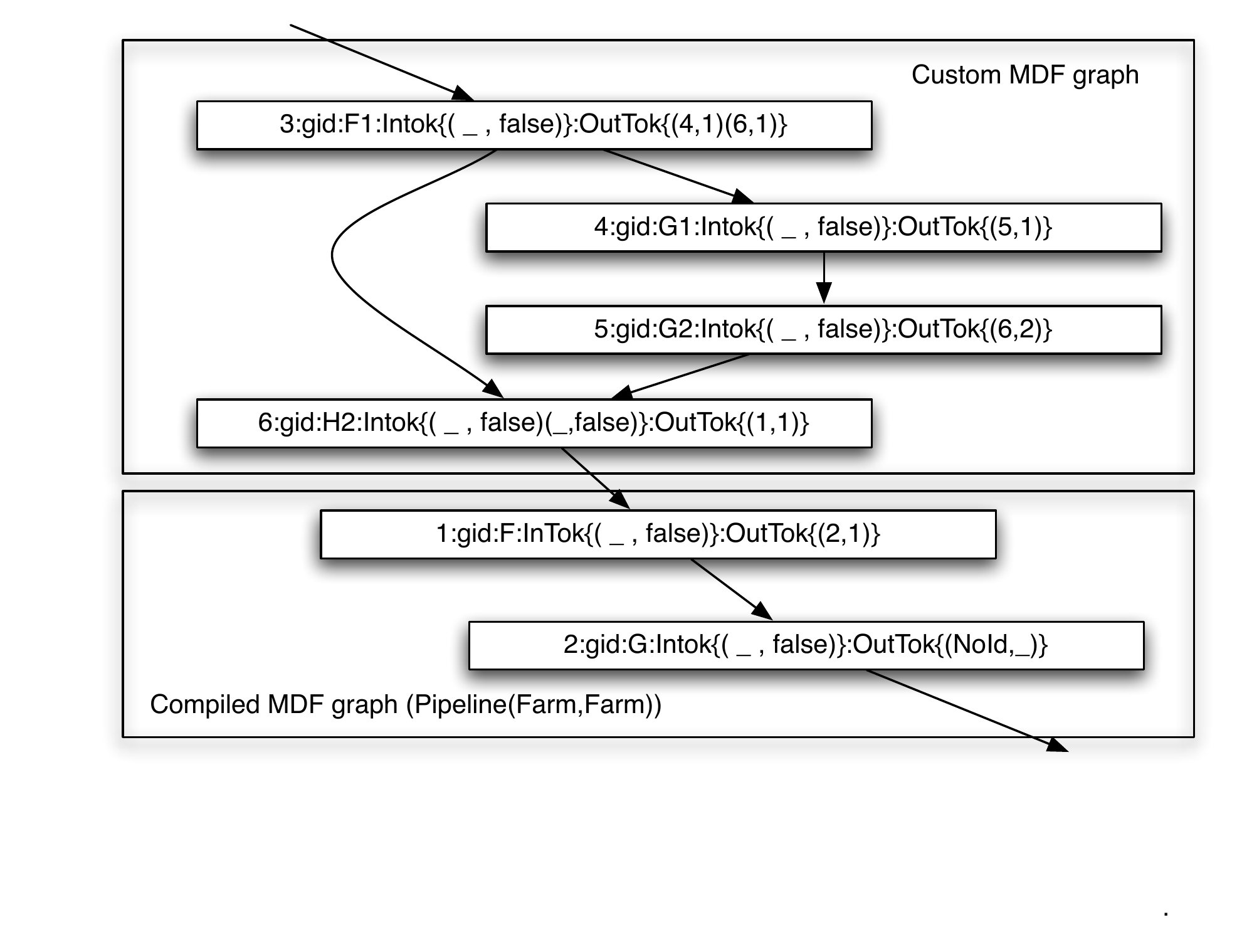}}
\vspace*{-5em}
\caption{Mixed sample MDF graph: the upper part comes from a programmer-defined MDF graph (it cannot be derived using primitive \muskel skeletons) and the lower part is actually coming from a three stage pipeline with two sequential stages (the second and the third one) and a parallel first stage (the programmer-defined one).}
\label{fig:grafoMuskel}
\end{figure}
To show how the whole process works, let us suppose we want to pre-process each input tasks in such a way that for each task $t_i$ a new task
\[t'_i = h_1(f_1(t_i), g_2(g_1(f_1(t_i))))\] is produced. This computation cannot be programmed using the stream parallel skeletons currently provided by the original \muskel. Then we want to process the preprocessed tasks through a two-stage pipeline, in order to produce the final result.
In this case the programmer can set up a new graph using a code similar to the one shown in Figure \ref{fig:code} and then used that new \textbf{ParCompute} object as the first stage of a two-stage pipeline whose second stage happens to be the postprocessing two-stage pipeline. When compiling the whole program, the outer pipeline is compiled first. As the first stage is a programmer-defined skeleton, its macro data-flow graph is directly taken from the programmer-supplied one. The second stage is compiled according to the (recursive) procedure previously described and eventually the (unique) last instruction of the first graph is modified in such a way it sends its only output token to the very first instruction in the second stage graph. The resulting graph is outlined in Figure \ref{fig:grafoMuskel}.

Making good usage of the mechanisms that allow to define new data-flow graphs, the programmer can arrange to express computations with arbitrary mixes of arbitrary data-flow graphs and graphs coming from the compilation of structured, stream parallel skeleton computations.  The execution of the resulting data-flow graph is supported by the \muskel distributed data-flow interpreter as the execution of any other data-flow graph derived from the compilation of a skeleton program. Therefore, the customized skeletons are efficiently executed as the skeletons ``bundled'' with \muskel. Indeed,  in data-flow based skeleton systems, as we already stated when we presented them, the  optimizations do not directly depends on the skeleton structure but on the data-flow engine capability of executing the macro data-flow instruction in an efficient way.  
 
%
%Actually, at the moment the \muskel\ prototype allows user-defined skeletons to be used as parameters of primitive \muskel\ skeletons, but not vice versa. This is only a matter of extending a little bit the compiler module, however. There is no conceptually difficult step behind.
In order to allow primitive \muskel skeleton usage as code to be executed in an instruction of a programmer-defined macro data-flow graph it is sufficient to compile ``on the fly'' the primitive skeleton and include the result (i.e. the macro data-flow graph) of this compilation in the programmer-defined macro data-flow graph.

\begin{figure}
\centerline{\includegraphics[scale=0.60]{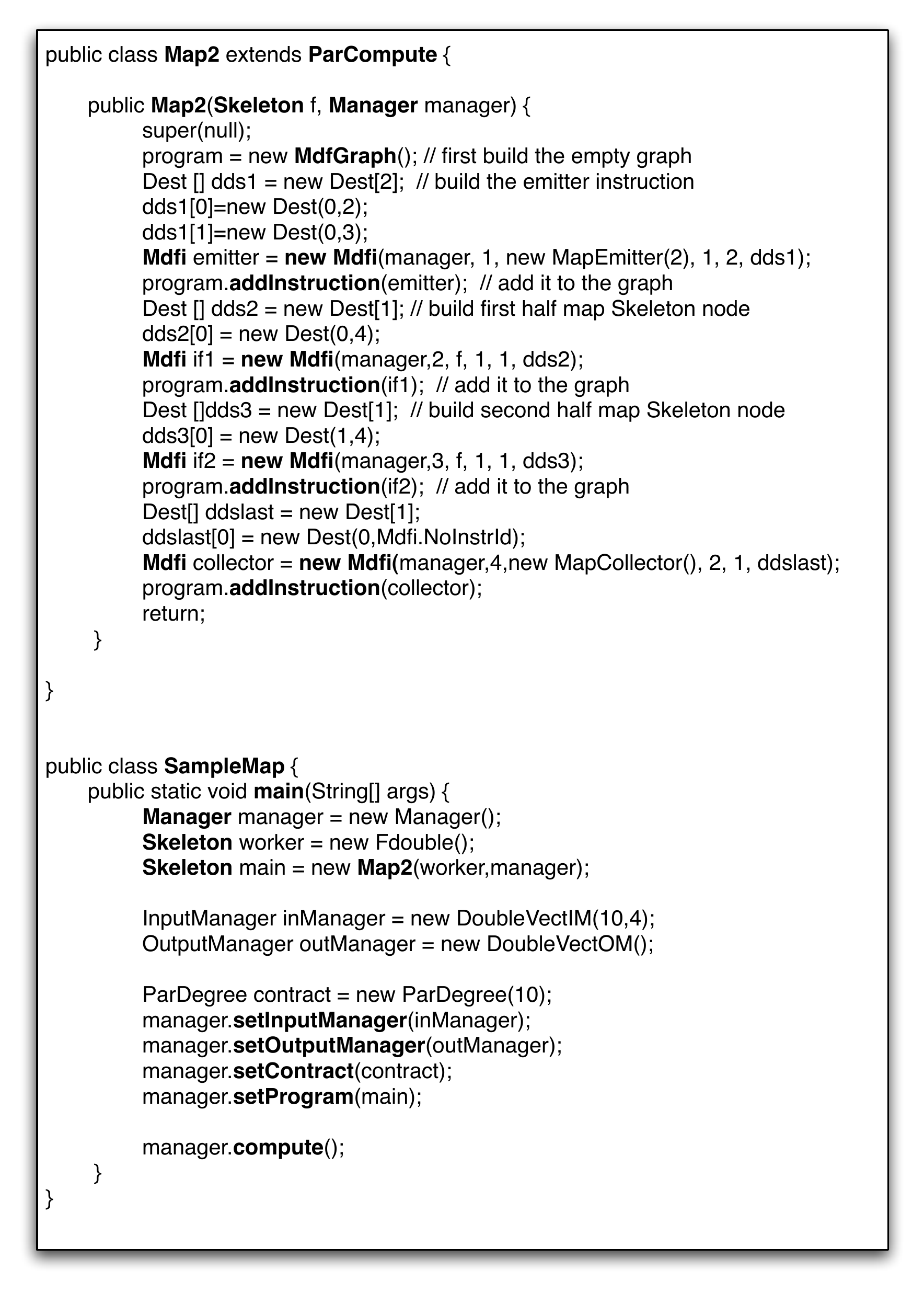}}
\caption{Introducing a new, programmer-defined skeleton: a map working on vectors and with a fixed, programmer-defined parallelism degree.}
\label{fig:map}
\end{figure}

As a final example, consider the code of Figure \ref{fig:map}.
This code actually shows how a new \textbf{Map2} skeleton, performing in parallel the same computation on all the portions of an input vector, can be defined and used. It's worth pointing out how programmer-defined skeletons, once properly debugged and fine-tuned, can simply be incorporated in the \muskel skeleton framework and used seamlessly, as the primitive \muskel ones, but for the fact (as show in the code) the constructor needs the manager as a parameter. This is needed just to be able to link together the macro data-flow graphs generated by the compiler and those supplied by the programmer. This feature has been released %(a problem, actually) will be probably released
by postponing  the data-flow graph creation to the moment the graph needs to be instantiated after the arrival of a new task to compute, as at that time all the information necessary to perform graph ``conjunction'' is available.

%
% \begin{figure}
% %\hspace*{-2ex}
% \begin{center}
% \includegraphics[scale=0.5]{figure/perfe}
% \includegraphics[scale=0.5]{figure/effe}
% \end{center}
% %\vspace*{-2em}
% \caption{Scalability of the \muskel prototype and effect of computation grain.}
% \label{fig:scala}
% \label{fig:grana}
% \end{figure}

\section{Experimental results}
\label{sec:results}
To validate our approach we conducted some test with our modified version of the \muskel framework. The original \muskel interpreter engine has been left basically unchanged, whereas the part supporting parallelism exploitation pattern programming has been changed to support linking of custom MDF graphs to the code produced by the compiler out of plain \muskel skeleton trees. We used our customized version for implementing an application that can not be (at least not easily) implemented using standard (i.e. without our proposed customization support) skeleton environments. 

\begin{figure}
%\hspace*{-2ex}
\begin{center}
\includegraphics[scale=0.7]{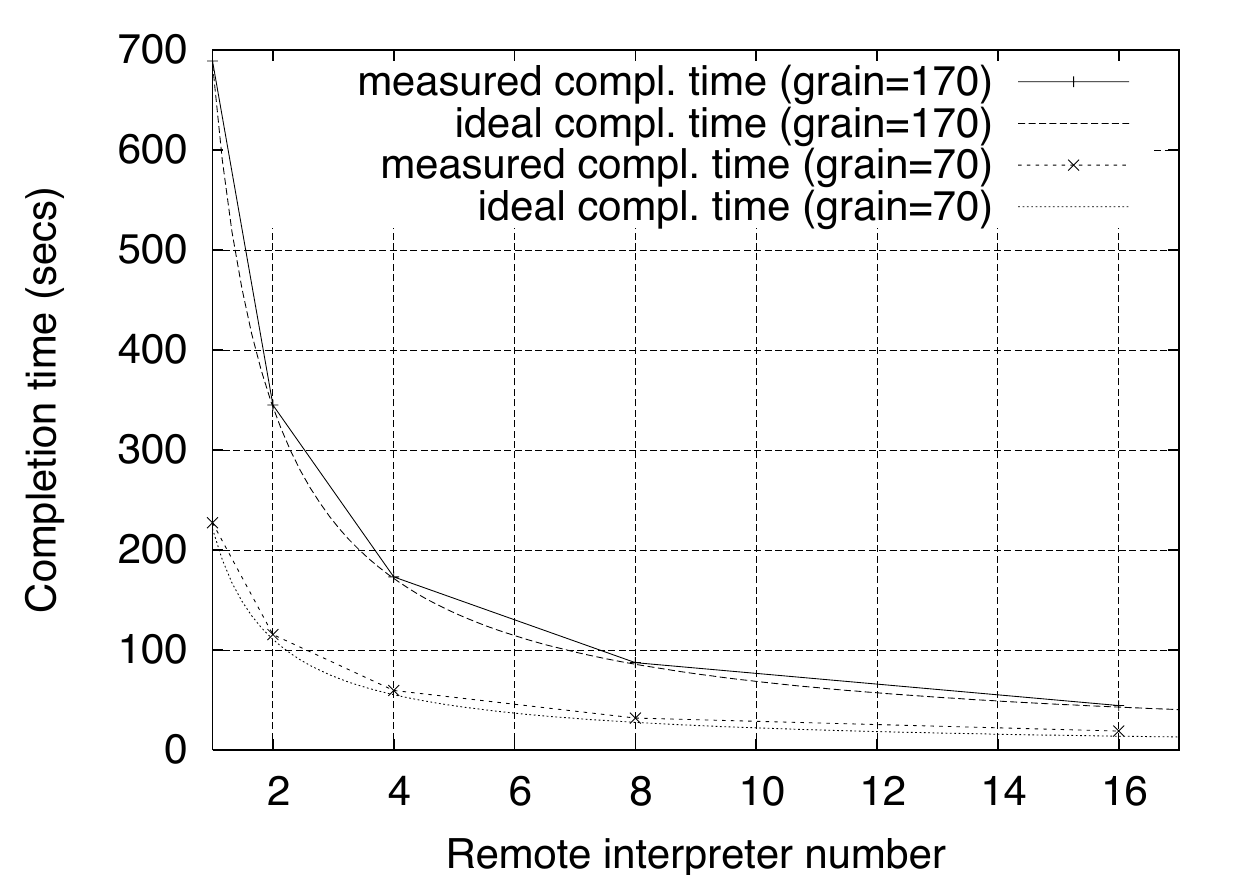}
\includegraphics[scale=0.7]{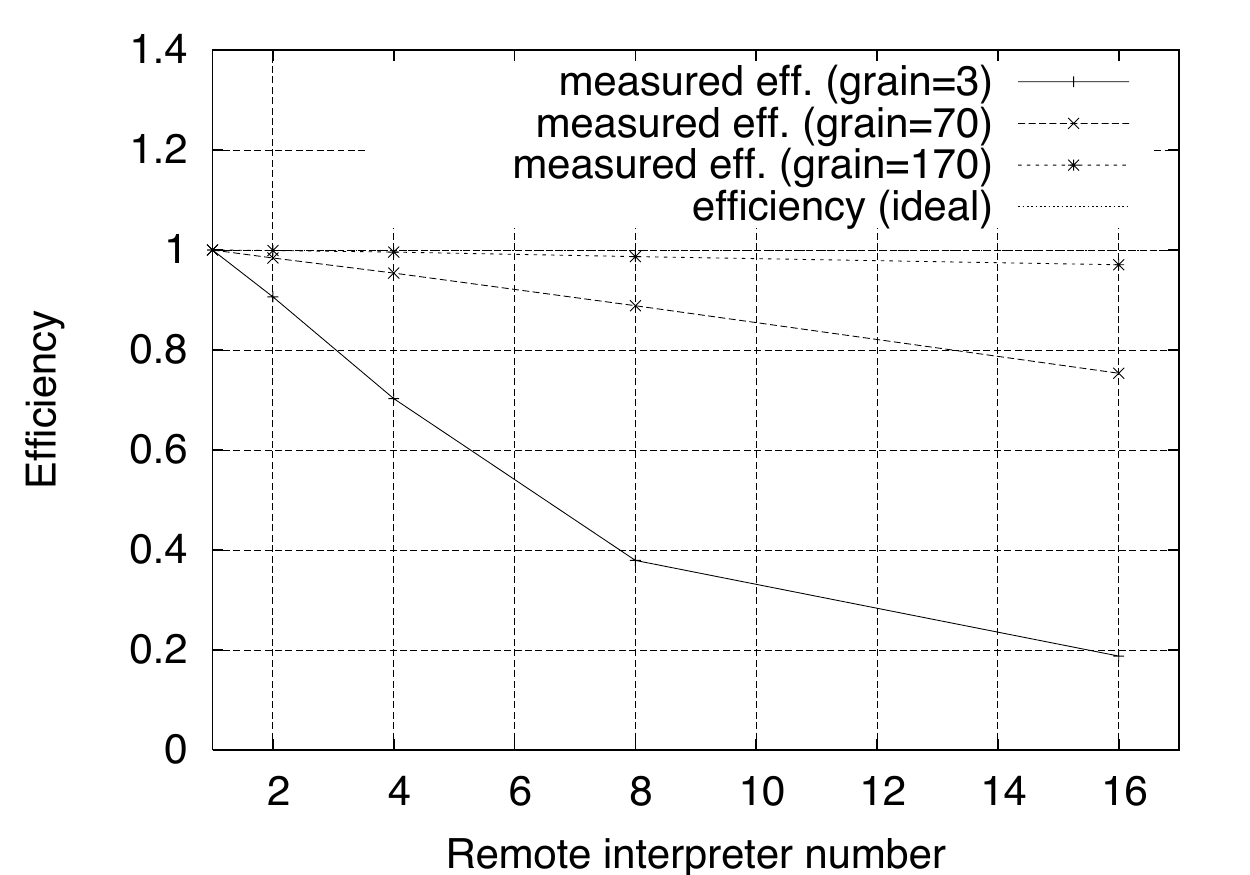}
\end{center}
%\vspace*{-2em}
\caption{Scalability of the \muskel prototype and effect of computation grain.}
\label{fig:scala}
\label{fig:grana}
\end{figure}

Figure \ref{fig:grana} summarizes the typical performance results of our enhanced interpreter. We ran several synthetic programs using the custom macro data-flow graph features introduced in \muskel. We designed the programs in such a way the macro data-flow instructions appearing in the graph had a precise ``average grain'' (i.e. average ration between the time spent by the remote interpreter to compute the macro data flow instruction sent to it, and the time spent in communicating data to the remote interpreter plus the time to retrieve the computation results). For each test-bed we passed as input parameters to the developed programs 1K input tasks. 

The results show that when the computational grain is small, \muskel does not scale well, even using a very small number of remote interpreter instances. Indeed, Figure \ref{fig:grana} clearly shows that when the computational grain is 3 the efficiency rapidly decreases, going under 0.7 even when only four computational resources are used. When the grain is 70 the efficiency goes under 0.8 only when the number of recruited computational resources is higher than 14. Finally, when the grain is high enough (about 200 times the time spent in communications actually spent in computation of MDF instructions) the efficiency is definitely close to the ideal one even using 16 or more machines. 

Despite the data shown refers to some synthetic computations, actual computations (e.g. image processing ones) achieved very similar results. This because the automatic load balancing mechanism implemented in the \muskel distributed interpreter, obtained by mean of auto scheduling techniques, perfectly optimized the execution of variable grain MDF instructions.
All the experiments have been performed on a Linux (kernel 2.4.22) RLX Pentium III blade architecture, with Fast Ethernet interconnection among the blades, equipped with Java 1.4.1\_01 run-time.
%
%\begin{figure}
%\hspace*{-2ex}
%\centerline{\includegraphics[scale=0.35]{figure/bini}}
%\vspace*{-2em}
%\caption{Effect of middleware: scalability of the \muskel prototype using plain RMI vs. the one using ProActive active objects.}
%\label{fig:proactive}
%\end{figure}

\begin{figure}
\centerline{\includegraphics[scale=0.33]{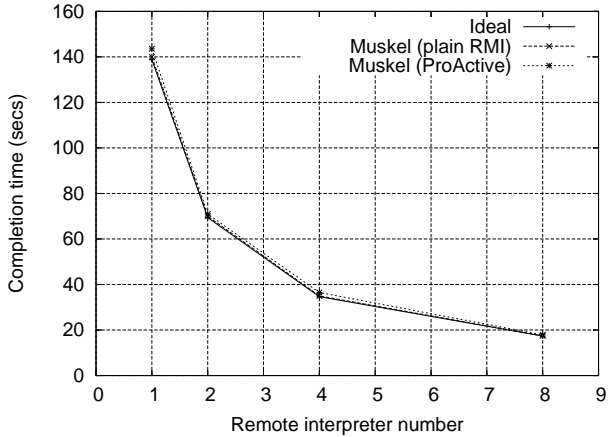}}
%\centerline{\includegraphics[scale=0.65]{figure/rate_speedup}}
\caption{Effect of middleware: scalability of the \muskel prototype using plain RMI vs. the one using ProActive active objects}%. Bottom figure: Effect of providing security in the distributed data flow interpreter: scalability of the \muskel prototype using plain TCP/IP sockets vs. the one using SSL for different computational grains.}
\label{fig:sec1}\label{fig:proactive}
\end{figure}

Despite measuring scalability of our modified \muskel framework, we also have taken into account the possibility to use different mechanisms to support distributed data-flow interpreter execution. In particular, we investigated the possibility of implementing the \muskel approach for skeleton customization on top of the ProActive framework \cite{proactive} both to be able to target a different set of architectures and  to demonstrate the ``portability'' of our approach, i.e. that it is a feasible and efficient solution not only when it exploits the \muskel data-flow interpreter.

For this purpose, we conducted some experiments aimed at verifying the overhead introduced by ProActive with respect to the plain Java RMI \muskel prototype, when using the secure shell (\textbf{ssh}) tunneling of the RMI protocol (feature natively provided by the ProActive framework). In particular, we modified the ``kernel'' of the data-flow interpreter of \muskel in order to make it able to exploit the ProActive active objects in place of plain RMI objects as remote data-flow interpreter instances. The results we achieved are summarized in Figure \ref{fig:proactive}. The figure plots the completion times for the very same program run on a Linux workstation cluster when using plain Java RMI and when using ProActive active objects to implement the remote data-flow interpreter instances. The macro data-flow instructions, in this case, have a grain comparable to the ``high grain'' of instructions of Figure \ref{fig:grana}. Experiments showed that ProActive active objects are slightly less efficient but the difference is negligible. In this case, the setup time of the remote data-flow interpreter instances was not considered in the overall completion time, being paid once and forall when the system is started up.

\newpage

\section*{Summarizing the Chapter}
\emph{
\hrule
\medskip
\noindent In this Chapter we discussed a methodology for extending algorithmic skeletons based parallel programming frameworks aimed at providing programmers with the possibility to freely customize the structure of their parallel applications. It is based on mechanisms allowing programmers to modify the data-flow graph derived from the compilation of skeleton based application.
In particular, we discussed how we modified the \muskel framework for parallel programming. The version we developed (collaborating with the team that developed the original \muskel) supports extendability of the skeleton set, as advocated by Cole in his ``manifesto'' paper \cite{cole:manifesto:02}.
In particular, we discussed how our modified \muskel supports the introduction of new skeletons, modeling parallelism exploitation patterns not originally covered by the primitive \muskel skeletons. %, and the introduction of non-functional features, i.e. features related to parallel program execution but not directly related to the functional computation of the application results.
%
%The former
This possibility is supported by allowing programmers to define new skeletons providing the arbitrary data-flow graph executed in the skeleton and by letting \muskel to seamlessly integrate such new skeletons in the primitive ones.
%
%The latter possibility is supported exploiting more innovative programming techniques such as annotations and aspect-oriented programming.
%
%This second part is actually under development, while the first one is already being available in the \muskel prototype.
%
We also presented experimental results validating our \muskel approach to extend and customize its skeleton set.
As far as we know, this is the most significant effort in the skeleton community to tackle problems deriving from a fixed skeleton set. Only Schaeffer and his group at the University of Alberta implemented a system were programmers can, in controlled ways, insert new parallelism exploitation patterns in the system \cite{bromling:parco:2001}, although the approach followed here is a bit different, in that programmers are encouraged to intervene directly in the run-time support implementation, to introduce new skeletons, while in our \muskel new skeletons may be introduced using the intermediate macro data-flow language as the skeleton ``assembly'' language.
\medskip
\hrule
}
%Eventually, we also discussed how relatively new programming techniques, including annotations and AOP, can be usefully exploited in \muskel to support details and features related to parallel program execution.

%Preliminary versions of \muskel have been released under GPL are currently available on the \muskel web site at \url{htpp://www.di.unipi.it/~marcod/muskel}. The new version, sporting the features discussed in this paper, is currently being developed. The support for new skeletons is already completed (and it is available, as a beta release, on the web site) and the other features will be released soon.
%\fussy

%\insertblankpage

%\line(1,0)(50mm) 
\chapter{Metaprogramming Run-time Optimizations}\label{muskelWorkflow}

% \begin{chapterabstract}
% We discuss how Java annotations can be used to provide the meta information needed to automatically transform plain Java programs into suitable parallel code that can be run on workstation clusters, networks and grids. Programmers are only required to decorate the methods that will eventually be executed in parallel with standard Java 1.5 annotations. Then these annotations are automatically processed and parallel byte code is derived. When the annotated program is started, it automatically retrieves the information about the executing platform and evaluates the information specified inside the annotations to transform the byte-code into a semantically equivalent multithreaded or multitask version, depending on the target architecture features. The results returned by the annotated methods, when invoked, are futures with a wait-by-necessity semantics.
% A PAL (\textit{Parallel Abstraction Layer}) prototype exploiting the annotation based parallelizing approach has been implemented in Java. PAL targets JJPF, an existing, skeleton based, JAVA/JINI programming environment, as Parallel Framework. The experiments made with the prototype are encouraging: the design of parallel applications has been greatly simplified and the performances obtained are the same of an application directly written in JJPF.
% \end{chapterabstract}
%\sloppy

\paragraph{Chapter road-map} %perspective
\emph{
This Chapter presents our efforts aimed at exploiting metaprogramming techniques for optimizing at run-time the execution of structured parallel applications. The approaches are based on the run-time generation of macro data-flow blocks from the application code. % starting from java source enriched with annotations or with aspects.
We start presenting the motivations (Section \ref{motivations}) of our contributions. Then we present PAL (Section \ref{PAL}), our first result in the field. PAL is a metaprogramming engine that transforms at run-time an annotated sequential java code in a parallel program, exploiting both programmer hints and executing platform information. We describe our PAL prototype implementation (Section \ref{PALimpl}) and the results of the tests we made with it (Section \ref{PALtests}). After we discuss the motivations that convinced us to integrate the PAL approach with our version of the \muskel framework (Section \ref{PALmotivations}). In the following section (\ref{metamuskel}) we describe the preliminary attempts we made integrating metaprogramming techniques in \muskel. % showing how aspect oriented programming can be exploited in order to normalize a skeleton code.
In Section \ref{muskworkflows} we present how we further enhanced \muskel making it able to exploit metaprogramming for run-time code optimizations. In particular, how it can be exploited to optimize the parallel execution of computations expressed as workflows. In Section \ref{sec:imple} we describe the implementation of workflows transformations and in Section \ref{sec:perfresults} we present the performance results obtained. Finally, we compare the two approaches (Section \ref{sec:gendifferences}) and we summarize the Chapter contributions.
}

\section{Our efforts in run-time optimization}\label{motivations}

In the previous chapter we described how the macro data-flow model can be exploited in order to allow the customization of algorithmic skeletons. We showed how we modified the \muskel parallel framework in order to provide programmers with mechanisms able to change skeletons structure. In this chapter we present the metaprogramming techniques we exploited both to ease the generation of the macro data-flow graph and to optimize at run-time the parallel execution of the macro data-flow blocks.

\subsection{Metaprogramming}
Code-generating programs are sometimes called metaprograms; writing such programs is called metaprogramming. Metaprograms do part of the work during compile-time that is otherwise done at run-time. Compile-time metaprogramming exploits information available at compile-time to generate temporary source code, which is merged by the compiler with the rest of the source code and then compiled. The goal of run-time metaprogramming, instead, is to achieve real-time code optimizations transforming or adapting the code whenever some information becomes available.

\subsubsection{Compile-time metaprogramming}
The most common metaprogramming tool is a compiler, which allows a programmer to write a relatively short program in a high-level language and uses it to write an equivalent assembly language or machine language program. Another still fairly common example of metaprogramming might be found in the use of Template Metaprogramming. Template metaprogramming is a metaprogramming technique in which templates are used by a compiler to generate temporary source code, which is merged by the compiler with the rest of the source code and then compiled. The output of these templates includes compile-time constants, data structures, and complete functions. The use of templates can be thought of as compile-time execution. The technique is used by a number of languages, the most well-known being C++, but also D, Eiffel, Haskell, ML and XL. The use of templates as a metaprogramming technique requires two distinct operations: a template must be defined, and a defined template must be instantiated. The template definition describes the generic form of the generated source code, and the instantiation causes a specific set of source code to be generated from the generic form in the template. Template metaprogramming is generally Turing-complete, meaning that any computation expressible by a computer program can be computed, in some form, by a template metaprogram. Templates are different from macros. A macro, which is also a compile-time language feature, generates code in-line using text manipulation and substitution. Macro systems often have limited compile-time process flow abilities and usually lack awareness of the semantics and type system of their companion language (an exception should be made with Lisp's macros, which are written in Lisp itself, and is not a simple text manipulation and substitution). Template metaprograms have no mutable variables that is, no variable can change value once it has been initialized, therefore template metaprogramming can be seen as a form of functional programming. In fact, many template implementations only implement flow control through recursion. Some common reasons to use templates is to implement generic programming (avoiding sections of code which are similar except for some minor variations) and especially to perform automatic compile-time optimization such as doing something once at compile-time rather than every time the program is run, for instance having the compiler unroll loops to eliminate jumps and loop count decrements whenever the program is executed. The main problem of this approach is the inefficient exploitation of the executing environment. Indeed to guarantee the code portability such optimizations are done in a generic way, for instance without exploiting specific CPU extension like SSE or 3DNow. To overwork it the application should be re-compiled once all the running architecture details are known.

\subsubsection{Run-time metaprogramming}
Run-time metaprogramming points at either the generation of programs specialized with respect to the running architecture or the adaptation of programs with respect to additional information provided by programmers, e.g. non-functional requirements. The metaprogramming related information (metadata) is processed by the metaprogramming run-time support. It exploits both such metadata and the environmental information to transforms the original code into an optimized one.
Nevertheless, this solution presents a major problem: the re-compilation overhead. Indeed, re-compile the whole application from scratch on each machine it is moved for execution is computationally expansive.
A viable solution consists in writing the applications using bytecode based languages, like Java and .NET. Indeed, their compilers do not translate the program into target machine language but translate it into an intermediate language (IL). The IL has greater expressiveness than the machine and the assembly languages and can be transformed in a machine-level program paying a small overhead.
Furthermore, there are other advantages in implementing application, especially the distributed ones, exploiting a virtual machine based language: e.g. the possibility to run programs across different platforms at the only cost of porting the execution environment and to achieve better security (the execution engine mediates all accesses to resources made by programs verifying that the system can not be compromised by the running application).

In the past, other programming languages with the same architecture, essentially p-code, have been proposed (see for instance the introduction of \cite{krall}) but Java has been the first to have a huge impact on programming mainstream. Java approach has been recognized as successful, indeed, since the 2002 also Microsoft introduced their virtual-machine based programming languages. They are based on the Common Language Infrastructure (CLI). The core of CLI is the virtual execution system also known as Common Language Runtime(CLR).
Both JVM \cite{JavaLang} and CLR \cite{ECMA335} implement a multi-threaded stack-based virtual machine, that offers many services such as dynamic loading, garbage collection, clearly the Just In Time (JIT) compilation and above all a noteworthy reflection support. Features like garbage collection raise the programming abstraction level whereas dynamic loading, JIT compilation and a native multi-thread support simplify the task of programming distributed and concurrent applications. Reflection support enables programs to read its own metadata. A program reflecting on itself extract metadata (from its representation expressed in terms of intermediate language) and using that metadata can modify its own behavior. Reflection support is useful to inspect the structure of types, to access fields and even to choose dynamically the methods to invoke. Exploiting reflection support programs can change their structure and their (byte) code.
The reflection support can be provided by the run-time system at different levels of complexity \cite{CLOS}:
\begin{itemize}
\item{Introspection} : the program can access to a representation of its own internal state. This support may range from knowing the type of values at run-time to having access to a representation of the whole source program.
\item{Intercession} : the representation of the state of the program can be changed at run-time. This may include the set of types used, values and the source code.
\end{itemize}
Both introspection and intercession require a mechanism, called reification, to expose the execution state of a program as data. The reification mechanism exposes an abstraction of some elements of the execution environment. These elements may include programming abstractions such as types or source code; they may also include other elements, like the evaluation stack (as in 3-LISP \cite{LISP}), that are not modeled by the language.
For compiled languages it could be harder to reflect elements of the source language: the object program runs on a machine that usually is far from the abstract machine of the source language. Enabling RTTI (Runtime Type Identification, a support that allows a program to have exact information about type of objects at run-time) in C++, for instance, requires that the run-time support contain additional code to keep track of types at run-time. Besides, the programmer would expect abstractions compatible with the structure of the programming language abstract machine (unless he is interested in manipulating the state of the machine that is target of the compilation).

\subsubsection{Custom metadata management}
The metadata readable through the advanced reflection supports are both the information about types (class, method, field names an hierarchies) and  about additional, non-functional attributes. %Such non-functional information are used for other purposes than mere execution.
A straightforward example is the Java serialization architecture: the programmer can declare the instances of a serializable class simply by implementing the \textbf{Serializable} interface, which in fact is an empty interface. Thus, two types that differ only for the implementation of the \textbf{Serializable} interface are indistinguishable from the execution (functional) standpoint. Besides, the serialization of the instances of non-serializable types will not be allowed by the serialization support. Clearly, this ``interface-based'' mechanism for the metadata specification is not flexible and can not be expressed at more fine level, for instance at method-level. This limitation leads to the development of Java annotations \cite{javaAnnotation}.
A Java annotation is a special syntax that adds metadata to Java source code. Annotations can be added to program elements such as classes, methods, fields, parameters, local variables, and packages. Unlike Javadoc tags, Java annotations are reflective in that they may be retained by the Java VM and made retrievable at run-time. The possibility to retain and retrieve this information at run-time makes the ``real'' difference between the Java annotations and the earlier annotation based approach. For instance, the OpenMP pragma based approach or the HPF annotation or consisting in simple directives to compiler driving the data decomposition optimization, approaches that are not designed to work with non-shared memory architectures.

The exploitation of Java annotations as a way to embed non-functional information is at the base of Attribute Oriented Programming \cite{AttributeOP1, AttributeOP2}.
Attribute Oriented Programmers use Java annotations to mark program elements (e.g. classes and methods) to indicate that they maintain the application-specific or domain-specific semantics. As an example, some programmers may define a ``logging'' attribute and associate it with a method to indicate the method should implement a logging function, while other programmers may define a ``web service'' attribute and associate it with a class to indicate the class should be implemented as a web service. Attributes aim the separation of concerns: application's core logic (or business logic) are clearly distinguished from application-specific or domain-specific semantics (e.g. logging and web service functions). By hiding the implementation details of those semantics from program code, attributes increase the level of programming abstraction and reduce programming complexity. The program elements associated with attributes are transformed in order to fit the programmers' requirements.

The effectiveness of the approach is demonstrated by its rapidly diffusion, indeed some very popular and widely used programming frameworks \cite{JBoss,Spring} adopted the Attribute Oriented Programming approach as a way to embed programmers' hints and requirements. There are also some scientific works exploiting annotations information to drive the application run-time transformation, for instance in \cite{1066964} authors propose a way to transform an annotated application in a multithreaded one and \cite{fraclet} describes a way to transform a POJO in a Fractal component simply transforming the code according to the programmer annotations.

In Section \ref{PAL} we describe how we exploited the Attribute Oriented Programming approach in our Parallel Abstraction Layer (PAL). PAL is a metaprogramming engine able to dynamically restructure parallel applications depending both on the information gathered at run-time about the running platform and on the hints specified inside the source code by programmers. 

A slightly different approach that aims to a clear separation between the application business code and application management information is the Aspect Oriented Programming (AOP) model. Whereas the Attribute Oriented Programming model separates the management code from the business one exploiting a language support, Aspect Oriented Programming model requires programmers provide additional files containing a set of rules which describe the actions to perform when the application execution flow reach certain points. The main actions performed consist in code injection and code substitution.
Some scientific works exploit AOP for code transformations.
Sobral et al. discussed the usage of AOP to support modular computing \cite{Sobral:aosd-acp4is06,SobralIpdps2006,10.1109/PDP.2007.20}. They use AOP techniques to separately solve partition, concurrency and distribution problems and eventually show how the related aspects can be used to provide a (kernel for a) general purpose, modular parallel computing framework.
Other authors \cite{aspect-mpi-c++} demonstrated that AOP can be efficiently exploited in conjunction with components and patterns to derive parallel applications for distributed memory systems. It highly relies on the ability of the programmer to find out the right places to exploit aspects.
In \cite{bruno04} another approach exploiting aspects to parallelize Java applications from the Java Grande forum using AspectJ is presented. Good results are shown in the paper, but the procedure used to exploit aspects requires entering the program details to find out possibilities for parallelization.

In the Sections \ref{metamuskel} and \ref{muskworkflows} we describe how we integrated the AOP approach in our next generation \muskel. In particular, how we exploited the AspectJ \cite{aspectj} tool to manage the generation of macro data-flow blocks, aimed at the parallelization of workflow computations.

Both the PAL and the AspectJ integration with \muskel approaches have been published, respectively in \cite{pal} and \cite{muskaspects:cg_book:08}. In both the cases
the authors collectively contributed to the paper.

%Separation of concerns is the ``concept behind'' both of Aspect Oriented Programming (AOP) \cite{AOP} and of Attribute Oriented Programming (@OP) \cite{AttributeOP1, AttributeOP2}. It allows to program separately and in an independent way the application functional code (aka business code) and the application non-functional code (the management code, actually). These metaprogramming techniques demonstrated to be an effective way to enforce such distinction. Indeed, they have been sucessfully used for several purposes, e.g. components generation \cite{fraclet, ejb3} or  application framework programming \cite{Spring}. As described in the rest of Chapter we exploited the separation of concerns enforced by AOP and @OP techniques for expressing non-functional information in structured parallel programming frameworks. More precisely, we exploited these techniques in proper structured programs to provide tools enabling code parallelization as enforcement of non-functional performance requirements.

\section{The PAL experience}\label{PAL}

%We fully subscribe the opinion ``\emph{...people know the application domain and can better decompose the problem, compilers can better manage data dependence and synchronization}'' \cite{grimshaw93mentat}. Indeed, the PAL approach to parallel grid programming relies on programmer knowledge to ``structure'' the parallel schema of application and then to the compiler/run time tool ability to efficiently implement the parallel schema conceived by the programmer. The general idea is outlined in Figure \ref{fig:idea}.

\begin{figure}[!ht]
\centering
\includegraphics[width=0.85\linewidth]{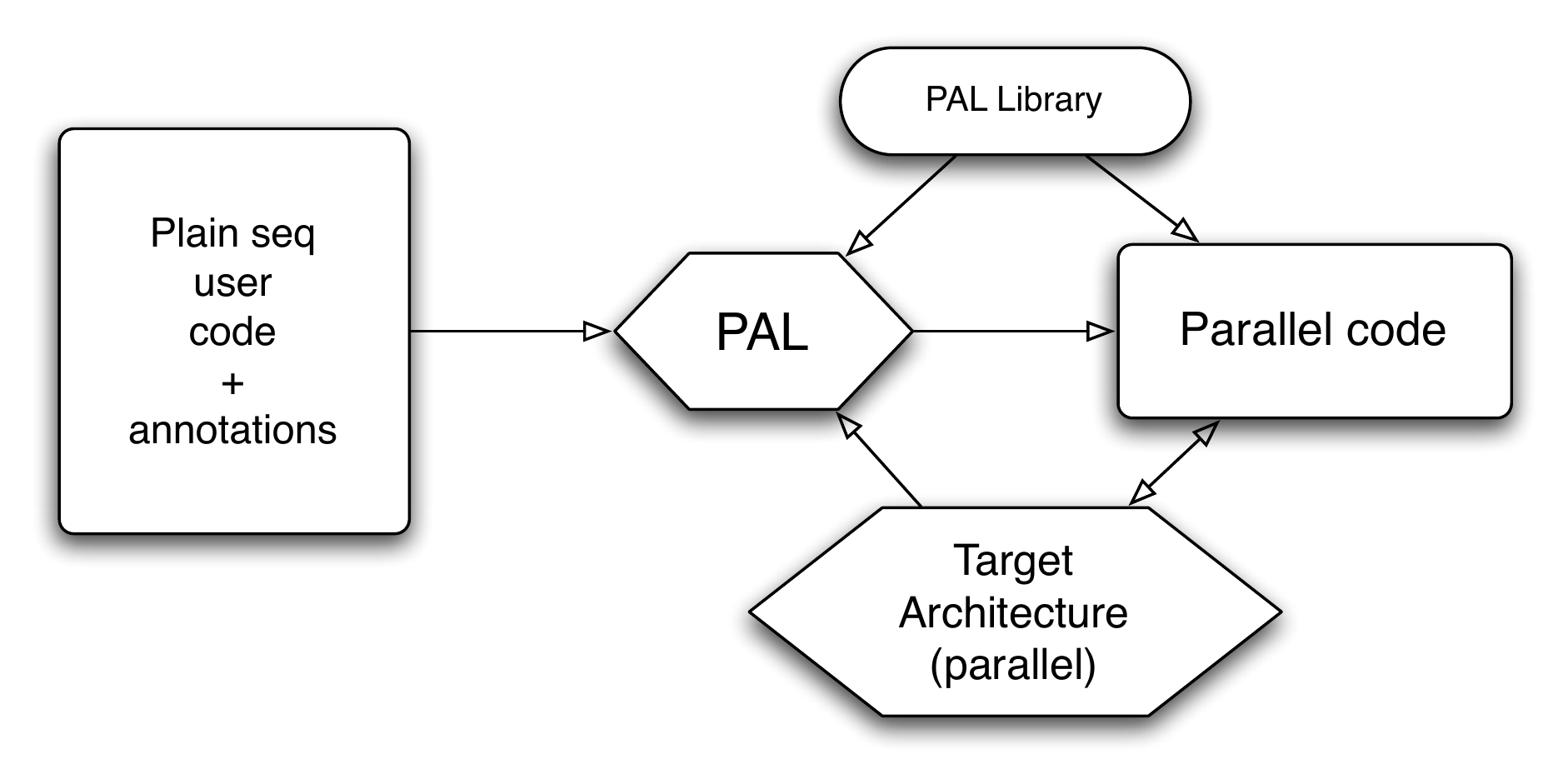}
\caption{PAL approach overview
}\label{fig:idea}
\end{figure}

\begin{figure}[!ht]
\centering
\includegraphics[width=0.95\linewidth]{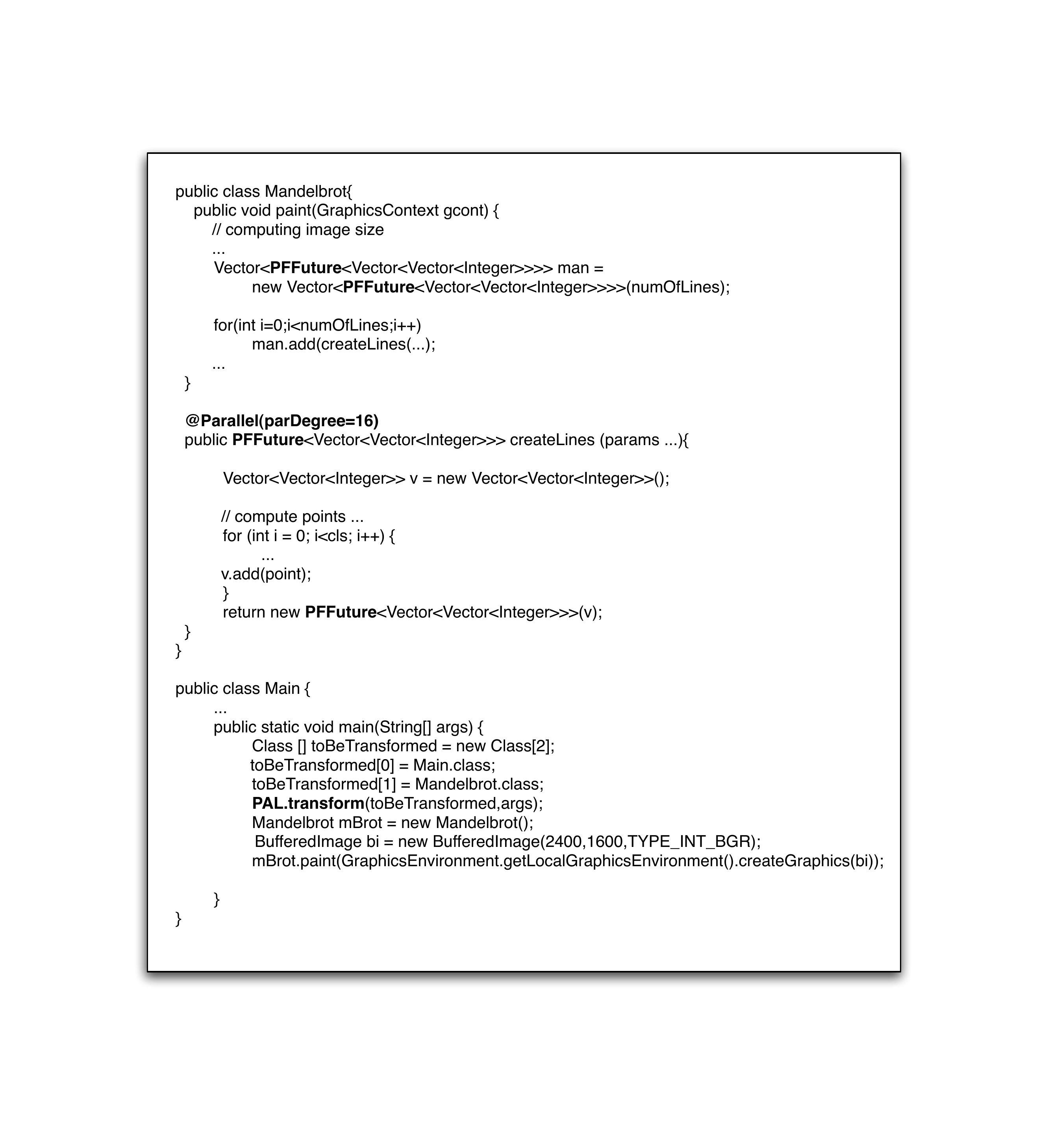}
\caption{Sample code using PAL
}\label{fig:pal}\label{fig:file-example}
\end{figure}

The Parallel Abstraction Layer is a general-purpose approach for implementing simple parallel applications that does not require complex application structuring by programmers. Programmers are only required to insert, in the source code, some hints, eventually exploited by the PAL run-time support to transform the application code. The transformation is aimed at in  enforcing an efficient parallel (even distributed) execution of the application. The general idea is outlined in Figure \ref{fig:idea}.
Programmers' hints consist in non-functional requirements, namely, requirements which specify criteria that can be used to judge the operation of a system, rather than specific behaviors. Examples of non-functional requirements includes: Efficiency, Price, Hardware Reliability, Software and tools availability and Parallelism degree. In PAL implementation they are specified through the annotation mechanisms provided by Java \cite{javaAnnotation}.
The PAL run-time support exploits the information conveyed in the annotations to transform the original program in a parallel one. The transformed program is optimized with respect to the target parallel/distributed architecture.

Programmers are required to give some kind of ``parallel structure'' to the code directly at the source code level, as it happens in the algorithmic skeleton case.
In our PAL implementation it can be done exploiting the java annotation mechanism. For instance, the farm semantics is obtained indicating which ``parts'' of code should be replicated and executed in parallel. A ``part'' is intended to be a piece of side-effect free code which input and output data are well-defined. Programmers are in charge of ensuring  the ``parts'' satisfy these requirements. Each java code ``part'' is transformed by the PAL in a macro data-flow block that can be dispatched for execution.

\begin{figure}[!ht]
\centering
\includegraphics[width=0.65\linewidth]{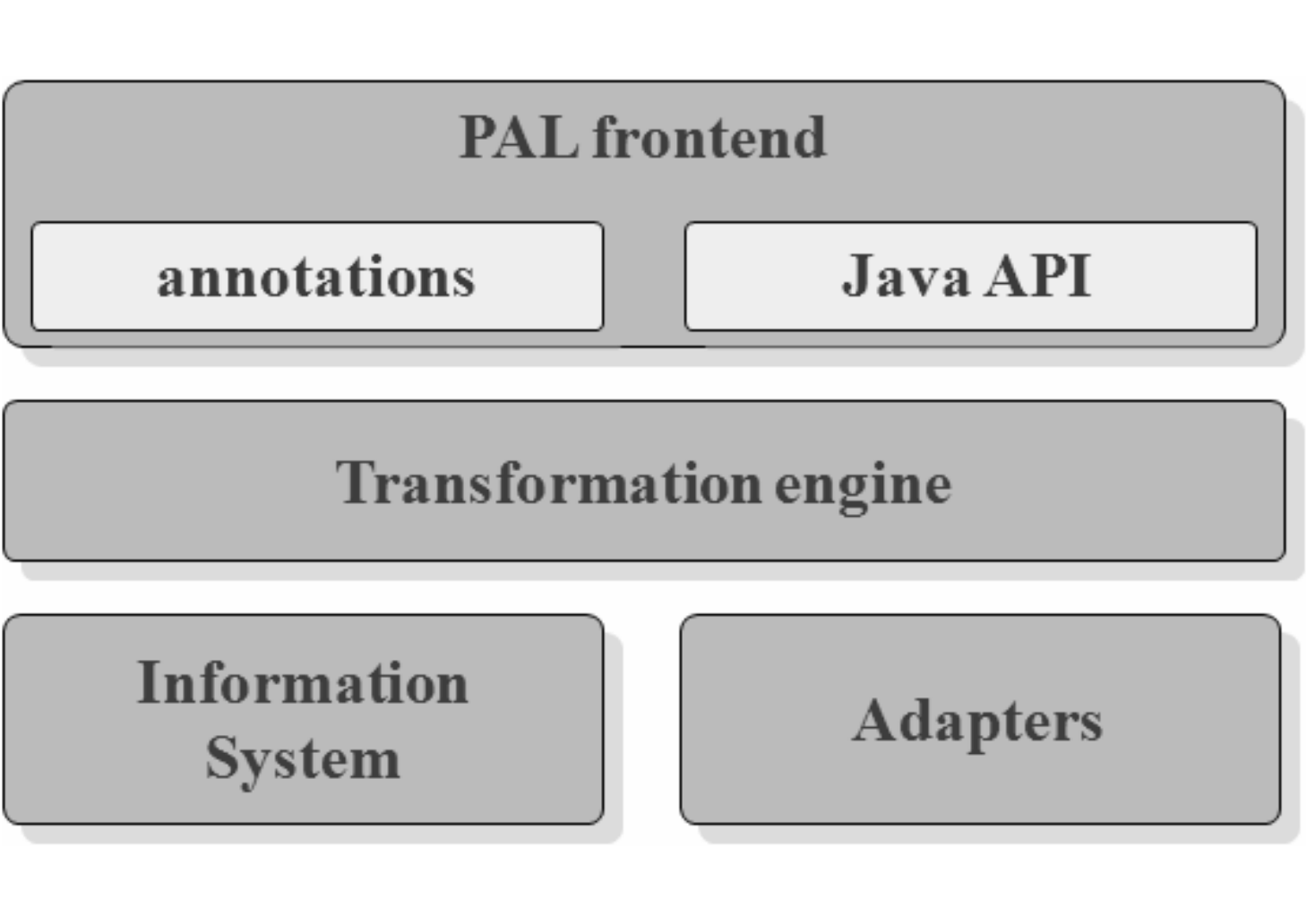}
\caption{PAL architecture}
\label{fig:PAL-arch}
\end{figure}

PAL has a multi-level software architecture. It is depicted in Figure \ref{fig:PAL-arch}. On top, there is PAL frontend, namely the annotations provided by PAL and the host language, Java in our PAL implementation. In the bottom layer, there are the adapters and the information system: the formers foster PAL during code transformation instructing it about how to structure the application code to make it parallel and compliant with a specific parallel framework.
The latter is a set of tools aimed at run-time information gathering. Finally, the middle layer is the real metaprogramming engine that uses the information gathered in order to decide which adapter exploit among the available to enforce the non-functional requirements expressed by the programmers through annotations.

Compared with traditional skeletal environments, PAL presents three additional advantages.
\begin{itemize}
\item First, annotations can be ignored and the semantics of the original sequential code is preserved. This means that the programmers' application code can be run through a classical sequential compiler (or interpreter) suite and debugged using normal debugging tools.
\item Second, annotations are processed at run-time, typically exploiting reflection properties of the hosting language. As a consequence, while handling annotations, a bunch of knowledge can be exploited which is not available at compile-time (kind of machines at hand, kind of interconnection network, etc.) and this can lead to more efficient parallel implementations of the user application.
\item Third, the knowledge concerning the kind of target architecture can be exploited leading to radically diverse implementation of the very same user code. As an example, if the run-time can figure out that the target architecture where the program is running happens to be a grid, it can transform the code in such a way possibly coarser grain parallelism is exploited. On the other hand, in case the run-time figures out that user asked to execute the code on a SMP target, a more efficient, possibly finer grain, multithreaded version of the code can be produced as the result of the annotation handling.
\end{itemize}

%In order to experiment the feasibility of the proposed approach, we considered the languages that natively support code annotations.
%Both Java and .NET frameworks provide an annotation mechanism. They also provide an intermediate language (IL) \cite{IL}, portable among different computer architecture (compile once -- run everywhere), and holding some information typically only available at source code level (e.g. code annotations) that can be used in the runtime for optimization purposes.

PAL enforces code optimizations via automatic application restructuring in order to exploit all the available application parallelism with respect to programmer's annotations (non-functional application requirements). The transformation process is done at run-time, which is at the time we have the information we need to optimize the restructuring process with respect to the available parallel tools and underlying resources. The code is transformed at bytecode level thus, it does not need to recompile the application source code on the target architecture. Hence, the transformation introduces only a small overhead for the code transformations.

%More in detail, we designed a \textit{Parallel Abstraction Layer} (PAL) filling the gap between the traditional and the parallel programming metaphor.
%
The generative \cite{genProgramming} metaprogramming engine of PAL gathers at run-time information on available parallel tools and computational resources. Then, it analyzes the  bytecode looking for programmer annotations (non-functional requirements) and transforms the annotated original code to an optimized, parallel one. The structure of the transformed bytecode depends on the selected parallel framework (clearly subjected to adapters availability) and on the presence and/or value of some non-functional requirements.

PAL exploits the available parallelism by asynchronously executing parts of the original code. The parts to be executed asynchronously are individuated by the  annotations specified by programmers.  In particular, in Java the most natural choice consists in individuating methods calls as the parts to be asynchronously executed.
%PAL translates the  byte-code of the part (method) annotated as parallelizable by structuring them as needed by the selected parallel tools/libraries exploiting the corresponding adapter.
Asynchronous execution of method code is based on the concept of \emph{future} \cite{caromel05theory, caromel04asynchronous}. When a method is called asynchronously it immediately returns a future, that is a stub ``empty'' object. The caller can then continue its own computations and access to the future object content (e.g. calling its methods) just when needed. If in the meanwhile the return value has already been computed, the call to reify the future succeeds immediately, otherwise it blocks until the actual return value is computed and then returns it.

In our PAL implementation, to indicate a method as ``parallelizable'' PAL programmers have simply to put a proper \textbf{@Parallel} annotation enriched with non-functional requirements, such as the required parallelism degree, on the line right before method declaration.
Exploiting the annotation mechanism allows to keep the PAL applications very similar to normal sequential applications, actually. Hence, Programmers may simply run the application through standard Java tools to verify it is functionally correct. PAL autonomically performs at run-time activities aimed at achieving the asynchronous and parallel execution of the PAL-annotated methods and at managing any consistency related problems, without any further programmer intervention.
The PAL approach also avoids the proliferation of source files and classes, that is a quite common situation in framework based programming, as it works transforming bytecode. Unfortunately, it raises several problems related to data sharing management. As an example, methods annotated with a  \textbf{@Parallel} should not access class fields: they may only access their own parameters and the local method variables. This is due to the impossibility to intercept all the accesses to non-private class fields. This limitation prevent the usage of static class fields as a way for sharing data among different instances of annotated method calls, making more complex the development of application in which the computational resources running the different annotated method calls need to exchange data during the method computation. It is worth to note that this is not a limitation of the approach but depends by the Java language. Indeed having a proper language support for detecting public field changes it would not be difficult to provide a proper annotation for managing the remote accesses to fields.

\subsection{PAL: implementation details}\label{sec:pro}\label{PALimpl}
We implemented a PAL prototype in Java 1.5, as Java provides a manageable intermediate language (Java bytecode \cite{javaVMSpec}) and natively supports code annotations, since version 1.5. Furthermore, it owns all the properties needed by our approach (e.g. type safety and security).
For this implementation we developed two distinct adapters. One for transforming the bytecode in a multithreaded one and another to transform the bytecode making it compliant with JJPF. In order to do this our PAL implementation makes better usage of ASM \cite{bruneton02asm}: a Java bytecode manipulation framework.

%To enable the PAL support, the programmer has to insert the \texttt{@Parallel} java annotation on the methods he wants to candidate for distributed/asynchronous invocation.  In addition, with PAL, a programmer can specify some non-functional requirements.
The current PAL prototype accepts only one kind of non-functional attribute that can be specified with the \textbf{@Parallel} annotation: \textbf{parDegree}. It denotes the number of processing elements to be used for the method execution. PAL uses such information to make a choice between the multithreaded and JJPF adapter. This choice is driven by the number of processors/cores available on the host machine: if the machine owns a sufficient number of processors the annotated bytecode directly compiled from user code is transformed in a semantically equivalent multithreaded version. Otherwise, PAL chooses to transform the compiled bytecode in a semantically equivalent JJPF version that uses several networked machines to execute the program.
PAL basically transforms code in such a way the annotated methods can be computed asynchronously. The original code is ``adapted'' using an adapter in order to be compliant with the parallel framework associated with the adapter. In our implementation, where the only available adapter for distributed computations is the JJPF one, the methods are adapted to be run on the remote JJPF servers displaced onto the processing elements. Conversely, the \textbf{main} code invoking the \textbf{@Parallel} methods is used to implement the ``client'' code, i.e. the application the user runs on its own local machine. This application eventually will interact with the remote JJPF servers according to proper JJPF mechanisms and protocols.
% INIZIO - aggiunto il 19/4
Method call parameters, the input data for the code to be executed asynchronously, are packaged in a ``task''. When a server receives a task to be computed, it removes its server-descriptor from the processing elements available for JJPF. When the task computation is completed the server re-inserts its descriptor from the available ones. In other words, when a annotated method is called an empty future is immediately returned, a ``task'' is generated and it is inserted into the JJPF queue; eventually it is sent to one among the available processing element, which remove itself from the available resources, computes the task and returns the result that JJPF finally put inside the proper future.
% FINE - aggiunto il 19/4
This implementation schema looks like very close to a classical master/slave implementation.

We could have developed an adapter for other parallel programming frameworks as targets. As an example, we could have used the Globus toolkit. However, JJPF is very compact and required a slightly more compact amount of code to be targeted, with respect to the Globus or other grid middleware frameworks. As the principles driving the generation of the parallel code are the same both using JJPF and other grid middleware frameworks, we preferred JJPF to be able to implement a proof-of-concept adapter prototype in a very short time.
%\begin{floatingfigure}[r]{55mm}
%\hspace*{-3em}\includegraphics[width=0.6\linewidth]{MandelbrotCode.pdf}
%%\includegraphics[width =0.4 \linewidth]{paltest.pdf}
%\caption{Sample code using PAL}\label{fig:file-example}
%\end{floatingfigure}

%
%To use JJPF, programmers must write their applications structuring them as composition of farm and pipeline patterns. JJPF basic architecture uses two components: the client, that is the user program, and the services, that are the distributed server instances that actually compute results out of input task data to execute client programs.  To exploit the JJPF features, programmers have to structure their applications dividing at least in two java class: one for the client side and one for the service side. The former drive the computation and holds the tasks to be computed. The latter must implements the \texttt{ProcessIf} interface: this requires the presence of methods to provide the input task data, to retrieve the result data and to compute result out of task data.  Using PAL, the programmers, can avoid this burden. In fact, the abstraction layer transforms the annotated byte-code in order to make it compliant with JJPF, dividing the code between the service part and client part.  In more detail, for every method annotated with the \texttt{@Parallel} keyword, the prototype create a class implementing the \texttt{ProcessIf} interface. The information needed to build such class is acquired analyzing the formal parameters and the return type of the annotated method.

As we already stated before, our current PAL prototype %therefore accepts plain Java programs with methods annotated as \verb1@Parallel1 and generates either multithreaded parallel code or parallel code suitable for the execution on a network of workstations running Java/JINI and JJPF. It
has some limitations, in particular, the only parameter passing semantics available for annotated methods is the \emph{deep-copy} one, and the program sequential semantics is not guaranteed if the class fields are accessed from inside the PAL-annotated methods.
%Finally, the prototype use only JJPF as parallel tool.
%However we want to remark that the PAL approach is general and the prototype can be easily adapted to generate code  compliant with every Java parallel tool which offers invisible deployment and invisible fault tolerance management.

%In order to enable the PAL features, programmers have only to add a few lines of code. 
Figure \ref{fig:file-example} shows an example of PAL prototype usage, namely a program computing the Mandelbrot set. The \textbf{Mandelbrot} class uses a \textbf{@Parallel} annotation to state that all the \textbf{createLines} calls should be computed in parallel, with a parallelism degree equal to \textbf{16}. Observe that, due to some Java limitations (see below), the programmer must specify \textbf{PFFuture} as return type, and consequently return an object of this type. \textbf{PFFuture} is a template defined by the PAL framework. It represents a container needed to enable the future mechanism. The type specified as argument is the original method return type. Initially, we tried to have to a more transparent mechanism for the future implementation, without any explicit Future declaration. It consisted in the run-time substitution of the return type with a PAL-type inheriting from the original one. In our idea, the PAL-type would have filtered any original type dereferentiation following the \emph{wait-by-necessity} \cite{caromel89wait} semantics. Unfortunately, we had to face two Java limitations that limit the current prototype to the current solution. These limitations regard the impossibility to extend some widely used Java BCL classes (String, Integer,...) because they are declared \textbf{final}, and the impossibility to intercept all non-private class field accesses. 

In the \textbf{Main} class, the programmer just asks to transform the \textbf{Main} class and the \textbf{Mandelbrot} ones with PAL, that is, to process the relevant PAL annotations and to produce an executable IL which exploits parallelism according to the features (hardware and software) of the target architecture where the \textbf{Main} itself is being run.

\subsection{Experimental results}\label{PALtests}
\label{sec:test}
To validate the PAL approach we ran some experiments with the current prototype we developed. In particular, the conducted experiments were aimed at evaluating the effectiveness of PAL approach. It has been evaluated measuring the overhead caused by raising the programming abstraction by means of PAL. 

We ran tests for each adapter developed, i.e. both for the multithread adapter and for the JJPF one. In other words, the tests were covering parallel transformations suiting both multiprocessor and cluster architectures. In the former case, we used, as computing resource for the test-bed, a hyper-threading bi-processors workstation (Dual Intel Xeon 2Ghz, Linux kernel 2.6). In the latter case, instead, we ran the transformed application on a blade cluster (24 machines single PentiumIII-800Mhz processor with multiple Fast Ethernet network, Linux kernel 2.4).
\begin{figure}[!ht]
    \centering \vspace*{-2em}
\includegraphics[width=1.0\linewidth]{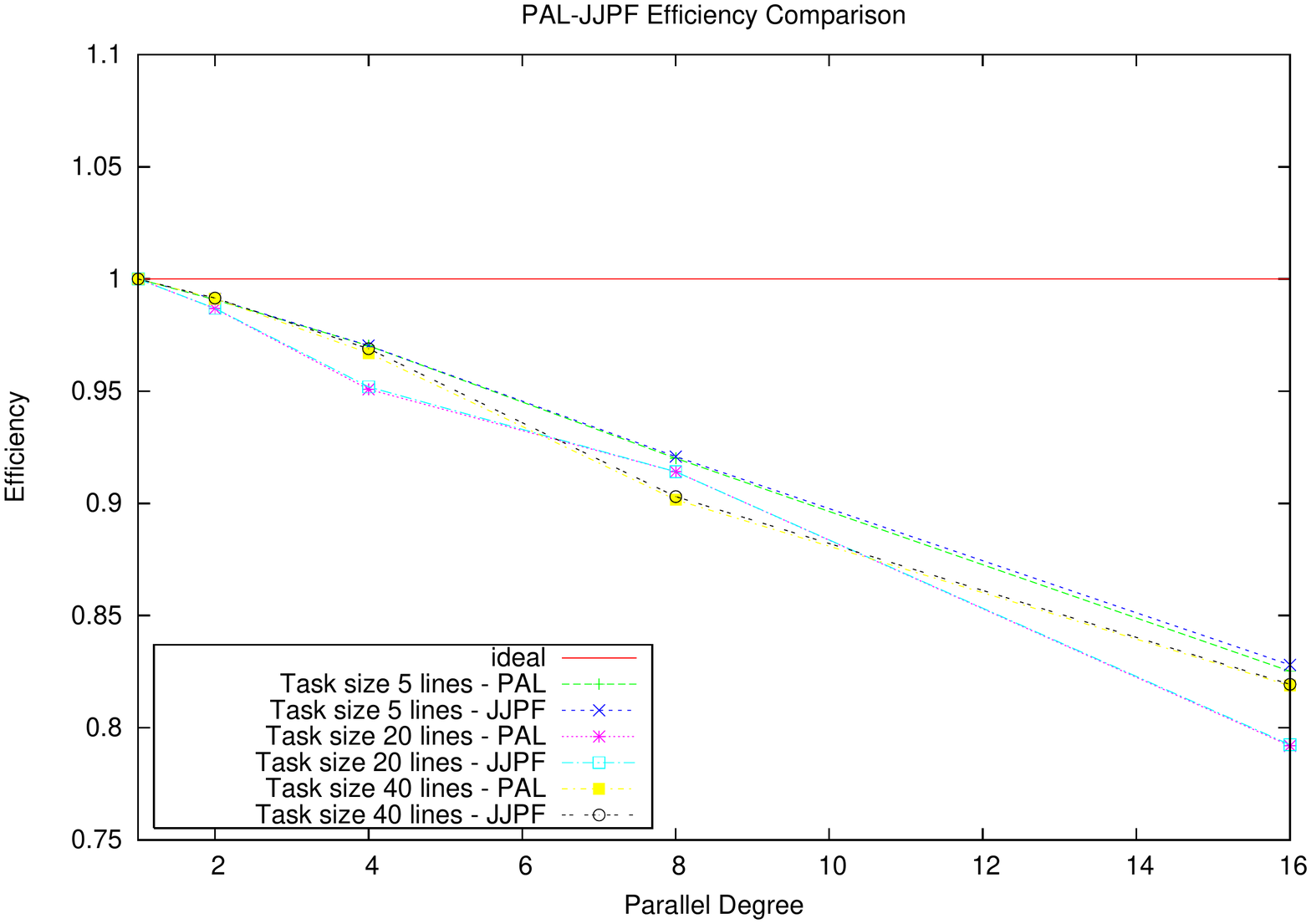}\vspace*{-2em}
    \caption{Mandelbrot computation: efficiency comparison with different image resolution, processing element number and task computational weight.}
    \label{fig:efficiency}
\end{figure}
In both cases, our test application was a fractal image generator, which computes sections of the Mandelbrot set. The Mandelbrot set is a set of points in the complex plane, the boundary of which forms a fractal. Mathematically, the Mandelbrot set can be defined as the set of complex $c$-values for which the orbit of 0 under iteration of the complex quadratic polynomial $x_{n+1} = x_{n^2} + c$ remains bounded. A complex number, $c$, is in the Mandelbrot set if, when starting with $x_0 = 0$ and applying the iteration repeatedly, the absolute value of $x_n$ never exceeds a certain number (that number depends on $c$) however large $n$ gets.
When computed and graphed on the complex plane, the Mandelbrot Set has an elaborate boundary, which does not simplify at any given magnification. This qualifies the boundary as a fractal.
We picked up Mandelbrot because it is a very popular benchmark for embarrassingly parallel computation. PAL addresses exactly these kinds of computations, as it only allows executing remotely methods not accessing shared (static) variables nor having any kind of side effects. On the one hand, this obviously represents a limitation, as PAL cannot compete, as an example, with other approaches supporting plain loop parallelization. On the other hand, huge amounts of embarrassingly parallel applications are executed on clusters, workstation networks and grids. Most of times, the implementation of these applications requires a significant programming effort, despite being ``easy'' embarrassingly parallel, far more consistent than the effort required to execute the same kind of application exploiting PAL.

To study in more detail the behavior of the transformed, parallel, version of the Mandelbrot application in several contexts, we ran the fractal generator setting different resolutions (600x400, 1200x800 and 2400x1600) and task computational weights, starting from 1 up to 40 lines at time. For each test-bed the total number of lines were fixed, hence when the task size (number of lines to compute) increases, the total number of tasks decreases.

The Mandelbrot application, when transformed exploiting the multithread adapter, has been executed only with \textbf{parDegree} parameter set to 1 or 2 (we used a bi-processor machine for the test-bed). Nevertheless, the multithreaded experiments achieved promising results, as the registered efficiency with parallel degree 2 is very close to the ideal one, for all the setting combinations (resolution and compute lines). Since in a multicore solution we have a lower communication impact than in a COW or grid solution, we can point out that this performance should be easily maintained with symmetric multiprocessors even with larger (with four, eight or more cores) processing elements.

After the test with the multithread adapter, we tested also the JJPF one for distributed architectures. We used the very same Mandelbrot source code. PAL transformed it exploiting the JJPF adapter in order to make it able to be executed on distributed workstation network. In this case, we achieved performances definitely close to the ones we achieved with hand written JJPF code (see  Figure \ref{fig:efficiency}). The Figure shows the result of the experiments with an image resolution of 2400x1600 (other results obtained using different image resolutions gave comparable results) when a different number of processing elements are used (i.e. different values specified to the \textbf{@Parallel(parDegree=...)} annotation).

These results demonstrate that PAL performance strictly depends on the parallel tool targeted by the PAL IL transformation techniques. Actually, the overhead introduced by PAL is negligible.

\subsection{Learning from PAL experience}\label{PALmotivations}%Conclusion and future work} \label{sec:conc}

Designing, developing and then testing PAL we are taught a lesson by exploiting generative metaprogramming techniques coupled with programmers high-level hints specified at source code level, it is possible to transform a java program that own some properties, enriched with some proper annotations, in a parallel program. The parallelization is obtained through the asynchronous and parallel execution of annotated methods. Annotated method code is transformed in a macro data-flow block that can be dispatched to be executed on the available computational resources.
This process executed at run-time directly at intermediate language level, allows to exploit the information available to parallelize the applications with respect both to the parallel tools available on the target execution environment and to the programmer supplied non-functional requirements. A run-time transformation allows to hide most of parallelization issues. The results we obtained are very encouraging and show that the overhead introduced by PAL is negligible. Nevertheless, the PAL prototype we developed has some limitations. The non-functional requirements are limited to the possibility to indicate the parallelism degree, the parameter passing semantic to PAL-annotated method is limited to deep-copy and the class fields are not accessible from PAL-annotated methods. Furthermore, the programmer has to include an explicit dereferentiation of objects returned by PAL-annotated methods. Finally, current PAL prototype allows only very simple forms of parallelization.

In a sense, PAL has been a proof of concept demonstrating the effectiveness of the approach. With this awareness in mind, we decided to exploit the gained experience to integrate some elements of the PAL approach in our modified muskel framework. The goal is to obtain a framework allowing programmers to develop customizable parallel structured applications which ``parts'' can be transformed in macro data-flow blocks optimized at run-time according to programmers directives and available hardware and software resources.

\section{Metaprogramming \muskel}\label{metamuskel}
\label{sec:aop}
PAL proved that, given the existence of a proper metaprogramming run-time support, annotations are a handy way both to indicate which parts of a program must run in parallel and to express non-functional requirements directly in the source code. Such information given as input to PAL metaprogramming engine can be actually exploited to optimize the original annotated code with respect to the running platform and the programmers' non-functional specifications.
Therefore, we decided to apply the main features of PAL approach to our modified \muskel implementation. Actually, adapting them to \muskel we changed a little bit the approach. Such a change is due to a few motivations. First of all because \muskel provides \emph{per se} a distributed macro data-flow executor whereas PAL exploits external tools for distributed program execution. Moreover, we would like to have a more flexible mechanism for macro data-flow block generation and management. Finally, we would like to exploit a standard tool for run-time code transformation instead of using ad-hoc tools.
%
% because our modified \muskel provides on its own the mechanisms to specify which parts have to be optimized at runtime and (possibly) run in parallel. As a consequence we exploited an alternative way to associate non-functional information to the program parts.
As a consequence we decided to use integrate in \muskel the AOP model and in particular the AspectJ framework.

The first step in this direction was exploiting AspectJ to implement aspect driven program normalization in \muskel. We already introduced normal form and code normalization in Section \ref{old-fashion}.
Let us to recall it briefly. Normalization consists in transforming an arbitrary \muskel program, whose structure is a generic skeleton tree, into a new, equivalent one, whose parallel structure is a farm with a worker made up of the sequential composition of the sequential skeletons appearing in the original skeleton tree taken left to right. This second program is the skeleton program normal form and happens to perform better (with respect to the service time) than the original one in the general case and in the same way in the worst case.

As an example, the code reported in the previous chapter in Figure \ref{fig:code} can be transformed into the equivalent normal form code:
\[
\texttt{Skeleton main = new Farm(new Seq(f,g));} \]
where \textbf{Seq} is basically a pipeline whose stages are executed sequentially on  a single processor.

\begin{figure}[!ht]
\centerline{\includegraphics[scale=0.55]{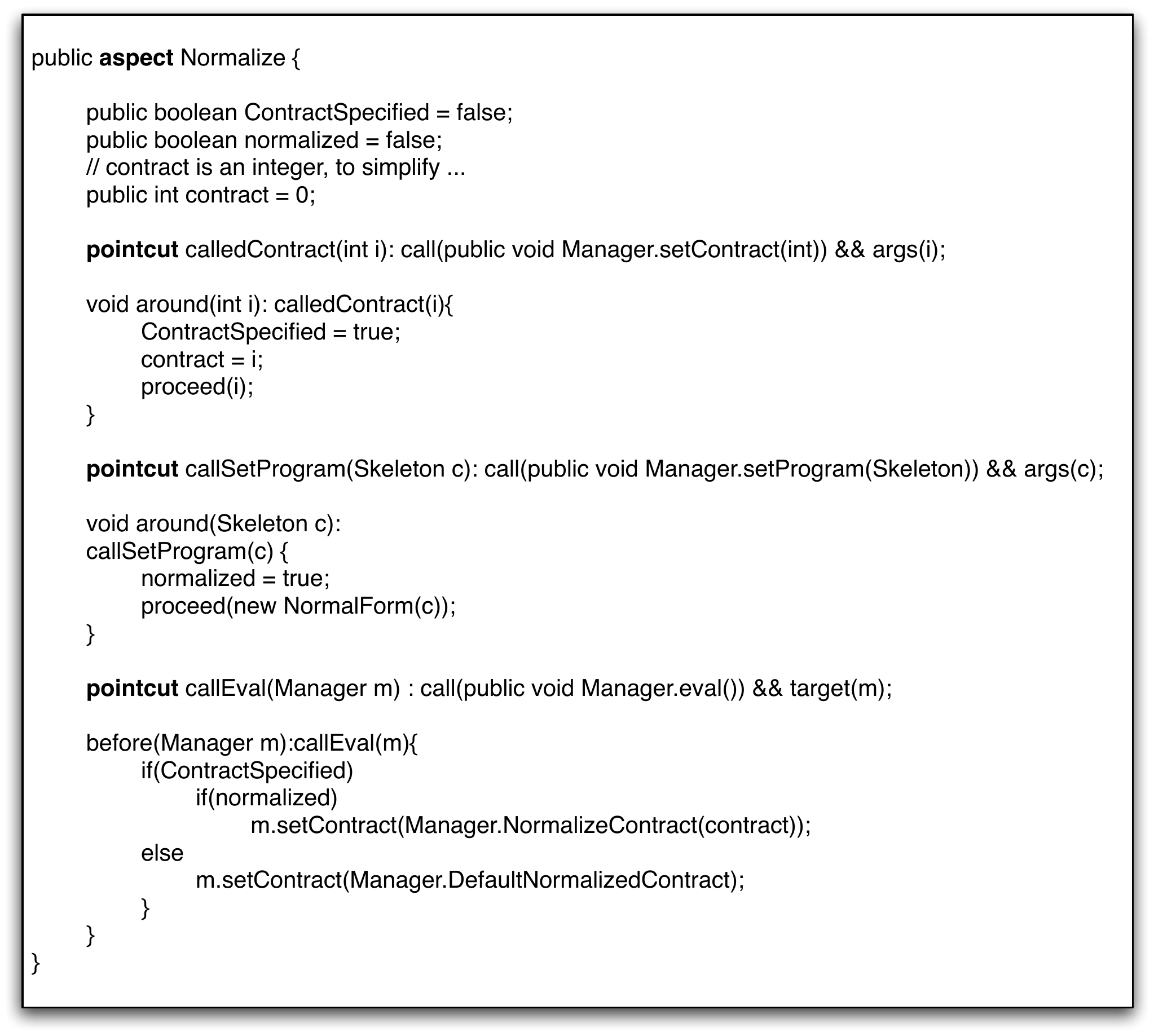}}
\caption{AspectJ code handling performance contracts in \muskel.}
\label{fig:contratto}
\end{figure}

\begin{figure}[!ht]
\centerline{\includegraphics[scale=0.55]{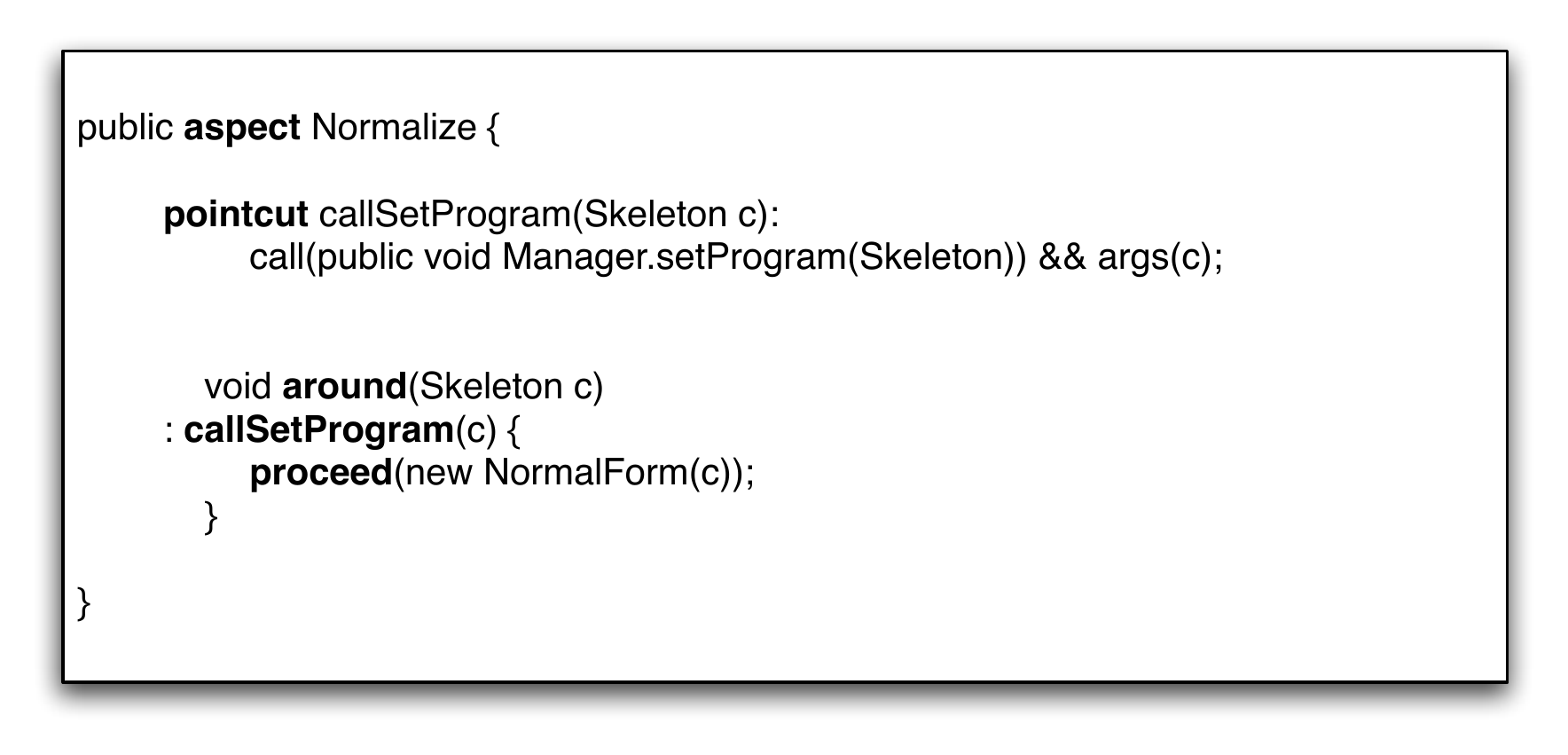}}
\caption{AspectJ code modeling normal form in \muskel.}
\label{fig:aop-nf}
\end{figure}

%As an example, let us consider our previous result on skeleton programs \textit{normal form}. Such result
%
Code normalization can be obtained explicitly inserting statements in the source code. This means that programmers must change the source code to use the normal form in place of the non-normal form version of the same program.
Exploiting AspectJ we defined a proper aspect dealing with normal form transformation by defining a pointcut on the execution of the \textbf{setProgram} \textbf{Manager} method and associating to the pointcut the action performing normal form transformation on the source code in the aspect, such as the one of Figure \ref{fig:aop-nf}.
As a consequence, the programmers can decide whether to use the original or the normal form version of the program just picking up the standard Java compiler or the AspectJ one. The fact the program is left unchanged means the programmer may debug the original bug and have the normal form one debugged too as a consequence, provided the AOP code in the normal form aspect is correct.
Moreover, exploiting aspects as discussed above, we handled also related features by means of proper aspects. In fact, in case the programmer provided a performance contract (a parallelism degree, in the simpler case) and then used the AspectJ compiler to ask normal form execution of the program, it turns out to be quite natural imagine a further aspect handling the performance contract consequently. Figure \ref{fig:contratto} shows the AspectJ aspect handling this aspect. In this case, contracts are stored as soon as they have been issued by the programmer, with the first pointcut, then, in when normalization has been required (second pointcut) and program parallel evaluation is required, the contract is handled consequently (third pointcut), that is, it is either left unchanged or a new contract is derived from the original one according to some normal form related procedure.
%
%In particular, we experimented the possibility to relief programmers of the need to specify farm skeletons at all. Instead of declaring farm skeletons, programmers may simply annotate as \texttt{@Parallel} the \texttt{Skeleton} objects and the run time support directly manages to transform calls to the \texttt{compute} methods of such objects into farms \cite{pal}. This is not a completely new technique, actually but it can be used to evaluate the clearness and effectiveness of the approach, compared both to the original \muskel\ farm handling and to a similar approach defining \texttt{Skeleton}\ objects to be computed in parallel in a farm by properly setting up a farm aspect with actions establishing task farm like computation patterns upon the invocation of the \texttt{Skeleton}\ \texttt{compute}\ method.

The second step consisted in testing the integration of \muskel with AspectJ to in a more complex scenario. Hence, we exploited the aspect oriented programming support integrated in \muskel in order to develop workflows which structure and processing are optimized at run-time.

\section{Workflows with \muskel}\label{muskworkflows}
Workflows represents a popular programming model for grid applications \cite{wfls}. In a workflow, programmers express the data dependencies that incurs among a set of blocks, possibly using a DAG. Each block processes input data to produce output data. Workflow schedulers arrange the computations for grid execution in such a way
\begin{itemize}
\item all the parallelism implicitly defined through the (absence of) dependencies in the DAG is exploited, and
\item available grid resources (processing elements) are efficiently used.
\end{itemize}
In a sense, a programming model that eases the development of efficient workflow applications can be successfully exploited for the development of many grid applications. For this reason, we conceived an approach aimed at the implementation of workflows on top of the \muskel distributed macro data-flow interpreter. We took into account the execution of workflows on a set of input data items. The set of input data items represents the program input stream. Each item on that stream will be submitted to a full workflow computation. The results of that computation will appear as a data items onto the program output stream.
Usually the workflows considered in grids are made of nodes that are computationally complex. Possibly parallel applications processing data contained in one or more input files to produce data in one or more output files \cite{wfls}.
We considered a very simple class of workflows: those whose DAG nodes are Java ``functions'' processing a generic Object input parameters to produce an Object output results.

%In \muskel, each skeleton program is translated in a macro data flow graph whose instructions (nodes) model large chunks of side effect free (i.e. functional) sequential Java code. ``Functional'' code is provided as classes implementing the \texttt{Compute} interface. This interface only includes a \texttt{Object compute (Object in)} method, which is the one use to wrap the sequential computation implementing the function. Each time a new data item is submitted to the program input stream, an instance of the data flow graph with the input data placed in the proper data flow tokens is submitted to the \muskel distributed interpreter for the evaluation. The distributed interpreter schedules fireable\footnote{a data flow instruction is \textit{fireable} iff all the input tokens it needs are present} macro data flow instructions for execution on the available processing resources and then it stores back result data tokens either in the proper positions of the graph (target macro data flow instructions) or on the program output stream (final result tokens only). %Eventually, output tokens (i.e. results of macro data flow instructions not directed to other macro data flow instructions) are delivered to the output stream. %\nota{md}{forse qualche dettaglio in piu' sul modo di implementare \muskel, o una figura}

\subsection{Aspects to implement workflows}
\label{sec:aspects}
%In a sense, the way \muskel implements skeleton programs on top of the macro data flow interpreter is definitely close to the way workflows are usually implemented on distributed architectures and grids.
%
%this is exactly what's happening when evaluating a workflow.
%: the input data is passed to the first node(s), once these nodes produce output data the results are passed to other nodes and so on, up to the moment workflow result data is eventually delivered to the user.
%
%As \muskel distributed macro data flow interpreter efficiency has already been demonstrated, we tried to exploit \muskel to implement workflow computations.
As already stated, we considered workflows processing stream of input data to produce stream of output data.
Actually, these are not classical workflows. As discussed in the following, however, classical workflows can be efficiently addressed as well as a side effect of the efficient implementation of stream parallel workflows.
This allows to express both parallelism implicit in the workflow definition (and therefore exploited within the computation of a single instance of the workflow) and stream parallelism (parallelism among distinct instances of workflow computation, relative to independent input data items).
In order to obtain a macro data-flow graph from the workflow abstract code, we exploited the AspectJ AOP framework \cite{aspectjnew}:
\begin{itemize}
\item Programmers express workflows as plain Java code, with the constraint the nodes of the workflow must be expressed using \textbf{Compute} object calls.
%\begin{verbatim}
%  Object resFirstNode = firstNode.compute(...);
%  Object resSecondNode = secondNode.compute(resFirstNode);
%\end{verbatim}
%\noindent
\item Programmers declare a \textbf{Manager} object passing it an \textbf{Iterator} providing the input tasks. The \textbf{Manager} object completely and transparently takes care of implementing stream parallelism using the \muskel distributed macro data-flow interpreter.
\item AOP pointcuts and advices are used to intercept the calls to the \textbf{compute} methods and to transform such calls into proper fireable macro data-flow instructions submitted to the \muskel distributed data-flow interpreter.
\end{itemize}

\begin{figure}[!ht]
\centerline{\includegraphics[width=0.95\linewidth]{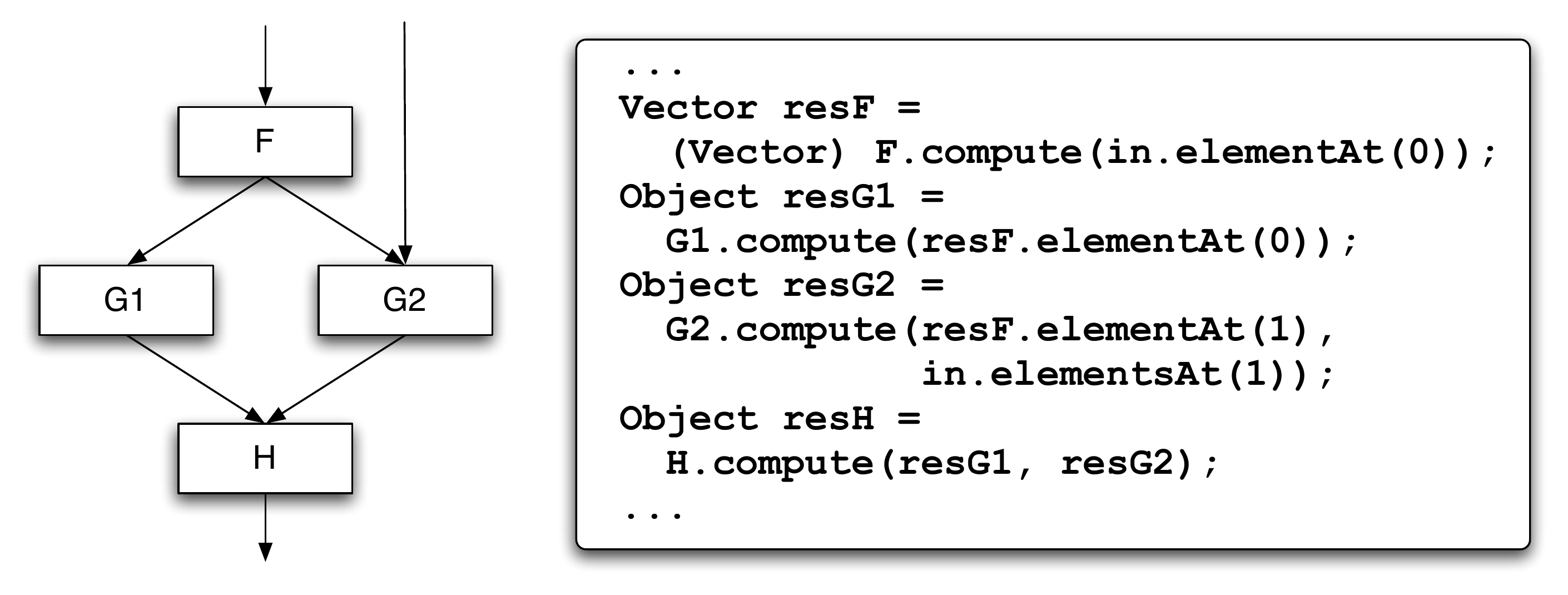}}
\caption{Sample workflow (left) and relative Java code (right)}
\label{fig:wf1}
\end{figure}

%As an example, a workflow node producing data for another workflow node can be modelled using the code:
Sample code used to model workflows is shown in Figure \ref{fig:wf1}. The right part of the Figure lists the Java code modeling the workflow graphically depicted in the left part of the Figure. Multiple results are modeled returning \textbf{Vector} objects and multiple input parameters are modeled with a ``vararg'' \textbf{compute} method\footnote{varargs have been introduced in Java 1.5 and allow to pass a variable number of arguments (of the same type) to a method; the arguments are referred to in the method body as array elements}.

More in detail, the calls to \textbf{compute} methods are transformed into the submission of a proper (already fireable) macro data-flow instruction to the muskel distributed macro data-flow interpreter modified in such a way a \textbf{Future} for the result is immediately returned.
If one of the input arguments of the \textbf{compute} call is a \textbf{Future}, the advice intercepting the \textbf{compute} method call takes care of waiting for its actual value to be computed before submitting the macro data-flow instruction to the interpreter.

\begin{figure}[!ht]
%\begin{wrapfigure}[20]{r}[0pt]{24em}
\centerline{\includegraphics[width=0.770\linewidth]{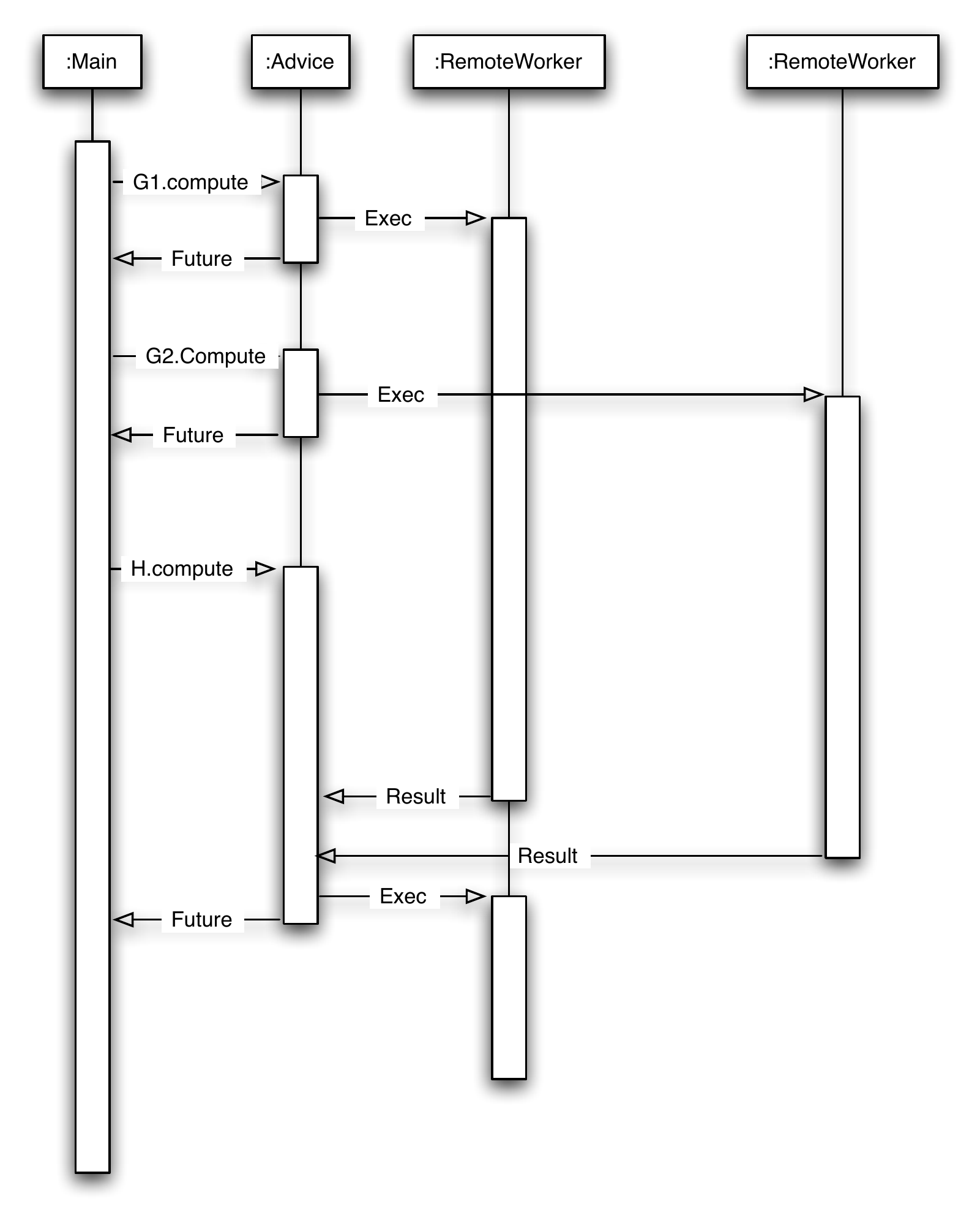}}
\caption{Transition diagram relative to the execution of part of the workflow of Figure \ref{fig:wf1}.}
\label{fig:par}
\end{figure}

As input \textbf{Future} actual values are only required by the advice right before the workflow node is started, parallelism implicit in the workflow is correctly delegated to the underlying \muskel interpreter. As an example, consider  the workflow of Figure \ref{fig:wf1}. The  functions \textbf{G1} and \textbf{G2} are evaluated (their evaluation is requested by the advice to \muskel interpreter) sequentially. However, as the first one immediately returns a \textbf{Future}, the second one (also returning a \textbf{Future}) will eventually run in parallel on a distinct remote processing element as outlined in Figure \ref{fig:par}. When the evaluation of the \textbf{H} node is requested, the advice intercepting the request will realize two futures are passed as input parameters and therefore it will wait before submitting the node evaluation request to the \muskel interpreter up to the moment the two actual values of the ``input'' \textbf{Future}s are available.
%\footnote{usage of \texttt{Future}s is clearly derived from our PAL experience \cite{pal:IW:06}}.
Overall, advices transforming calls to \textbf{compute} methods into fireable macro data-flow instructions act as the data-flow \textit{matching unit}, according to classical data-flow jargon.

%\end{wrapfigure}

The approach suggested here to implement workflows on top of the \muskel macro data-flow interpreter presents at least two significant advantages:
%a)
\begin{itemize}
\item
the whole, already existing, efficient and assessed \muskel macro data-flow interpreter structure is fully exploited. The \muskel interpreter takes completely care of ensuring load balancing, fault tolerance (w.r.t. remote resource faults) and security;
%b)
\item
programmers are only asked to express workflows with elementary Java code,
%\footnote{we are currently finalizing the customization of \muskel macro data flow design graphic interface in such a way it can automatically produce the actual Java workflow code from the GUI},
possibly spending some time wrapping workflow node code in \textbf{Compute} objects and declaring a \textbf{Manager} object which is used to supply input data, retrieve output data, control non functional features (e.g. parallelism degree in the execution of the workflow) and to ask the evaluation of the workflow code.
\item As in PAL, transformation can be easily disabled. This means that the programmers' application code can be run through a classical sequential compiler/interpreter suite and debugged using normal debugging tools.
\end{itemize}

\subsection{Aspects with \muskel: implementation details}
\label{sec:imple}
In order to be able to express workflows, the programmer must write one class per workflow node. The class has to implement the \textbf{Compute} interface, which is a very simple interface such as:

{\scriptsize
\begin{verbatim}
public interface Compute extends Serializable{
  public Object compute(Object... params);
}
\end{verbatim}
}
The \texttt{compute} method is assumed to compute the workflow node results (the returned \textbf{Object}) out of the input parameters \textbf{params}.
Then the workflow can be described in a class implementing the \textbf{Workflow} interface, which is defined as follows:

{\scriptsize
\begin{verbatim}
public interface Workflow  {
  public Object doWorkflow(Object param);
}
\end{verbatim}
}

As an example, a workflow %such as the one sketched in Fig. \ref{fig:grafo}
can be described by the class:

{\scriptsize
\begin{verbatim}
public class WorkFlow1 implements Workflow {
  public Object doWorkflow(Object task) {		
    Vector resF = (Vector) F.compute(((Vector)task).elementAt(0));
    Object resG1 = G1.compute(resF.elementAt(0));
    Object resG2 = G2.compute( resF.elementAt(1),
                               ((Vector)task).elementAt(1) );
    Object resH = H.compute(resG1, resG2);
	return resH;
  }
}
\end{verbatim}
}

The code style here is quite close to the style used when programming plain Java applications.
%The casts to \texttt{Vector} types for the output objects of the workflow nodes are needed to model multiple output arity of workflow nodes.
%We are currently developing a GUI tool that allows to generate automatically the code shown above. Fig. \ref{fig:grafo} (left) shows a snapshot relative to the usage of the GUI tool used to prepare the workflow shown above. The GUI tool generates XML code (such as the one outlined in Fig. \ref{fig:grafo} (right)) that is subsequently processed to produce the \texttt{Workflow1} class above.

%\begin{figure}
%\centerline{\includegraphics[width=0.50\linewidth]{figure/IG2}\includegraphics[width=0.480\linewidth]{figure/XMLcode3}}
%\caption{GUI used to setup the workflow in Fig. \ref{fig:wf1} (left) and excerpt of the XML code used to represent the workflow (right)}
%\label{fig:grafo}
%\end{figure}

%\paragraph{Executing workflows on \muskel}
We capture the execution of the \textbf{Compute} calls in the workflow exploiting aspects. \ The pointcut is defined on the calls of the \textbf{compute} method of any object implementing \textbf{Compute}:

{\scriptsize
\begin{verbatim}
pointcut computeRemotely(Object param[], itfs.Compute code) :
  call(Object itfs.Compute.compute(Object ... )) &&
  !within(execEngine.Engine) &&
  args(param) && target(code) ;
\end{verbatim}
}

\noindent The advice invoked on the pointcut is an \textbf{around} advice such as:

{\scriptsize
\begin{verbatim}
execEngine.Engine eng = new execEngine.Engine();

Future around(Object param[], itfs.Compute code)
 :computeRemotely(param, code) {
  for(int i=0; i<param.length; i++) {
    // reifing each parameter right before call
    if(param[i] instanceof Future) {
      param[i] = ((Future) param[i]).getValue();
    }
  }
  // deliver fireable instruction
  Object future = eng.exec(codice, param);

  // and return the corresponding Future object
  return future;
}
\end{verbatim}
}

It arranges to collect the \textbf{Compute} class name and the input parameters and creates a macro data-flow instruction, which is submitted to the distributed \muskel macro data-flow interpreter via the predefined \textbf{execEngine} object instance declared in the aspect class.
Input tokens to the macro data-flow instruction that are \textbf{Future} instances rather than plain reified objects, are eventually reified \textit{on the fly} within the advice.
Eventually, a \textbf{Future} object is returned. It can be eventually used to retrieve the actual data computed by the distributed interpreter during the \textbf{compute} call.
In particular, \textbf{Future} interface provides two methods: a \textbf{getValue()} method to get the actual value of the \textbf{Future}, possibly waiting for the completion of the corresponding computation, and a boolean \textbf{isReady()} method to test whether the computation producing the actual value of the \textbf{Future} is already terminated %\footnote{this is used to support asynchronous calls}.

%{\footnotesize
%\begin{verbatim}
%public interface Future {
%  public Object getValue();  // return the reified object (possibly waits)
%  public boolean isReady();  // true is the object can be reified immediately
%}
%\end{verbatim}
%}

%

%Actually, the code just shown executes all the workflow nodes sequentially. In our case, nodes G1 and G2 that can be potentially \nota{md}{da controllare con quanto detto prima} executed in parallel are actually executed one after the other.
As a whole, the procedure just described models an asynchronous execution of the macro data-flow instructions implementing the workflow nodes. It allows to fully exploit the parallelism intrinsic to the workflow, by properly using \textbf{Future}s.

%
%\nota{md}{va controllato nel codice dell'aspetto se possiamo mettere l'attesa ...}
%Then, the code generated out of the GUI generated XML code modeling the workflow substitutes plain usage of node results by proper calls retrieving their actual values, possibly incurring in delays, waiting for their actual computation.
%We do not show the code of the aspects here, but we will do in the final version of the paper.
%The aspect implementation just discussed allows to fully exploit the parallelism intrinsic to the workflow.

%\paragraph{Exploiting stream parallelism}
As already stated, we are interested not only in the exploitation of parallelism within the evaluation of a single workflow instance, but also in exploiting the parallelism among different instances of workflows run on distinct input data sets.
%
%In order to exploit stream parallelism on the workflow,
%we managed to implement two distinct solutions: the former, more close to the current \muskel philosophy (i.e. exploiting the manager concept: managers in \muskel are declared by the user, specialized with I/O directives and performance contracts and completely manage all the aspects related to parallel program execution) and, the latter, more close to normal programmers (e.g. sequential Java application developers) viewpoint (i.e. exploiting AOP techniques to transparently manage parallel program execution).
 %
%i.e. to allow parallel execution of workflow on several independent input items,
%\begin{itemize}
%\item
%In the former solution,
In order to support stream parallelism, we provide the programmer with a \textbf{StreamIterator} manager. This manager takes as parameters an \textbf{Iterator} (providing the input data sets to be processed by the \textbf{Workflow}) and a \textbf{Workflow}. It provides a method to compute the whole bunch of inputs, as well as a method to get an \textbf{Iterator} that can be used to retrieve workflow results.
Using the \textbf{StreamIterator} manager, the \textbf{main} code relative to our example can therefore be expressed as follows:

{\scriptsize
\begin{verbatim}
 public static void main(String[] args) {
  // workflow to be used (userdef)
  Workflow wf = new WorkFlow1();

  // provide the input tasks  via an iterator (userdef)
  InTaskIterator intIt =
       new InTaskIterator();

  // declare the manager
  Manager mgr = new StreamIterator(wf,intIt);

  // start parallel computation
  mgr.go();

  // get access to result iterator
  Iterator resIt = mgr.getResultIterator();

  // while there are more results ...
  while(resIt.hasNext()) {
    // get one and
    Object result =  resIt.next();

    // process it (userdef)
    ...

  }
}
\end{verbatim}
}

The main task of the \textbf{StreamIterator} manager is to invoke execution of the parameter \textbf{Workflow} instances on all the input data sets provided by the \textbf{Iterator}.
This is achieved exploiting a proper \textbf{Thread} pool and activating one thread in the pool for each independent workflow computation.
Then, the AOP procedure illustrated above intercepts the calls to \textbf{compute} methods and arrange to run them in parallel through the \muskel distributed macro data-flow interpreter.

\subsection{Experiments}\label{sec:perfresults}
\label{sec:exp}
%\nota{md}{dettagli su overhead per pointcut/advice/aspetti e scalabilita' e load balancing per muskel secco}
In order to prove the effectiveness of the approach, we tested it making some experiments on a distributed computing architecture (a network of workstations, actually). We directly used Java (version 1.5) accessible via plain secure shell (\textbf{ssh/scp}) rather than with other more sophisticated grid middleware. 
It is worth to point out that the tests have not been conducted to evaluate the scalability of plain \muskel, that has actually already been demonstrated, as discussed in \cite{muskel:qos:pdp:05}. Rather, the tests have been performed in order to give an estimation of the overhead introduced by aspectj transformations.

%\begin{wrapfigure}[15]{r}[0pt]{24em}
\begin{figure}
\centerline{\includegraphics[width=0.95\linewidth]{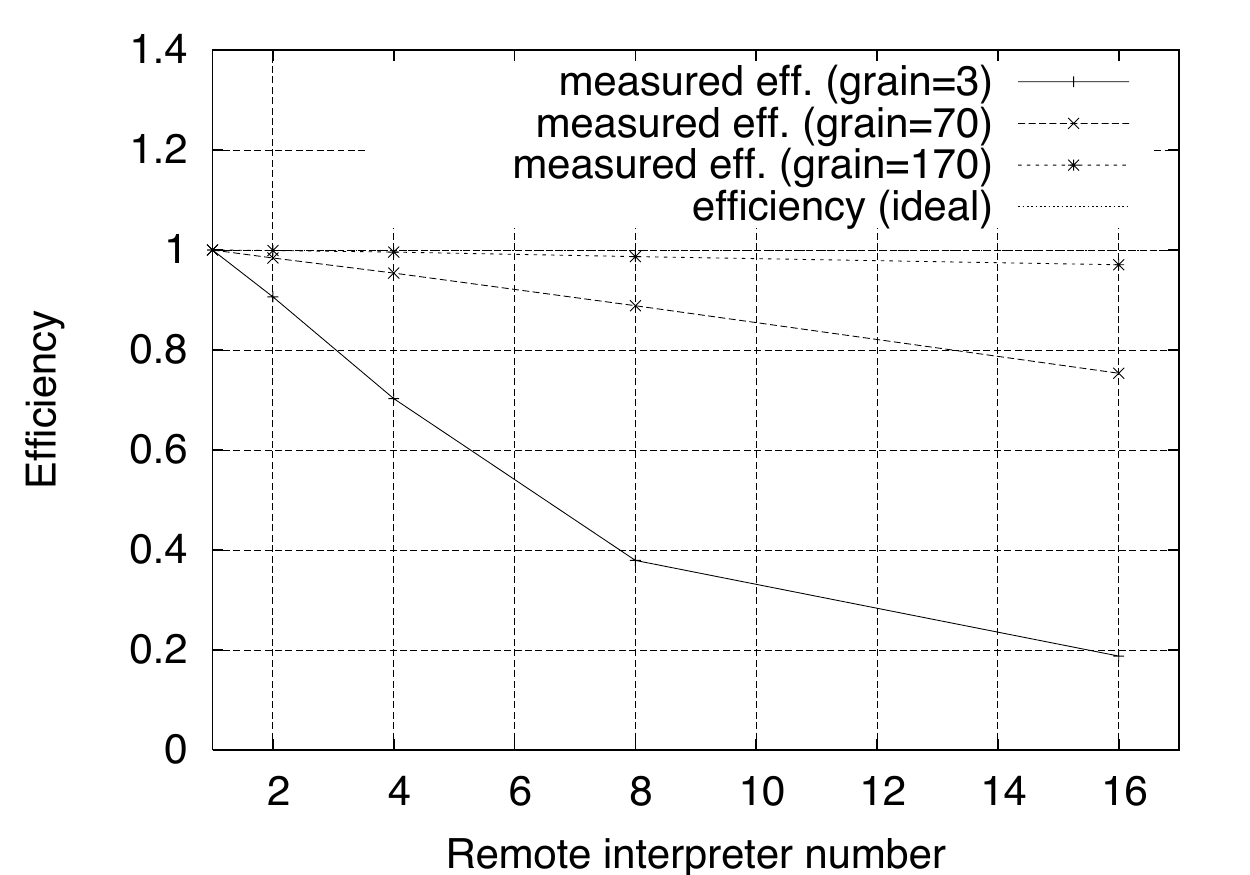}}
\caption{Efficiency of the muskel/aspect workflow prototype}
\label{fig:perf}
%\end{wrapfigure}
\end{figure}
In fact, the only difference between plain \muskel and the system proposed here, able to execute workflows on top of \muskel, lies in the way the fireable instructions are provided to the distributed data-flow interpreter of \muskel. 
Actually, in plain \muskel, fireable instructions are retrieved from a compiled representation of a data-flow graph. In particular, each time a new token arrives to a macro data-flow instruction in the graph (either from the input stream or as the result of the distributed computation of another macro data-flow instruction) the target data-flow instruction is checked for ``fireability'' and, possibly, delivered to the distributed macro data-flow interpreter. The time spent is in the sub-micro second range (only considering net time, not taking into account time spent to copy parameters in memory during the interpreter call).
When executing workflows according to the approach discussed here, instead, fireable instructions are generated by means of the aspectj tool. In particular, they come from the ``advice'' invoked on the ``pointcut'' intercepting the \textbf{compute} calls. In order to estimate the overhead introduced by using these Aspect Oriented Techniques we measured the time spent to intercept the \textbf{compute} calls and to transform them in macro data-flow blocks.  The measurement results are shown in the following table (times are in milliseconds):

\begin{center}
\begin{tabular}{l|r||l|r} \hline \hline
\textit{Average} & 23.09 &
\textit{Minimum} & 19 \\ \hline
\textit{Standard deviation} & 3.01 &
\textit{Maximum} & 27 \\ \hline
\hline
\end{tabular}
\end{center}

\noindent These values are relative to an Intel dual-core machine (2 GHz Core 2 Duo machine), running Mac OS/X 10.4, Java 1.5.0\_07, AspectJ 1.5.4 with AspectJ tools 1.4.2 and Eclipse 3.2.2. On the same machine, delivering a fireable instruction to the macro data-flow interpreter with plain \muskel requires a time average of 0.004 milliseconds. The difference in the times is not surprising: in the former case, we go through pure meta programming tools and we ``interpret'' each call, while in the latter we use plain (compiled) Java to handle each one of the calls.

Therefore, we can conclude the average 23 milliseconds represent the pure overhead spent each time a new fireable instruction has to be computed (i.e. each time one of the workflow \textbf{Compute} nodes is computed).
The time spent in \textbf{Future} reification (i.e. filling the object placeholder with the computed value, once available), instead, is negligible (this not taking into account the time spent to wait for actual production of \textbf{Future} values, of course).
This allows us to conclude that the parallel execution of workflows on top of \muskel slightly increases the grain required to achieve almost ideal scalability.

In fact, Figure \ref{fig:perf} shows how with suitable grain %\footnote{the grain $G ={T_{compute}}/{T_{communication}}$ is the time spent to compute a workflow node (i.e. a fireable macro data-flow instruction) on a machine divided by the time spent to sent the parameters and to receive the results to and from the machine}
of the workflow nodes (i.e. of the \textbf{Compute} functions) efficiency close to the ideal one is achieved.

%: the 23 msecs paid as overhead each time we find a fireable instruction make slightly lower the curves in Fig. \ref{fig:perf}.

%The workflow execution exploiting \muskel and aspects has been experimented on networks of Linux workstations with Java 1.5, accessible through \texttt{ssh/scp}. We proved scalability of the workflow execution when the workflow nodes have a suitable grain.\footnote{the grain $G ={T_{compute}}/{T_{communication}}$ is the time spent to compute a workflow node on a machine divided by the time spent to sent the parameters and to receive the results to and from the machine}
%The typical results are those shown in Fig. \ref{fig:perf}, showing the efficiency of the execution of the same workflow, processing a bunch of $1K$ input data sets on up to 16 processing elements.

%%\nota{md}{misuro l'overhead introdotto dagli aspetti rispetto al tempo richiesto per sottomettere una istruzione fireable e basta}
%We measured the overhead introduced by using Aspects to transform \verb1Compute1 calls into macro data flow instructions and to get back results, with respect to what happens in the standard \muskel interpreter.
%

\section{Differences between the two approaches} \label{sec:gendifferences}
%It derives from the approach we first suggested in \cite{pal:IW:06} where we used annotations to denote side effect free method calls  in plain Java codeto be executed in parallel.
%However, the approach here is more interesting in that the whole management of \verb1Future1 objects is delegated to the aspects rather than being exposed to the application programmers. \nota{MD}{che ne dici, si puo' affermare questo? per distinguere dal lavoro presentato sempre a coregrid nel 2006}
%However, the approach discussed in this work presents several peculiarities.:

As we already stated before, both PAL and AspectJ enriched \muskel (AEM) were conceived, designed and implemented to provide a proof-of-concept of our metaprogramming approach to structured parallel programming. Actually, they enforce code parallelization via a hints-driven code transformation. Hints are provided by programmers in the form of java annotations (PAL) and AspectJ rules AEM. Even if the two frameworks attain the same idea, they are slightly different. The main differences between the two frameworks are:

\begin{itemize}
\item In AEM there is a sharp-cut distinction between the ``control'' and ``business'' code, actually contained in separate files, whereas with PAL programmers write business code and annotations (that behaves as control code) inside the same file.
\item PAL was conceived to exploit method-level parallelism: through a simple program enrichment process, programmers choose which Java methods-call should be transformed in asynchronous ones, i.e. PAL allows to add parallelism to legacy java code with a minimal intervention.
Instead, in AEM programmers have to implement their application as a workflow.
\item PAL provides a fixed number of annotations (hence a very limited number of action can be performed) that an adapter-based architecture exploits to transform bytecode at run-time. The transformation process depends, in a way, on the adapter used. In AEM the code transformation policies implementation is based on AspectJ, the most widely diffused tool for aspect oriented programming, which offers a rich set of mechanisms for customizing the ``aspectization'' process. As a consequence, the programmers can customize/optimize/change the transformation process by simply modifying the aspects (without a direct code update).
\end{itemize}

\newpage

\section*{Summarizing the Chapter}\label{sec:conclu}
\emph{
\hrule
\medskip
In this Chapter we presented two results, about the exploitation of metaprogramming techniques in structured parallel programming environment. We exploited those techniques in order to generate and optimize at run-time macro data-flow blocks without directly dealing with their low-level management.
First we presented a new technique for high-level parallel programming based on the introduction of a \emph{Parallel Abstraction Layer} (PAL). PAL does not introduce a new parallel programming model, but actually exploits the programmer knowledge provided through annotations to restructure at run-time the application,  hiding most of parallelization issues, once it notice the information about the running platform.
This process is executed directly at intermediate language level. This allows to have a portable code transformation mechanism without paying a complete code recompilation for each change in the code.
In order to have a proof-of-concept of the approach we developed a PAL Java prototype and we used it to perform some experiments. The results are very encouraging and show that the overhead introduced by PAL is negligible, while keeping the programmer effort to parallelize the code negligible.
Then we presented the other result we obtained integrating the AspectJ framework with our modified \muskel. We described how AOP techniques can be seamlessly used to transform a very basic kind of workflows in such a way they can be executed on distributed target architectures through the \muskel macro data-flow interpreter. How AOP techniques allow to completely separate the concerns relative to parallelism exploitation and application functional core. In particular, the same application code used to perform functional debugging on a single, sequential machine may be turned into parallel code by adding aspects, compiling it through AspectJ and then running it on the \muskel run-time support.
The way used to write workflow code is quite basic Java programming. Workflow components must implement a simple interface, and programmers are explicitly required to provide them as side effect free sequential components.
The experiments conducted show that the approach is perfectly feasible and that actual speedups can be achieved provided that the workflow nodes are medium to coarse grain.
\medskip
\hrule
}

\insertblankpage

\chapter{Behavioural Skeletons}\label{mdf_as_components}

\paragraph{Chapter road-map} %perspective
\emph{
In this chapter we present  Behavioural Skeletons, an approach, we contribute to conceive and validate, aimed at providing programmers with the ability to implement autonomic grid component-based applications that completely take care of the parallelism exploitation details by simply instantiating existing skeletons and by providing suitable, functional parameters. The model has been specifically conceived to enable code reuse and dynamicity handling. We start describing (Section \ref{sec:introComponents}) how component-based application can ease the task of developing grid applications. Then we outline the Grid Component Model (Section \ref{sec:GCMintro}) with respect to its autonomic features. %Besides, we sketch (iii) the difficulties in describing of adaptive applications (Section \ref{sec:descr:behaviour}).
After we present the Behavioural Skeletons model (Section \ref{sec:BeSke}), a set of noteworthy Behavioural Skeletons (Section \ref{sec:BeSkeSet}) and their implementation (Section \ref{sec:BeSkeImpl}).
%Moreover, we provide a concise description of the (viii) Orc notation (Section \ref{orcsummary}) that we used to (ix) formal describe the set of Behavioural Skeletons we designed (Section \ref{sec:skelDesc}).
At the end of chapter we describe a set of experiment we conducted to validate the Behavioural Skeletons model (Section \ref{sec:BeSkeExp}).
}
%\end{chapterabstract}

%------------ Article text -------------------
\section{Components to simplify Grid programming}\label{sec:introComponents}

% Grid computing potentially enables the exploitation of aggregate
% software and hardware resources beyond the current availability
% threshold in a single site. The challenge of grid computing stems from
% the fact that %On the other hand,
% the ensemble of grid platforms is
% characterized by heterogeneity, dynamism, and uncertainty
% \cite{grads:overview}.

Developing grid applications is even more difficult than programming traditional parallel applications. This is due to several factors as, the heterogeneity of resources, their worldwide distribution, their dynamic recruiting and releasing. Indeed, when programming Grid applications neither the target platforms nor their status are fixed \cite{grads:overview}.

As a consequence, grid applications need to dynamically adapt to the features of the underlying architecture in order to be efficient and/or high performance \cite{advske:pc:06}.
In recent years, several research initiatives exploiting component technology \cite{gcm:coregrid:07} have investigated the area of component adaptation, i.e. the process of changing the component for use in different contexts. This process can be either static or dynamic.

The basic use of static adaptation covers straightforward but popular methodologies, such as \emph{copy-paste}, and \emph{OO inheritance}. A more advanced usage covers the case in which adaptation happens at run-time.  These systems enable dynamically defined adaptation by allowing adaptations, in the form of code, scripts or rules, to be added, removed or modified at run-time \cite{ac:superimpostion:99}. Among them is worth to distinguish the systems where all possible adaptation cases have been specified at compile-time, but the conditions determining the actual adaptation at any point in time can be dynamically changed \cite{reconf:adaptcomp:05}.
%but here all possible adaptation cases
%must have been specified at compile time.  These systems require that
%all possible adaptations must be known a priori and must be coded into
%the application \cite{comp:static:adaptation:00,reconf:adaptcomp:05}.
%
% A second class of systems enables dynamically defined adaptation by
% allowing adaptations, in the form of code, scripts or rules, to be
% added, removed or modified at run-time.
Dynamically adaptable systems rely on a clear separation of concerns between adaptation and application logic. This approach has
%which has already appeared in the context of
%distributed computing \cite{superimposition:CSP:88}~ and component
%engineering %\cite{ac:superimpostion:99},
recently gained increased impetus in the grid community, especially via its formalization in terms of the \emph{Autonomic Computing} (AC) paradigm \cite{ngg3:06,reinefeld:dagstuhl:2004,assist:qos:euromicro:06}. The AC term is emblematic of a vast \emph{hierarchy} of self-governing systems, many of which consist of many interacting, self-governing components that in turn comprise a number of interacting,
self-governing components at the next level down \cite{AC:vision:2003}. An autonomic component will typically consist of one or more managed components coupled with a single autonomic manager that controls them.  To pursue its goal, the manager may trigger an adaptation of the managed components to react to a run-time change of application QoS requirements or to the platform status.
% Overall, depicting a component assembly with a graph, one or
% more managers may cooperatively induce transformation of the graph
% (e.g. adding a stage in a linear workflow) and/or in the alteration of
% some attribute of nodes (e.g. numeric precision of a computation
% embedded in a component).
%

In this regard, an assembly of self-managed components implements, via their managers, a distributed algorithm that manages the entire application. Several existing programming frameworks aim to ease this task by providing a set of mechanisms to dynamically install reactive rules within autonomic managers. These rules are typically specified as a collection of \texttt{when-}\emph{event}\texttt{-if-} \emph{cond}\texttt{-then-}\emph{act} clauses, where \emph{event} is raised by the monitoring of component internal or external activity (e.g. the component server interface received a request, and the platform running a component exceeded a threshold load, respectively); \emph{cond} is an expression over component internal attributes (e.g. component life-cycle status); \emph{act} represents an adaptation action (e.g. create, destroy a component, wire, unwire components, notify events to another component's manager). Several programming frameworks implement
variants of this general idea, including ASSIST \cite{van:assist:02,advske:pc:06}, AutoMate \cite{ac:automate:06}, SAFRAN \cite{ac:safran:06}, and finally the forthcoming CoreGrid Component Model (GCM) \cite{gcm:coregrid:07}. The latter two are derived from a common ancestor, i.e. the Fractal hierarchical component model
\cite{fractal:spec}. All the named frameworks, except SAFRAN, are targeted to distributed applications on grids.  %\ma{AOP?}
\label{sec:relwork}

Though such programming frameworks considerably ease the development of an autonomic application for the grid (to various degrees), they rely fully on the application
programmer's expertise for the set-up of the management code, which can be quite difficult to write since it may involve the management of black-box components, and, notably, is tailored for the particular component or assembly of them. As a result, the introduction of dynamic adaptivity and self-management might enable the management of grid %heterogeneity,
dynamism, and uncertainty aspects but, at the same time, decreases the component reuse potential since it further specializes components with application specific
management code.

From the point of view of issues to address for designing and developing next generation structured parallel programming systems, this is a big problem. Indeed, if on the one hand making components adaptive addresses the issue of handling dynamicity (issue number VII), on the other hand it impairs the code reuse (issue number V).
In this chapter we cope with this problem proposing \emph{Behavioural Skeletons} as a novel way to describe autonomic components in the GCM framework. We contributed significantly to their conception, design and implementation together with other researchers, co-authored of the papers \cite{pdp08:beske, heraklion-beske} in which we presented this model. My personal contribution has mainly concerned the definition of the task farm Behavioural Skeleton as well as the implementation of that skeleton within GridCOMP.

Behavioural Skeletons aim to describe recurring patterns of component assemblies that can be (either statically or dynamically) equipped with correct and effective management strategies with respect to a given management goal. Behavioural Skeletons help the application designer to i) design component assemblies that can be effectively reused, and ii) cope with management complexity by providing a component with an explicit context with respect to top-down design (i.e. component nesting).

% Section~\ref{sec:gcm} briefly recaps the  basic design principles of GCM;
% Sec.~\ref{sec:gcm} focuses on the architectural description of
% adaptive applications; \ma{...}

\section{GCM: the Grid Component Model}\label{sec:GCMintro}
\label{sec:gcm}
GCM is a hierarchical component model explicitly designed to support component-based autonomic applications in highly dynamic and heterogeneous distributed platforms, such as grids. It is currently under development by the partners of the EU CoreGRID Network of Excellence\footnote{\texttt{http://www.coregrid.net}}. A companion EU STREP project, GridCOMP \footnote{\texttt{http://gridcomp.ercim.org}} is going to complete the development of an open source implementation of GCM (preliminary versions are already available for download as embedded modules in the ProActive middleware suite)\footnote{\texttt{http://www-sop.inria.fr/oasis/ProActive}}.
GCM builds on the Fractal component model \cite{fractal:spec} and exhibits three prominent features: hierarchical composition, collective interactions and autonomic management. We participate to both the projects (CoreGrid \& GridComp) and collaborate for the design and development of GCM, in particular in the context of autonomic management.
The full specification of GCM can be found in \cite{gcm:coregrid:07}.

\paragraph{Hierarchical composition} As in fractal, a GCM component is composed of
two main parts: the \emph{membrane} and the \emph{content}. The
membrane is an abstract entity that embodies the control
behavior associated with a component, including the mediation of
incoming and outgoing invocations of content entities. The content may
 include either the code directly implementing functional component behavior (\emph{primitive})
or other components (\emph{composite}). In the latter case, the included components are referred as the \textit{inner components}. GCM components, as Fractal ones, can be
hierarchically nested to any level. Component nesting represents the
\emph{implemented\_by} relationship. Composite components are first class citizens in GCM and, once designed and implemented, they cannot be distinguished from primitive, non-composite ones.

\paragraph{Collective interactions}
The Grid Component Model allows component interactions to take place with several distinct
mechanisms. In addition to classical ``RPC-like'' use/provide ports (or
client/server interfaces), GCM allows data, stream and event ports to be
used in component interaction. Both static and dynamic wiring between
dual interfaces is supported. Each interface may expose several
\emph{operations} of different types. Furthermore, collective interaction
patterns (communication mechanisms) are also supported. In particular,
composite components may benefit from customizable one-to-many and many-to-one
functional interfaces to distribute requests arriving to one
component's port to many inner components
and gather requests from many inner components to a single
outgoing port.

\paragraph{Autonomic management}
Autonomic management aims to attack the
complexity which entangles the management of complex systems (as applications for Grids are) by equipping their parts with self-management facilities
\cite{AC:vision:2003}. GCM is therefore
assumed to provide several levels of autonomic managers in components,
that take care of the non-functional features of the component
programs. GCM components thus have  two kinds of interfaces:
functional and non-functional ones. The functional interfaces host
all those ports concerned with implementation of the functional features of
the component.
The non-functional interfaces host all those ports needed to support the component management activity in the implementation of the
non-functional  features, i.e. all those features contributing to the efficiency of the component in obtaining the expected (functional) results but not directly involved in result computation. Each GCM
component therefore contains an \emph{Autonomic Manager} (AM),
interacting with other managers in other components via the component
non-functional interfaces. The AM implements the autonomic cycle via
a simple program based on the reactive rules described above. In this,
the AM leverages on component controllers for the \emph{event}
monitoring and the execution of reconfiguration \emph{actions}.
In GCM, the latter controller is called the \emph{Autonomic Behaviour
  Controller} (ABC). This controller exposes server-only non-functional
interfaces, which can be accessed either from the AM or an external
component that logically surrogates the AM strategy. From the point of view of autonomic features, the GCM components exhibiting just the ABC are called \emph{passive}, whereas the GCM components exhibiting both the ABC and the AM are called \emph{active}.

\section{Describing Adaptive Applications}
\label{sec:descr:behaviour}
The architecture of a component-based application is usually described via an ADL (Architecture Description Language) text, which enumerates the components and describes their relationships via the \emph{used-by} relationship. In a hierarchical component model, such as the GCM, the ADL describes also the \emph{implemented-by} relationship, which represents the component nesting.
%Thanks to this, the whole application itself may be
%represented by a single composite component.

However, the ADL supplies a static vision of an application, which is not fully satisfactory for an application exhibiting autonomic behavior since it may autonomously change behavior during its execution. Such change may be of several types:
\smallskip
\begin{itemize}
\item \emph{Component lifecycle.} Components can be started or stopped.
\item \emph{Component relationships.} The used-by and/or implemented-by relationships among components are changed. This may involve component creation/destruction, and component wiring alteration.
\item \emph{Component attributes.} A refinement of the behavior of some components (which does not involve structural changes) is required, usually over a pre-determined parametric functionality.
\end{itemize}
\smallskip
In the most general case, an autonomic application may evolve along adaption steps that involve one or more changes belonging to these three classes. In this regard, the ADL just represents a snapshot of the launch time configuration.

% At any point in time, a GCM autonomic application can be described
% by a labeled tree, where the leaves represent primitive
% components and the root represents  the whole application. In this
% tree, the edges represent the implemented-by relationship, while a node
% label represents component attributes. These attributes may also be used to
% encode the used-by relationship since each component can be wired only
% with its siblings. In particular, a label of a node can be used to encode
% the used-by relationship of its children. In
% this regard, an autonomic application adaptation can be described by
% a transformation of the tree involving both structural modification
% (addition/removal of nodes and their relationships) and label
% rewriting. These transformations, that do not necessarily involve the
% whole application  tree since different autonomic components may
% evolve independently, can be formalized via sub-tree rewriting.

%\subsection{Management is complex}
%An autonomic GCM component is defined by its functional and
%non-functional behaviour.
%The latter, in particular, may be implemented via
%the above mentioned reactive rules within the component AM.
%The AM, following its own management logic, may trigger an adaptation
%of nested components as described in
%Sec.~\ref{sec:descr:behaviour}.
The  evolution of a component is driven by its AM, which may request management action with the AM at the next level up in order to deal with management issues it cannot  solve locally. Overall, it is a part of a distributed system that cooperatively manages the entire application.

In the general case, the management code executing in the AM of a component depends both on the component's functional behavior and on the goal of the management. The AM should also be able to cooperate with other AMs, which are unknown at design time due to the nature of component-based design. Currently, programming frameworks supporting the AC paradigm (such as the ones mentioned in Section \ref{sec:relwork}) just provide mechanisms to implement management code. This approach has several
 disadvantages, especially when applied to a hierarchical component model:
\smallskip
\begin{itemize}
\item The management code is difficult to develop and to test since the context in which it should work may be unknown.
\item The management code is tailored to the particular instance of the management elements (inner components), further restricting the component reusability possible.
\end{itemize}

\section{Behavioural Skeletons}\label{sec:BeSke}
Behavioural Skeletons aim to abstract parametric paradigms of the GCM components assembly, each of them specialized to solve one or more management goals belonging to the classical AC classes, i.e. configuration, optimization, healing and protection.

They represent a specialization of the algorithmic
skeleton concept for component management.
Behavioural Skeletons, as algorithmic skeletons,  represent  patterns
of parallel computations (which are expressed in GCM as graphs of
components), but  in addition they exploit skeletons' inherent
semantics to design sound self-management schemes of parallel
components.

As a byproduct, Behavioural Skeletons allow categorization of GCM designers and programmers into three classes. They are, in increasing degree of expertise and decreasing cardinality:
\begin{enumerate}
\item \textit{GCM users}: they use Behavioural Skeletons together with their pre-defined AM strategy. In many cases they should just instantiate a skeleton with inner components, and get as result a composite component exhibiting one or more self-management behaviors.
\item \textit{GCM expert users}: they use Behavioural Skeletons overriding the AM management strategy. However, the specialization does not involve the ABC and thus does not require specific knowledge about the GCM membrane implementation.
\item \textit{GCM skeleton designers}: they introduce new Behavioural Skeletons or classes of them. To this end, the design and development of a brand new ABC might be required. This may involve the definition of new interfaces for the ABC, the implementation of the ABC itself, together with its wiring with other controllers, and the design and wiring of new interceptors. Obviously, this requires quite a deep knowledge of the particular GCM implementation.
\end{enumerate}

Due to the hierarchical nature of GCM, Behavioural
Skeletons can be identified with a composite component with no loss of
generality (identifying skeletons as particular higher-order
components \cite{gorlatch:hoc:dagstuhl:05}). %Since component composition is defined independently from behavioural skeletons, they do not represent the exclusive means of expressing applications, but can be freely mixed with non-skeletal components.
% component that
% \smallskip
% \begin{itemize}
% \item exposes a description of its functional behaviour;
% \item establishes a parametric orchestration schema of inner components;
% \item may carry constraints that inner components are required to comply with;
% \item may carry a number of pre-defined plans aiming to cope with a given self-management goal.
% \end{itemize}

Since skeletons are fully-fledged GCM components, they can be wired and nested via standard GCM mechanisms. From the implementation viewpoint, a Behavioural Skeleton is a partially defined composite component, i.e. a component with placeholders, which may be used to instantiate the skeleton.  As sketched in Figure \ref{fig:ABC}, there are three classes of placeholders:
\begin{enumerate}
\item The functional interfaces \textsf{S} and \textsf{C} that are GCM membrane controllers (thus objects).
%These interfaces can be chosen from
%  existing GCM collective interfaces (i.e. scatter/gather), or
%  specified and implemented by leveraging on GCM interface
%  configuration mechanisms.
\item The AM that is a particular inner component. It includes the management plan, its goal, and exported non-functional interfaces.
\item Inner component \textsf{W}, implementing the functional behavior.
\end{enumerate}
\smallskip
The orchestration of the inner components is implicitly defined by the skeleton type. In order to instantiate the skeleton, placeholders should be filled with suitable entities. Observe that just entities in the former two classes are skeleton specific. Indeed, the placeholders of the third class, representing the inner components implementing the functional behavior, are filled with user-defined components. The entities part of the first two classes characterize the composite component as a higher order one orchestrating the entities of the third class; like traditional skeletons are higher order functions taking as parameter user specified functions.

Behavioural Skeletons usage helps designers in two main ways. First, the application designer benefits from a library of skeletons, each of them carrying several pre-defined, efficient self-management strategies. Then, the component/application designer is provided with a framework that helps both the design of new skeletons and their implementation.

In both cases two features of Behavioural Skeletons are exploited:
on the one hand, the skeletons exhibit an explicit higher-order functional semantics that delimits the skeleton usage and definition domain. On the other hand, the skeletons describe parametric interaction patterns and can be designed in such a way that parameters affect non-functional behavior but are invariant for functional behavior.

\section{A Basic Set of  Behavioural Skeletons}\label{sec:BeSkeSet}
Here we present a basic set of Behavioural Skeletons. Despite their simplicity, they cover a significant set of parallel computations of common usage.

\begin{figure}[ht]
\centerline{\includegraphics[scale=0.6]{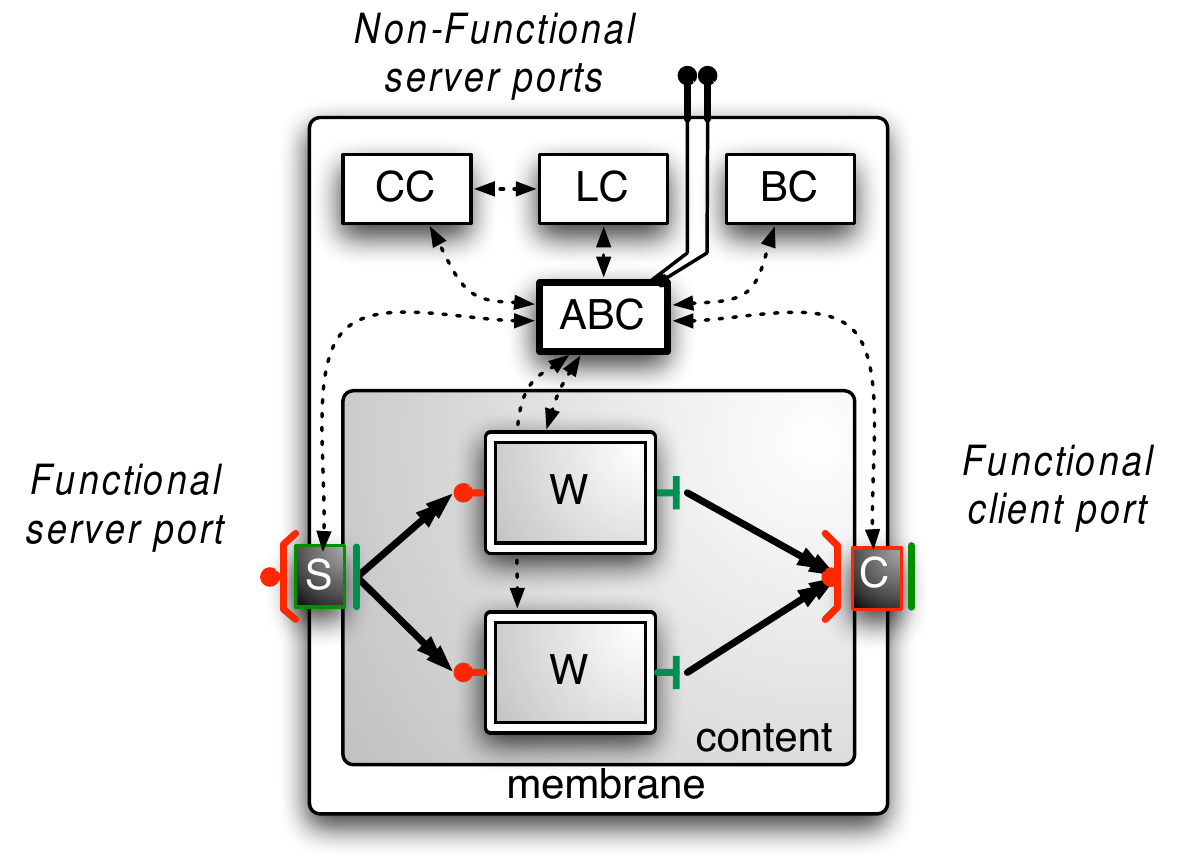}}
\caption{GCM: membrane and content (CC is the content controller, LC the lifecycle controller and  BC is the binding controller).}
\label{fig:ABC}
\end{figure}

The presented Behavioural Skeletons springs from the idea of \emph{functional replication}.
%that may come in either the \emph{stateless} or \emph{stateful}
%flavour.
Let us assume these skeletons have two functional interfaces: a one-to-many stream server \textsf{S}, and a many-to-one client stream interface \textsf{C} (see Figure \ref{fig:ABC}). The skeleton accepts requests on the server interface; and dispatches them to a number of instances of an inner
component \textsf{W}, which may propagate results outside the skeleton via \textsf{C} interface.
% that collects requests from \textsf{W}s.
Assume that replicas of \textsf{W} can safely lose the internal state between different calls. For example, the component has just a transient internal state and/or stores persistent data via an external database component.
%
% one-to-many server port \textsf{S} and a many-to-one
% client port \textsf{C}. They are connected to instances of the same
% worker component \textsf{W}. The idea is sketched in Fig.~\ref{fig:ske}.
%
\paragraph{Farm}
A task farm processes a stream of tasks $\{x_0, \ldots, x_m\}$ producing a stream of results $\{f(x_0), \ldots, f(x_m)\}$. The computation of $f(x_i)$ is independent of the computation of $f(x_j)$ for any $i \ne j$ (the task farm parallel pattern is often referred to as the ``embarrassingly parallel'' pattern).  The items of the input stream are available at different times, in general: item $x_i$ is available $t \ge 0$ time units after item $x_{i-1}$ was available. Also, in the general case, it is not required that the output stream keeps the same ordering as the input stream, i.e. item $f(x_i)$ may be placed in the output stream in position $j \ne i$.
In this case, in our farm Behavioural Skeleton, a stream of tasks is absorbed by a \emph{unicast} \textsf{S}. Then each task is computed by one instance of \textsf{W} and the result is sent to \textsf{C}, which collects results according to a  \emph{from-any} policy.
This skeleton can be equipped with a self-optimizing policy as the number of \textsf{W} can be dynamically changed in a sound way since they are stateless.
The typical QoS goal is to keep a given limit (possibly dynamically
changing) of served requests in a time frame. Therefore, the AM just checks the average time tasks need to traverse the skeleton, and possibly reacts by creating/destroying instances of \textsf{W}, and
wiring/unwiring them to/from the interfaces.
%A stream of tasks is absorbed by a \emph{unicast} \textsf{S}, each task is computed by one instance of \textsf{W} and sent to \textsf{G}, which collect tasks \emph{from-any}. This skeleton can be equipped with a self-optimizing policy because the number of \textsf{W}s can be dynamically changed in a sound way since they are stateless. The typical QoS goal is to keep a given limit (possibly dynamically changing) of served requests in a time frame. The AM just checks the average time tasks need to traverse the skeleton, and eventually reacts by creating/destroying instances of \textsf{W}s, and wiring/unwiring them to/from the interfaces.
\paragraph{Data-Parallel}
%A stream of tasks is absorbed by a \emph{scatter} \textsf{S}; each task is split in (possibly overlapping) partitions, which are distributed to replicas of \textsf{W} to be computed. Results are \emph{gathered} and assembled by \textsf{G} in a single item. As in the previous case, the number of \textsf{W}s can be dynamically changed (between different requests) in a sound way since they are stateless. As in the previous case, the skeleton can be equipped with a self-configuration goal, i.e. resource balancing and tuning (e.g. disk space, load, memory usage), that can be achieved by changing the partition-worker mapping in \textsf{S} (and \textsf{C}, accordingly).
the task farm Behavioural Skeleton can be conveniently and easily adapted to cover other common patterns of parallel computation. For example, data parallel computations can be captured by simply modifying the behavior associated with the \textsf{S} and \textsf{C} interfaces. In a data parallel computation a stream of tasks is absorbed by a
\emph{scatter} \textsf{S}. Each of the tasks appearing is split into (possibly overlapping) partitions, which are distributed to  replicas of \textsf{W} to be computed. The results computed by the \textsf{W} are \emph{gathered} and assembled by \textsf{C} in a single item, which is eventually delivered onto the output stream.
As in the previous case, the
number of \textsf{W} can be dynamically changed (between different
requests) in a sound way since they are stateless. In addition to the
previous case, the skeleton can be equipped with a self-configuration
goal, e.g. resource balancing and tuning (e.g. disk space,
load, memory usage), that can be achieved by changing the
partition-worker mapping in \textsf{S} (and \textsf{C}, accordingly).

\smallskip

The task farm (and data parallel) Behavioural Skeleton just outlined can be easily modified to the case in which the \textsf{S} is an RPC interface. In this case, the \textsf{C} interface can be either an RPC interface or missing.
Also, the stateless functional replication idea can be extended to the stateful case by requiring inner components \textsf{W} to expose suitable methods to serialize, read and write the internal state. A suitable manipulation of the serialized state enables the reconfiguration of workers (also in the data-parallel scenario \cite{advske:pc:06}).
%\paragraph{Active-Replication} A stream of tasks is absorbed by a \emph{broadcast} \textsf{S}, which sends identical copies to the \textsf{W}s. Results are sent to \textsf{G}, which \textsf{reduces} them. This paradigm can be equipped with a self-healing policy because it can deal with \textsf{W}s that do not answer, produce an approximate or wrong answer by means of a result reduction function  (e.g. by means of averaging or voting on results).

% The presented behavioural skeletons can be easily adapted to the case that \textsf{S} is a RPC interface. In this case, the \textsf{C} interface can be either a RPC interface or missing. Also, the  functional replication idea can be extended to the stateful case by requiring the inner components \textsf{W}s to expose suitable methods to serialize, read and write the internal state. A suitable manipulation of the serialized state enables the reconfiguration of workers (also in the data-parallel scenario \cite{advske:pc:06}).

Anyway, in order to achieve self-healing goals some additional requirements on the GCM implementation level should be enforced. They are related to the implementation of GCM mechanisms, such as component membranes and their parts (e.g. interfaces) and messaging system. At the level of interest, they are primitive mechanisms, in which correctness and robustness should be enforced ex-ante, at least to achieve some of the described management policies.

The process of identification of other skeletons may benefit from the work done within the software engineering community, which identified some common adaptation
paradigms, such as \emph{proxies}
%\cite{ac:interpositionfilters:03}
\cite{ac:proxy:04}, which may be interposed between interacting components to change their interaction relationships; and dynamic \emph{wrappers} \cite{ac:wrapping:01}. Both of these can be used for self-protection purposes. As an example, a couple of encrypting proxies can be used to secure a communication between components. Wrapping can be used to hide one or more interfaces whether a component is deployed into an untrusted platform.
% probably, right before the end of 5.1, a better hint can be given on
% the AC aspects of proxies and wrappers. I don't know if we can add a
% single statement stating that wrappers can be used to secure
% computations  on-the-fly in case the manager realizes the
% communication media involved are not secure, and in that case Ws in
% the string of workers you introduced in farm, data parallel etc. can
% be transformed substituting plain Ws with wrapped ones ...

% ; and \emph{superimposition}
% \cite{ac:superimpostion:99}, which enable the imposition of predefined but
% composable and configurable functionality on individual
% components. To the best of our knowledge, a similar effort has not
% yet been made for behaviour of the management code.

%
%
%

\section{Autonomic Components: \\design and implementation}
%\label{sec:imple} 
\label{sec:BeSkeImpl}
The two main characteristics of autonomic components are the ability to
self-manage and to cooperate with other autonomic components to
achieve a common goal, such as guaranteeing a given behavior of an entire
component-based application. In the light of this, viewing  the
management of a single component as an atomic feature enables design of
its management (to a certain extent) in isolation.
The management of a single component is therefore considered a
\emph{logically centralized} activity. Components will be able to
interact with other components according to well-defined protocols
described by management \emph{interaction patterns}, which are
established by the component model.

\subsection{The management of a GCM component}
The management of a single component is characterized by its ability to make non-trivial decisions.
Thus GCM components are differentiated as being \emph{passive} or \emph{active}, with the following meanings:
\begin{description}
\item[Passive] A component exposes non-functional
  operations enabling introspection
  (state and sensors) and dynamic reconfiguration.
   These operations exhibit a parametric but
  deterministic behavior. The operation semantics
  is not underpinned by a decision making process (i.e. does not
  implement any optimization strategy), but can only be constrained by
  specific pre-conditions that, when not satisfied, may nullify an
  operation request. All components should implement at least a
  reflection mechanism that may be queried about the list and the type of
  exposed operations.
\item[Active] A component exhibits self-managing behavior, that is
  a further set of autonomic capabilities built on top of passive level functionality. The process
  incarnates the autonomic management process: monitor, analyze, plan,
  execute. The \emph{monitoring} phase is supported by introspective operations,
  while the \emph{executing} phase is supported by re-configuring
  operations described above.
\end{description}

In the architecture of GCM components, these two features are
implemented within the Autonomic Behaviour Controller (ABC) and
Autonomic Manager (AM), respectively. Since the management is a
logically centralized activity, a single copy of each of them can
appear in a component. Notice that, this does not prevent a
parallel implementation of them for different reasons, such as
fault-tolerance or performance.  A passive component implements
just the ABC, whereas an active component implements both the ABC and
the AM.  The following relationship holds\\

\[ \SubType{\SubType{Comp}{PassiveComp}}{ActiveComp} \]

\noindent
where \code{<:} is a subtyping relation. This is described in the GCM
specification by increasing values of conformance levels
\cite{gcm:coregrid:07}.

\paragraph{GCM Passive Autonomic Components}

The ABC and the AM represent two successive levels of abstraction of
component management. As mentioned above, the ABC implements operations
for component reconfiguration and monitoring. The design of these operations is strictly related to membrane structure and implementation, and therefore the choice of implementing the ABC as a controller in the membrane was the more obvious and natural.
Within the membrane, the ABC can access all the services
exposed by sub-component controllers, such as that related to life cycle and
binding,  in order to implement correct reconfiguration
protocols. In general, these protocols depend on component structure
and behavior. However, in the case of Behavioural Skeletons they depend
almost solely on the skeleton family and not on the particular
skeleton. In this regard, the ABC effectively abstracts out management
operations for Behavioural Skeletons.

As we presented Behavioural Skeletons based on the idea of functional replication, we show the details of these skeletons.
In this case, the reconfiguration operations require the
addition/removal of workers as well as the tuning of distribution/collection strategies used to distribute and collect tasks and results to and from the workers.
The worker addition and/or  removal operations can be used to change
the parallelism degree of the component as well to remap workers on
different processing elements and/or platforms. The distribution/collection tuning operations can be used to throttle and
balance  the resource usage of workers, such as CPU, memory and IO.
The introspection operations involve  querying component status
with respect to one or more pre-defined QoS metrics. The component status
is generally obtained as a harmonized measure involving component
status and inner component status.

In the following we describe in some detail the implementation of a
reconfiguration and an introspection operation.

\paragraph{\texttt{add\_worker(k)}}
\label{sec:addw}
\emph{Semantics:} Add $k$ workers to a skeleton based on the functional replication.
\begin{enumerate}
\item \emph{Stop.} The ABC requires the \emph{Lifecycle Controller} (LC) to
  stop all the components. To
  this end, the LC retrieves from the \emph{Content Controller} (CC) the list
  of inner components \serifd{W}{1} $\cdots$ \serifd{W}{n}, and then
  issues a \texttt{stop}  on them.
\item \emph{Type Inspection.} All the \serifd{W}{1} $\cdots$ \serifd{W}{n} have
  the same type. The ABC retrieves from the CC the list of inner components
  \serifd{W}{1} $\cdots$ \serifd{W}{n}, then retrieves TypeOf(\serifd{W}{1}).
\item \emph{New.} One or more new inner components of type
   TypeOf(\serifd{W}{1}) are created.
\item \emph{Bind.} The component server interface \serif{S} is wired to newly
  created  \serifd{W}{n+1} $\cdots$ \serifd{W}{n+k} inner
  components via the \emph{Binding Controller} (BC). \serifd{W}{n+1}
  $\cdots$ \serifd{W}{n+k}, in turn,
  wire their client interfaces to the component
  collective client interface \serif{C}. The process requires the inspection of
  the types of the interfaces of \serifd{W}{1} that is used again as a
  template for all \serifd{W}{i}.
\item \emph{Restart.} The ABC requires the LC to re-start all the components.
\item \emph{Return.} Return a failure code if some of the previous operations
  failed (e.g. inner components do not implement stop/start
  operations); return success otherwise.
\end{enumerate}

\paragraph{\texttt{get\_measure(m)}}
\emph{Semantics:} Query the component about the current status of the
measure $m$, which may depend on the status of the inner components
(possibly involving other measures) and the membrane status.\\
\emph{Examples:} Transactions per unit time, load balancing, number
of up-and-running workers, etc.
\begin{enumerate}
\item \emph{Collect Workers' Measures.} The ABC retrieves from the CC the list of inner components \serifd{W}{1} $\cdots$ \serifd{W}{n}, then
  issues a \texttt{get\_measure(m)}  on each.
\item \emph{Collect Membrane Measures.} The ABC queries membrane sensors relating
  to the particular metric $m$.
\item \emph{Harmonize Measures.} Measures acquired from workers and from
  the membrane are harmonized by using a $m$-dependent function (e.g.
  average, maximum, etc.).
\item \emph{Return.} Return a failure code if some of the previous operations
  failed (e.g. sensor not implemented in inner components); return
  monitor information otherwise.
\end{enumerate}

\begin{figure*}[ht]
\centerline{\includegraphics[scale=0.6]{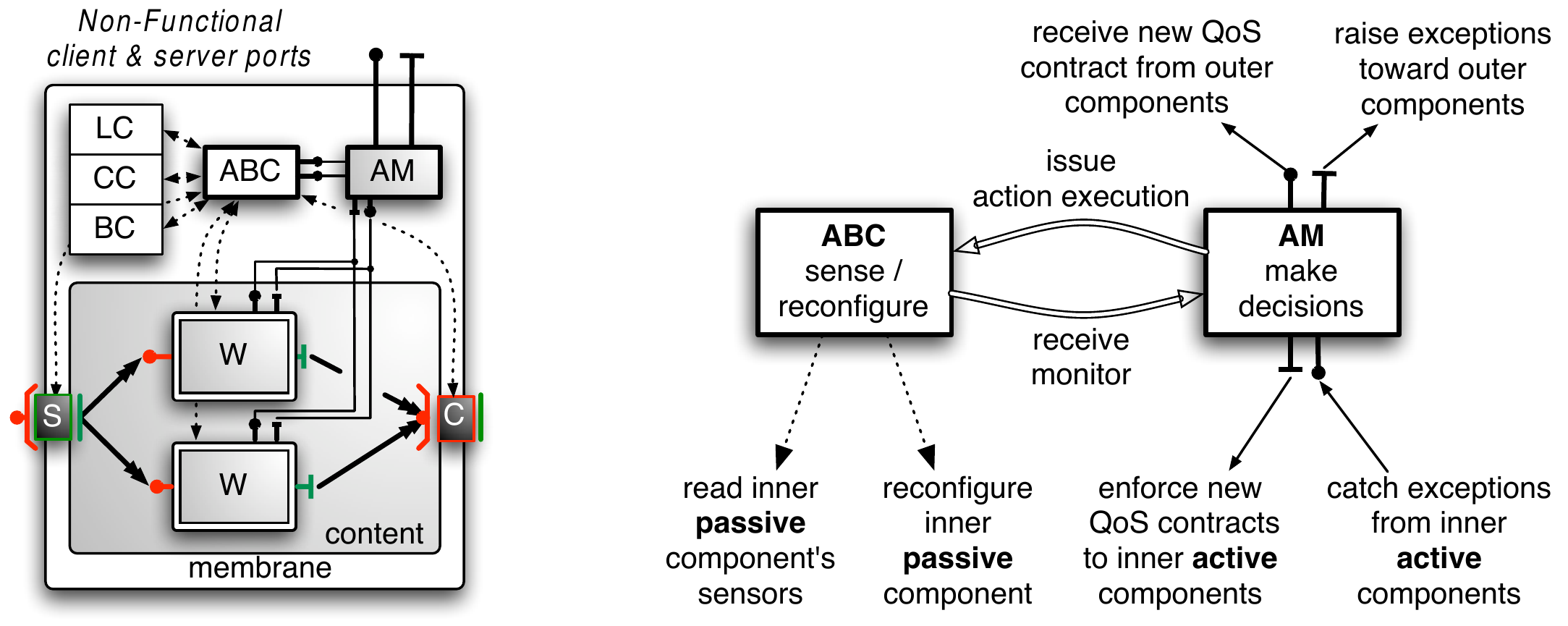}}
\caption{Left) GCM active component architecture. Right) ABC and AM interaction.}
\label{fig:AM}
\end{figure*}

\paragraph{GCM Active Autonomic components}
The operations implemented in the ABC can be arbitrarily complex;
however, they do not involve any decision making process. In general,
each of them implements a protocol that is a simple list
of actions. On the contrary, the AM is expected to enforce a
contractually specified QoS. To this end the AM should decide
\emph{if} a reconfiguration is needed, and if so, \emph{which}
reconfiguration plan can re-establish contract validity
\cite{adaptivity:parco:05}. Furthermore, as we shall see in
Section \ref{sec:coopman}, the AM should also determine if the contract
violation is
due to the managed component or is the byproduct of other components'
malfunction.  The architecture of an active GCM component is shown in
Figure \ref{fig:AM}.

The AM accepts a QoS contract\footnote{the notion of \textit{QoS contract} is still the subject of further investigations and possible refinements. The one discussed here is the bare minimum necessary to discuss AM behavior and implementation.}, which is currently defined as pair
$\langle V, E \rangle$, where $V$ is a set of variables
representing the measures the AM can evaluate (via the ABC), and
$E$ is a mathematical
expression over these variables that might include the $\min$ and $\max$
operator over a finite domain. The set of $V$ determines the
minimum set of measures the AM should be able to monitor to accept the
contract. The $E$ encodes the constraints and goal the AM is
required to pursue. This encoding can be realized in many different
ways provided $E$ can be evaluated in finite time and possibly
quite efficiently.

Having accepted a QoS contract, the AM iteratively checks its validity,
and in the case that it appears broken, evaluates a number of pre-defined
reconfiguration plans.
Each reconfiguration plan consists of a sequence of actions (to be
executed via the ABC), and a QoS forecast formula. This formula allows
 the value of a subset of $V$ after the reconfiguration to be forecast.
The AM instantiates in turn all reconfiguration plans
obtaining, for each plan, a set of forecast values. A plan is marked
as \emph{valid} if the set of $V$ updated with forecast values satisfies
the QoS contract. Among the valid plans, the AM heuristically chooses the
reconfiguration plan to be executed. If no reconfiguration plan is
valid, an exception is raised.

As is clear, the main difficulty in the AM definition is the
specification of a reconfiguration plan. In the general case, the reconfiguration
plans, and especially their forecast formula, are strictly related to
the behavior of a particular component. As discussed in Section \ref{sec:descr:behaviour},
Behavioural Skeletons enable the definition of reusable reconfiguration
plans by categorizing and restricting component behavior in families
and skeletons.

\subsection{Cooperative management}
\label{sec:coopman}
The ultimate goal of QoS management is to guarantee programmer intentions
despite software  and environmental instabilities and
malfunctions. To this end, the management of a whole system should be
coordinated to achieve a common goal. In general, we envisage a
component-based system as a graph, whose nodes are components, and
edges are relations among them, such as data
dependency, management, geographic locality, etc. Different relations
can be kept distinct by a proper labeling of edges. Here we
restrict the focus to two relations which are of particular interest for GCM:
 \textit{used\_by} and the \textit{implemented\_by} (see
Section \ref{sec:descr:behaviour}). Since the GCM is a hierarchical
model, the nesting relation naturally defines the \textit{implemented\_by}
relationship.  In particular, the application structure along the nesting
relation describes a tree whose nodes represent components (leaves are
primitive components) and edges represent their nesting. In this
case, the management of a composite component C is cooperatively
performed by the \lab{AM}{C} of the component itself and the
\lab{AM}{C_i} of the
child components $C_i, i=1..n$. In the case where inner components are
passive, the cooperation is really one of control by the outer
component: services exposed by the \lab{ABC}{C_i} are called by
the \lab{ABC}{C}.

Conceptually, non-functional properties modeling run-time behavior
of the whole hierarchy can be synthesized in a bottom-up fashion: the
behavior of a composite component depends on the behavior of its nested
components. Management actions and QoS contracts should be
projected along the tree in a top-down fashion: the
users usually would like to declare a global goal they expect from an
application. This matches the idea of submitting a contract at the
root of tree. A fully autonomic system should automatically split the
global goal into sub-goals that should then be forced on inner components.

On the whole, each GCM component enforces local decisions. When a
contract violation is detected, its AM tries autonomously to re-establish
the contract to a valid status by re-configuring its membrane or inner
components. In the event that it cannot (no valid plan), it raises an event
to its father component, thus increasing the extent of the
reconfiguration. The overall behavior enforces the maximum locality
of reconfigurations, which is a highly desirable property in a
distributed system, since it eases the mapping of components onto the
network of platforms that usually exhibit a hierarchical nature in
terms of uniformity of resources and latency/bandwidth of networks
(cluster of clusters).

Observe that cooperation between components is unavoidable even in
very simplistic applications. Let us consider an example:

\begin{figure}[ht]
\centerline{\includegraphics[scale=0.6]{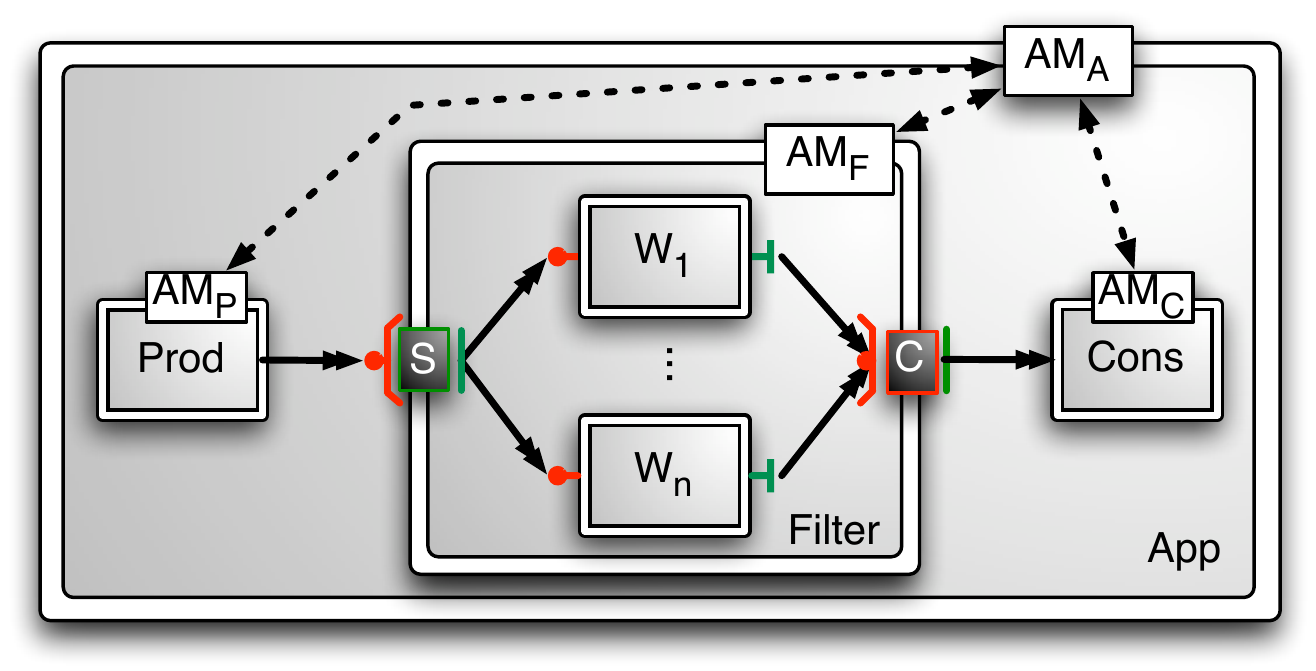}}
\caption{Producer-filter-consumer with parallel filter (farm skeleton).}
\label{fig:pipe}
\end{figure}

% \begin{figure*}
% \centerline{\includegraphics[width=0.48\linewidth]{stop}\hfill\includegraphics[width=0.48\linewidth]{restart.pdf}}
% \caption{ABC reconfiguration overhead. Left) Stop. Right) Restart.}
% \label{fig:overhead}
% \end{figure*}

\paragraph{Producer-filter-consumer} Let us assume that the application
sketched in Figure \ref{fig:pipe} has the final goal to generate,
render, and display a video with a given minimum number of frames/sec
($FPS>k$). The contract is split into three identical contracts since
the property should be enforced on all stages in order to hold globally.
The rendering (filter) has been parallelized since it is
the most CPU-demanding stage. Two common problems of such
applications are a transient overload of platform where \serifd{W}{1}
$\cdots$ \serifd{W}{n} are running,  or an increased complexity of
scene to be rendered. These events may lead to a violation of QoS
contract at the \lab{AM}{F}. In this case, it may increase the number of workers
(mapped on fresh machines) to deal with the insufficient aggregate power of
already running resources. In many cases this will \emph{locally} solve the
problem. However, a slightly more sophisticated contract should
consider also the input and output channels. In particular the filter
stage might be not rendering enough frames because it does not receive
enough scenes to render. In this case the \lab{AM}{F} can detect the
local violation, but cannot locally solve the problem. As a matter of
fact, no plan involving a change of parallelism degree can solve this
problem. \lab{AM}{F} can just signal the problem to a higher level \lab{AM}{A},
which can try to remap the input channel to a faster link, or simply
signal to the end user that the contract is not satisfied.

\section{Experiments}\label{sec:BeSkeExp}
\label{sec:experiments}

\begin{figure}
\centerline{\includegraphics[width=0.8\linewidth]{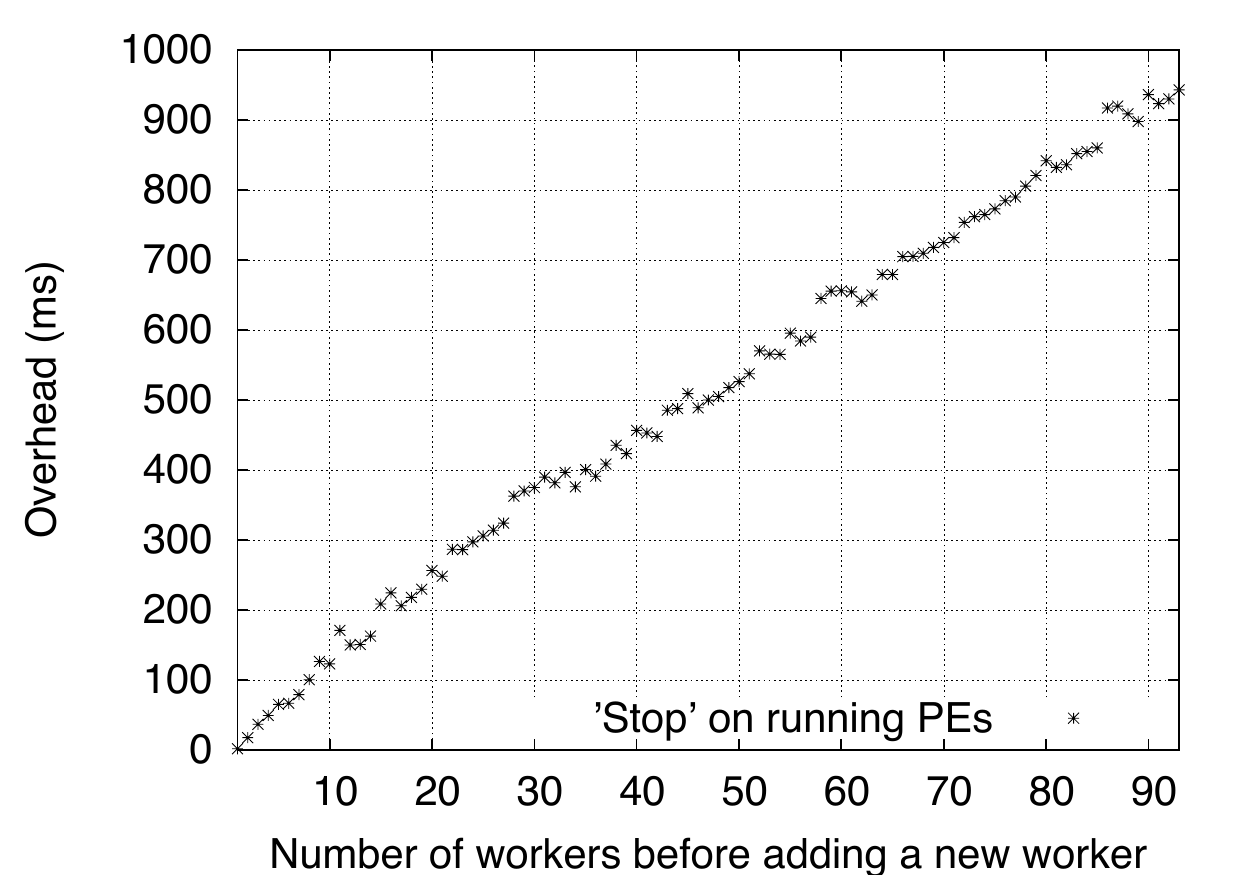}}
\caption{Reconfiguration overhead: Stop.}
\label{fig:overhead:stop}
\end{figure}

\begin{figure}
\centerline{\includegraphics[width=0.8\linewidth]{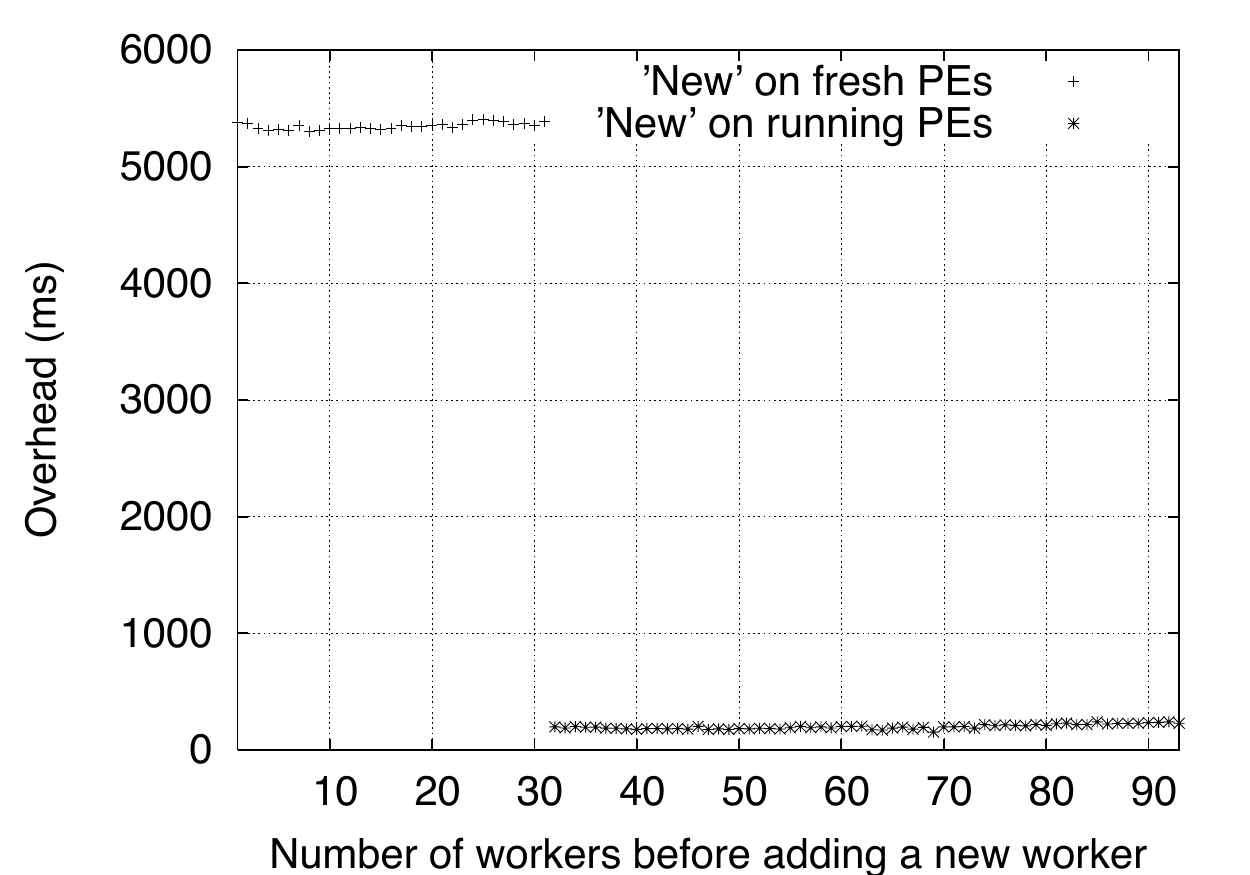}}
\caption{Reconfiguration overhead: New.}
\label{fig:overhead:new}
\end{figure}

\begin{figure}
\centerline{\includegraphics[width=0.8\linewidth]{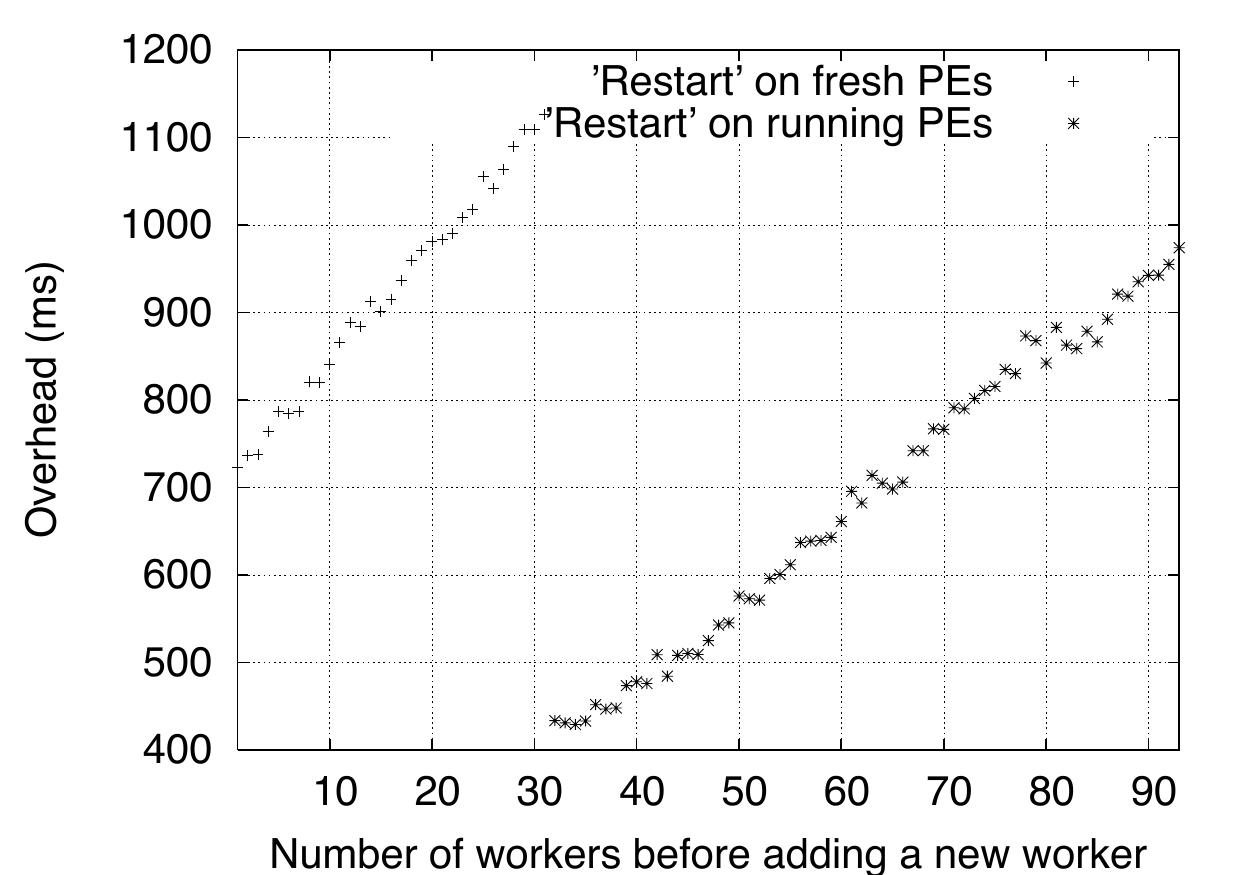}}
\caption{Reconfiguration overhead: Restart.}
\label{fig:overhead:restart}
\end{figure}

In order to validate the Behavioural Skeletons approach, we conducted some experiments with the current prototype of the GCM. It is under development in the GridCOMP STREP project \cite{gridcomp}. The prototype, which is being developed on top of ProActive middleware \cite{proactive}, includes almost all of the features described in this chapter.
% (alpha quality).
All the experimental data are measured on the application shown in Figure \ref{fig:pipe} that we already presented in the previous section. It basically is a three-stages pipeline in which the second stage consists in a farm of workers processing the images coming from the first stage, and delivering them to the third stage. The experiments mainly aim to assess the overhead due to management and reconfiguration of GCM components. For the sake of reproducibility, the experiments have been run on a cluster instead of a more heterogeneous grid. The cluster includes 31 nodes (1 Intel P3@800MHz core per node) wired with a  fast Ethernet. Workers are allocated in the cluster in a round robin fashion with up to 3 workers per node (for a total of 93 workers). Note however, the very same experimental code can run on any distributed platform supported by the ProActive middleware.

Figures \ref{fig:overhead:stop}, \ref{fig:overhead:new}, and \ref{fig:overhead:restart} respectively show the time spent on the farm Behavioural Skeleton (filter) for the \emph{stop}, \emph{new} and \emph{restart} Autonomic Behavioural Controller (ABC) services described in Section \ref{sec:addw}.
This time consists in application overhead, since in current implementation none of the
workers can accept new tasks during the process. In the figures, a point  
$k$ in the X-axis describes the overhead due to \emph{stop/new/restart}  
in the adaptation of the running program from a $k$ to \mbox{$k+1$} worker  
configuration.
As highlighted by the curves in Figure \ref{fig:overhead:stop} and
\ref{fig:overhead:restart} the overhead of \emph{stop} and
\emph{restart} is linear with respect to the number of workers
involved in the operations. This is mainly due to a linear time
barrier within the Life cycle Controller (LCC), which is an inherent part of the
underlying ProActive middleware. Indeed, in the current implementation the LCC sequentially stops all the workers. Note that adaptation process does not
strictly require such a barrier. Both stopping all the workers and
linear time synchronization are peculiarities of the current GCM
implementation on top of the ProActive middleware, and not of the farm
Behavioural Skeleton, which can be implemented avoiding both
problems. In addition, the creation of a new
worker can be executed, at least in principle, outside the critical path by using a
speculative creation.

%we encouraged the ProActive team to get rid of the linear barrier, actually, and we are quite confident this issue can be forgot soon).

Figure~\ref{fig:overhead:new} shows the time spent for the \emph{new} Autonomic Behavioural Controller (ABC)
operation (see Section \ref{sec:addw}). Again, in this case, the time is
overhead. The experiment measures the time required for the creation of a single worker, and
thus the times measured are almost independent of the number of
workers pre-existing the new one.

As highlighted by the Figure \ref{fig:overhead:new} and
\ref{fig:overhead:restart} the overhead of the \emph{new} and
\emph{restart} operations is much higher in the case where a fresh
platform is involved (number of workers less than 32). The difference
is mainly due to the additional time for Java remote class loading. In fact, when a worker is created, if the classes it needs are not present (in the machine that is running it), they are copied locally then loaded in the cluster node main memory and compiled. Clearly, performing such operations require time, hundreds of milliseconds. Rather, if the classes are already present, already loaded in main memory or even already compiled in machine target code by the Java JIT, performing these reconfiguration operations is noticeably cheaper.

\begin{figure}
\centerline{\includegraphics[width=0.7\linewidth]{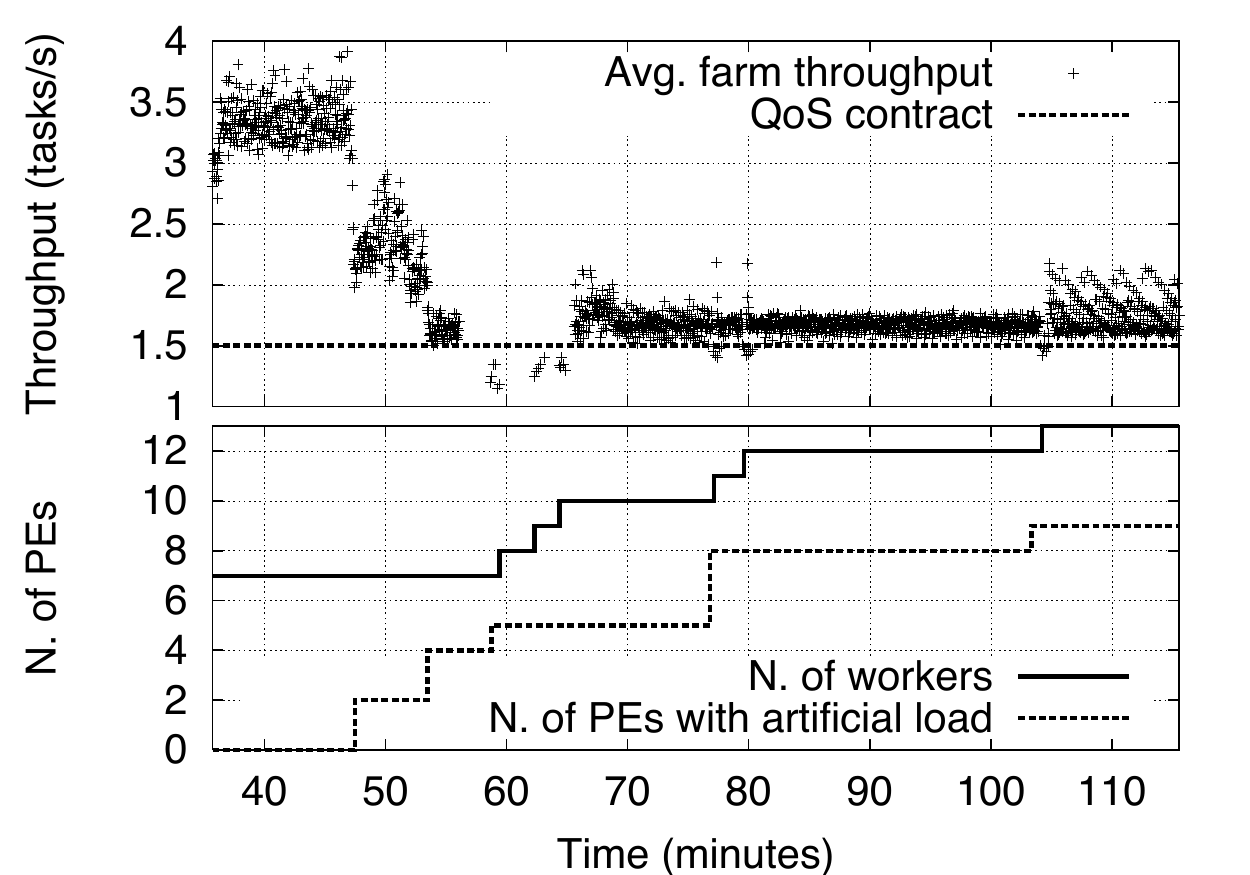}}
\caption{Self-optimization experiment.}
\label{fig:adapt}
\end{figure}

The results of the last experiment are presented in Figure \ref{fig:adapt}. It describes
the behavior of the application over quite a long run (two hours, approximately) that includes
several self-triggered reconfigurations. In this case the
application is provided with a Quality of Service (QoS) contract that enforces the
production of a minimum of $1.5$ results per second (tasks/s). During
the run, an increasing number of platforms are externally overloaded
with an artificial load (we started the compilation of some complex software written in C++). The top half of the figure
reports the measured average throughput of the filter stage (the second, actually), and the
QoS contract. The bottom half of the figure reports the number of
overloaded machines along the run, and the corresponding increase of
workers of the filter stage. Initially the throughput of the filter
stage is abundantly higher than requested ($\sim 3.5$ tasks/s); but it
decreases when more machines are overloaded. As soon as the contract
is violated, the Autonomic Manager reacts by adding more workers.

\newpage

\section*{Summarizing the Chapter}
\emph{
\hrule
\medskip
The challenge of autonomicity in the context of component-based development of grid software is substantial. Building into components autonomic capability typically impairs their reusability. In this Chapter we proposed Behavioural Skeletons as a compromise: being skeletons they support reuse, while their parameterization allows the controlled adaptivity needed to achieve dynamic adjustment of QoS while preserving functionality.
% that are a novel abstraction of
% self-management components able to support component reuse in the grid
% environment, and to ease the development of correct management
% strategies in the autonomic computing framework.
We also presented a significant set of skeletons and we discussed how Behavioural Skeletons can be implemented in the framework of the GCM component model. Behavioural Skeletons provide the programmer with the ability to implement autonomic managers completely taking care of the parallelism exploitation details by simply instantiating existing skeletons and by providing suitable, functional parameters.
%
%Moreover we provided the formal Orc functional behaviour description and self-management strategies.
%
%
%\label{sec:conclu}
%
%
%
Finally, we discussed the experimental results achieved when running an application exploiting instances of our Behavioural Skeletons and we showed how the skeletons used may take  decisions at the appropriate time to maintain the application behavior within the limits stated by the user with a specific performance contract.
The whole experiments have been performed using GCM components and Behavioural Skeletons, as being designed and implemented in the framework of the CoreGRID and GridCOMP projects.
To our knowledge,  no other similar results are available yet.
%As the behavioural skeleton approach has been proven feasible and effective, we are currently working to provide further skeletons, to refine the implementation and to perform more experiments involving the use case applications identified in the framework of the GridCOMP project, that include data intensive as well as transaction processing applications.
\medskip
\hrule
}

%\insertblankpage
%\fussy

\chapter{Conclusions}\label{thesis_concl}

Over the years, a lot of models and tools for parallel programming have been proposed. This great deal of efforts is mainly due to the difficulties in coordinating several, possibly hundreds or thousands, activities in an easy way but allowing an efficient exploitation of computational resources. In fact, to date does not exist a universal approach working better than others in every situation. Actually, there are several good approaches based on different perspectives and abstraction levels. Nevertheless, starting from the second half of nineties, with the advent of computational Grids, parallel programming difficulties became greater and greater and also the most promising approaches trail along. Indeed, programming the Grids is even more difficult than traditional parallel programming. This is because the computers belonging to a Grid can be heterogeneous, separated by firewalls, unsafe  %connected to low-bandwidth networks
 and managed by different administration policies. To address these additional difficulties most of the models and tools conceived and developed for parallel programming have to be re-thought and adapted.
In particular, Structured Parallel Programming models, and the derived environment have been proved to be very effective approach for programming parallel applications, but some well-known issues prevent them from achieving significant popularity in the wider parallel and grid programming community.

In this thesis we presented an organic set of tools and models conceived, designed and developed or properly modified to address most of these issues.

We started discussing how we modified the \muskel framework for supporting the issue related to the lack of extendability of the skeleton systems. We discussed how our customized \muskel supports the introduction of new skeletons, modeling parallelism exploitation patterns not originally covered by the primitive \muskel skeletons.
This possibility is supported by allowing \muskel users (the programmers) to define new skeletons providing the arbitrary data flow graph executed in the skeleton and by letting our \muskel version to seamlessly integrate such new skeletons with the primitive ones.
We also presented experimental results validating our \muskel approach to extend and customize its skeleton set. We ran several test programs using the custom features introduced in \muskel. When grain is small, \muskel does not scale well, even using a very small number of remote interpreter instances. When the computational grain is high enough %(about 200 times the time spent in communications actually spent in computation of macro data-flow instructions)
 the efficiency is definitely close to the ideal one. Despite the data shown in this thesis refer to synthetic computations, the tests we conducted using actual computations %(e.g. image processing ones) we made
  achieved very similar results. This because the automatic load balancing mechanism implemented in the \muskel distributed interpreter through auto scheduling perfectly optimized the execution of variable grain macro data-flow instructions.
As far as we know, this is the most significant effort in the skeleton community to tackle problems deriving from a fixed skeleton set. Only Schaeffer and his group at the University of Alberta implemented a system were programmers can, in controlled ways, insert new parallelism exploitation patterns in the system \cite{bromling:parco:2001}, although the approach followed here is a bit different, in that programmers are encouraged to intervene directly in the run-time support implementation, to introduce new skeletons, while in \muskel new skeletons may be introduced using the intermediate macro data flow language as the skeleton ``assembly'' language.
Unfortunately, programmers using this approach, in order to program unstructured parallel application, have to interact directly with data-flow graph. It requires to programmers to reason in terms of program-blocks instead of a monolithic program.
In order to ease the generation of macro data-flow blocks and in general to provide mechanism easing the use of structured parallel programming environment, we exploited some \textit{metaprogramming} techniques.

We exploited some metaprogramming techniques based both on Aspect Oriented Programming (AOP) and on Attribute Oriented Programming ($@$OP). We showed how these techniques can be seamlessly exploited to transform sequential applications into parallel ones. In particular, we showed how annotations and aspect can be exploited to drive the sequential application transformation into a macro data-flow graph that can be executed on distributed architectures.
The exploitation of $@$OP and AOP techniques allows to completely separate the concerns relative to parallelism exploitation and application functional code. In particular, the same application code used to perform functional debugging on a single, sequential machine may be easily turned into parallel code.
To validate the $@$OP approach we implemented PAL, a java annotation based metaprogramming framework that restructures applications at bytecode-level at run-time in order to make them parallel. PAL transformations depend on: i) the resources available at run-time, ii) the hints provided by programmers and iii) the available \textit{adapters}. An adapter is a specialized entity that instructs the PAL transformation engine to drive the code transformation depending on the available parallel tools and frameworks. Experimental results show that the PAL is an effective and efficient approach for handling resource heterogeneity and dynamicity. Actually, run-time code transformation brings to a very good exploitation of computational resources. For this implementation we developed two distinct adapters. The first  adapter we developed foster the  bytecode transformation making the original code a multithreaded one. The other adapter supports the bytecode transformation that makes the original code compliant with JJPF, a structured parallel programming framework we developed some years ago.
PAL demonstrated that, given the existence of a proper metaprogramming run-time support, annotations are a handy way both to indicate which parts of a program must run in parallel and to express non-functional requirements directly in the source code.
Therefore, we decided to apply the main features of PAL approach to our modified \muskel implementation. Actually, adapting them to \muskel we changed a little bit the approach. Such a change is due to a few motivations. First of all because \muskel provides \emph{per se} a distributed macro data-flow executor whereas PAL exploits external tools for distributed program execution. Moreover, we would like to have a more flexible mechanism for macro data-flow block generation and management. Finally, we would like to exploit a standard tool for run-time code transformation instead of using ad-hoc tools, like the one we developed for PAL.
As a consequence we decided to use integrate in \muskel the AOP model and in particular the AspectJ framework.
The integration has been performed in two steps, in the first step we integrated the AOP mechanisms in order to achieve very simple code transformation.
The second step consisted in testing the integration of \muskel with AspectJ to in a more complex scenario. Hence, we exploited the aspect oriented programming support we integrated in \muskel in order to develop workflows which structure and processing are optimized at run-time.
In order to prove the effectiveness of the approach in \muskel, we conducted some experiments on a network of workstations.
The only difference between plain \muskel and the system proposed here to execute workflows lies in the way fireable instructions are provided to the distributed data-flow interpreter of \muskel.
Indeed, in plain \muskel, fireable instructions are taken from a compiled representation of a data-flow graph. Each time a new token arrives to a macro data-flow instruction in the graph the target data-flow instruction is checked for ``fireability'' and, possibly, delivered to the distributed macro data-flow interpreter. The time spent is in the sub-micro second range (net time, not taking into account time spent to copy parameters in memory during the interpreter call). When executing workflows according to the approach discussed here, instead, fireable instructions is generated at run-time by the AOP engine. We measured the overhead when exploiting the AOP approach, it is approximately 23 milliseconds per workflow node.
%
% Summarizing, we can state that both exploit AOP and @OP
% The experiments conducted show that the approach is perfectly feasible and that actual speedups can be achieved provided that the macro data-flow blocks are medium to coarse grain.
%In the final version of the paper, if accepted, we will provide: a) complete description of the proposed approach, b) experimental results validating the approach (scalability, efficiency), c) description of the GUI used to program workflows, d) sample workflow code and e) complete reference section. \nota{md}{va citato che in questo modo il debug sequenziale del workflow e' immediato: basta spegnere gli aspetti}
%

These two results presented are feasible approaches for programming  cluster or networks of workstation but are not suitable for computational Grids, where component models are preferable. This is due to several motivations we described in deep in this thesis.
Provide parallel programming models for Grids are important because they are becoming the dominant type of parallel architectures. Moreover, due to their heterogeneous and distributed nature, they represent a very good test-bed for testing parallel programming models dealing with dynamicity handling.
The \muskel framework, handle dynamicity exploiting the \emph{Application Manager}: an entity that observes the behavior of the parallel application and in case of problems reacts aiming to fix them. This approach has proved to be effective. Nevertheless, some of the implementation choices done when \muskel was developed limit its exploitation on Grids.
Therefore, we decided to generalize and extend the \muskel \textit{Application Manager} approach to make it suitable for components models, in order to be able to port the approach in existing component models.
We ported the \muskel approach in the Grid Component Model. Actually, the \textit{Application Manager} approach form the base of the autonomic features of GCM: each self-optimizing GCM component contains an \textit{Application Manager} that in GCM is called \textit{Autonomic Manager}.
Nevertheless, \textit{Autonomic Manager} rely fully on the application programmer's expertise for the setup of the management code, which can be quite difficult to write since it is tailored for the particular component or assembly of them. As a result, the introduction of dynamic adaptivity might enable the management of grid dynamism but, at the same time, decreases the component reuse potential since it further specializes components with application specific management code.
In order to address this problem, we proposed the \emph{Behavioural Skeletons} as a novel way to describe autonomic components in the GCM framework. Behavioural Skeletons aim to describe recurring patterns of component assemblies that can be equipped with correct and effective management strategies with respect to a given management goal.
The Behavioural Skeletons model provides a way for handling dynamicity, supporting reuse both of functional and non-functional code.
We presented a significant set of skeletons and we discussed how behavioural skeletons can be implemented in the framework of the GCM component model. Behavioural skeletons provide the programmer with the ability to implement autonomic managers completely taking care of the parallelism exploitation details by simply instantiating existing skeletons and by providing suitable, functional parameters.
To validate our Behavioural Skeletons we conducted some experiments with the current prototype of the GCM that is currently under development in the GridCOMP STREP project \cite{gridcomp}.
We discussed the experimental results achieved when running an application exploiting instances of our Behavioural Skeletons and we showed how the skeletons used may take  decisions at the appropriate time to maintain the application behaviour within the limits stated by the user with a specific performance contract.

\section*{Future Works}
New efforts for future work can be invested in different directions, as suggested by the results offered by this thesis.

Concerning the macro data-flow based skeleton customizations, new mechanisms for modifying the macro data-flow graph can be conceived, possibly simpler than the existing one. Just as a note, currently we are developing a graphic tool that allows programmers (\muskel users) to design their macro data-flow graphs and then compile them directly to Java code as required by \muskel.

Several other annotations and aspects can be designed and implemented for easing the run-time generation of macro data-flow blocks. Possibly supporting several types of non-functional requirements. Regarding PAL, many adapters, even more complex than existing one can be developed. In particular, adapters for widely-used frameworks for Grid programming, like Globus or ProActive. Another interesting possibility can be the porting of the adapters model in our customized \muskel, perhaps making possible the transformation, at run-time, of the macro data-flow blocks generated by \muskel in GCM components.

In this thesis we presented a reduced set of Behavioural Skeletons, other skeletons can be conceived, designed and implemented. As an example, a Behavioural Skeleton supporting the non-functional replication management for easing the development of fault-tolerant component applications. Furthermore, a lot of research can be conducted on the distributed (cooperative) self-management of component applications, in particular regarding to the methodologies for splitting the user specified QoS contracts. 

\insertblankpage

%\appendix

%\include{appendix/A/appA}

%{\small
\sloppy
\bibliographystyle{plain}
\renewcommand{\bibname}{References}
\bibliography{bibliografie/database}

\begin{thebibliography}{100}

\bibitem{ada-language}
Ada programming language specification.
\newblock http://www.open-std.org/jtc1/sc22/wg9/.

\bibitem{grid5000}
Grid5000 project webpage.
\newblock www.grid5000.fr/.

\bibitem{gridcomp}
Gridcomp eu strep project website.
\newblock http://gridcomp.ercim.org.

\bibitem{jini}
Jini website.
\newblock http://www.jini.org.

\bibitem{cca}
Common component architecture forum home page.
\newblock www.acl.lanl.gov/cca/, 2003.

\bibitem{aspectj}
Aspectj home page, 2007.
\newblock http://www.eclipse.org/aspectj/.

\bibitem{ECMA335}
ECMA 335.
\newblock Common languageinfrastructure(cli).
\newblock http://www.ecma.ch/ecma1/STAND/ecma-335.htm, 2001.

\bibitem{javaAnnotation}
Java specification requests 175: A metadata facility for the java programming
  language.
\newblock http://www.jcp.org, September 2004.

\bibitem{547755}
Harold Abelson and Gerald~J. Sussman.
\newblock {\em Structure and Interpretation of Computer Programs}.
\newblock MIT Press, Cambridge, MA, USA, 1996.

\bibitem{ackerman}
W.B. Ackerman.
\newblock Dataflow languages.
\newblock {\em Computer}, 15(2):50--69, 1982.

\bibitem{aldinuc:sem:parco2003}
M.~Aldinucci and M.~Danelutto.
\newblock An operational semantics for skeletons.
\newblock In G.R. Joubert, W.E. Nagel, F.J. Peters, and W.V. Walter, editors,
  {\em Parallel Computing: Software Technology, Algorithms, Architect ures and
  Applications}, Advances in Parallel Computing. Elsevier, The Netherland,
  2004.

\bibitem{teti-fgcs}
M.~Aldinucci, M.~Danelutto, and P.~Teti.
\newblock An advanced environment supporting structured parallel programming in
  {Java}.
\newblock {\em Future Generation Computer Systems}, 19(5):611--626, 2003.
\newblock Elsevier.

\bibitem{adaptivity:parco:05}
Marco Aldinucci, Francoise Andr{\'e}, J{\'e}r{\'e}my Buisson, Sonia Campa,
  Massimo Coppola, Marco Danelutto, and Corrado Zoccolo.
\newblock Parallel program/component adaptivity management.
\newblock In G.~R. Joubert, W.~E. Nagel, F.~J. Peters, O.~Plata, P.~Tirado, and
  E.~Zapata, editors, {\em Parallel Computing: Current \& Future Issues of
  High-End Computing, Proc. of {PARCO 2005}}, volume~33 of {\em NIC}, pages
  89--96, Germany, December 2005. Research Centre J{\"u}lich.

\bibitem{assist:parco:03}
Marco Aldinucci, Sonia Campa, Pierpaolo Ciullo, Massimo Coppola, Marco
  Danelutto, Paolo Pesciullesi, Roberto Ravazzolo, Massimo Torquati, Marco
  Vanneschi, and Corrado Zoccolo.
\newblock A framework for experimenting with structure parallel programming
  environment design.
\newblock In G.~R. Joubert, W.~E. Nagel, F.~J. Peters, and W.~V. Walter,
  editors, {\em Parallel Computing: Software Technology, Algorithms,
  Architectures and Applications, PARCO 2003}, volume~13 of {\em Advances in
  Parallel Computing}, pages 617--624, Dresden, Germany, 2004. Elsevier.

\bibitem{assist:imp:europar:03}
Marco Aldinucci, Sonia Campa, Pierpaolo Ciullo, Massimo Coppola, Silvia Magini,
  Paolo Pesciullesi, Laura Potiti, Roberto Ravazzolo, Massimo Torquati, Marco
  Vanneschi, and Corrado Zoccolo.
\newblock The implementation of {ASSIST}, an environment for parallel and
  distributed programming.
\newblock In H.~Kosch, L.~B{\"o}sz{\"o}rm{\'e}nyi, and H.~Hellwagner, editors,
  {\em Proceedings of the 9th International Euro-Par Conference}, volume 2790
  of {\em Lecture Notes in Computer Science}, pages 712--721, Klagenfurt,
  Austria, August 2003. Springer Verlag.

\bibitem{heraklion-beske}
Marco Aldinucci, Sonia Campa, Marco Danelutto, Patrizio Dazzi, Peter
  Kilpatrick, Domenico Laforenza, and Nicola Tonellotto.
\newblock Behavioural skeletons for component autonomic management on grids.
\newblock In Marco Danelutto, Paraskevi Frangopoulou, and Vladimir Getov,
  editors, {\em Making Grids Work}, CoreGRID. Springer Verlag, June 2008.

\bibitem{pdp08:beske}
Marco Aldinucci, Sonia Campa, Marco Danelutto, Marco Vanneschi, Peter
  Kilpatrick, Patrizio Dazzi, Domenico Laforenza, and Nicola Tonellotto.
\newblock Behavioral skeletons in gcm: automatic management of grid components.
\newblock In Julien Bourgeois and Didier El~Baz, editors, {\em Proceedings of
  the 16th International Euromicro Conference on Parallel, Distributed and
  Network-based Processing}, pages 54--63. IEEE Computer Society Press,
  Toulose, France, February 2008.

\bibitem{pdcs:nf:99}
Marco Aldinucci and Marco Danelutto.
\newblock Stream parallel skeleton optimization.
\newblock In {\em Proceedings of the PDCS: International Conference on Parallel
  and Distributed Computing and Systems}, pages 955--962, Cambridge,
  Massachusetts, USA, November 1999. IASTED, ACTA press.

\bibitem{advske:pc:06}
Marco Aldinucci and Marco Danelutto.
\newblock Algorithmic skeletons meeting grids.
\newblock {\em Parallel Computing}, 32(7):449--462, 2006.

\bibitem{lithium:sem:CLSS}
Marco Aldinucci and Marco Danelutto.
\newblock Skeleton based parallel programming: functional and parallel semantic
  in a single shot.
\newblock {\em Computer Languages, Systems and Structures}, 2006.

\bibitem{muskelJournal}
Marco Aldinucci, Marco Danelutto, and Patrizio Dazzi.
\newblock Muskel: an expandable skeleton environment.
\newblock {\em Scalable Computing: Practice and Experience}, 8(4):325--341,
  December 2007.

\bibitem{assist:qos:euromicro:06}
Marco Aldinucci, Marco Danelutto, and Marco Vanneschi.
\newblock Autonomic {QoS} in {ASSIST} grid-aware components.
\newblock In Beniamino~Di Martino and Salvatore Venticinque, editors, {\em
  Proceedings of the 14th International Euromicro Conference on Parallel,
  Distributed and Network-based Processing}, pages 221--230, Montb{\'e}liard,
  France, February 2006. IEEE.

\bibitem{reconf:adaptcomp:05}
F.~Andr{\'e}, J.~Buisson, and J.-L. Pazat.
\newblock Dynamic adaptation of parallel codes: toward self-adaptable
  components for the {Grid}.
\newblock In V.~Getov and T.~Kielmann, editors, {\em Proceedings of the
  International Workshop on Component Models and Systems for Grid
  Applications}, CoreGRID series, ICS '04, Saint-Malo, France, January 2005.
  Springer Verlag.

\bibitem{reinefeld:dagstuhl:2004}
A.~Andrzejak, A.~Reinefeld, F.~Schintke, and T.~Sch{\"u}tt.
\newblock On adaptability in grid systems.
\newblock In V.~Getov, D.~Laforenza, and A.~Reinefeld, editors, {\em Future
  Generation Grids}, CoreGRID series. Springer Verlag, November 2005.

\bibitem{armstrong99toward}
Rob Armstrong, Dennis Gannon, Al~Geist, Katarzyna Keahey, Scott~R. Kohn, Lois
  McInnes, Steve~R. Parker, and Brent~A. Smolinski.
\newblock Toward a common component architecture for high-performance
  scientific computing.
\newblock In {\em {HPDC}}, 1999.

\bibitem{CullerArvind}
Arvind and D.E Culler.
\newblock Dataflow architectures.
\newblock Technical report, Annual Reviews of Computer Science, 1986.

\bibitem{50455}
Arvind and K.~Ekanadham.
\newblock Future scientific programming on parallel machines.
\newblock {\em Journal of Parallel and Distributed Computing}, 5(5):460--493,
  1988.

\bibitem{Asanovic:EECS-2006-183}
Krste Asanovic, Ras Bodik, Bryan~Christopher Catanzaro, Joseph~James Gebis,
  Parry Husbands, Kurt Keutzer, David~A. Patterson, William~Lester Plishker,
  John Shalf, Samuel~Webb Williams, and Katherine~A. Yelick.
\newblock The landscape of parallel computing research: A view from berkeley.
\newblock Technical Report UCB/EECS-2006-183, EECS Department, University of
  California, Berkeley, Dec 2006.

\bibitem{Darlington1996}
P.~Au, J.~Darlington, M.~Ghanem, Y.~Guo, H.W. To, and J.~Yang.
\newblock Co-ordinating heterogeneous parallel computation.
\newblock In L.~Bouge, P.~Fraigniaud, A.~Mignotte, and Y.~Robert, editors, {\em
  Europar '96}, pages 601--614. Springer Verlag, 1996.

\bibitem{orlando-grosso}
Bruno Bacci, Marco Danelutto, Salvatore Orlando, Susanna Pelagatti, and Marco
  Vanneschi.
\newblock {P$^3$L:} a structured high level programming language and its
  structured support.
\newblock {\em Concurrency Practice and Experience}, 7(3):225--255, May 1995.

\bibitem{p3l}
Bruno Bacci, Marco Danelutto, Salvatore Orlando, Susanna Pelagatti, and Marco
  Vanneschi.
\newblock P3l: A structured high level programming language and its structured
  support.
\newblock {\em Concurrency: Practice and Experience}, 7(3):225--255, May 1995.

\bibitem{skie:PC:1999}
Bruno Bacci, Marco Danelutto, Susanna Pelagatti, and Marco Vanneschi.
\newblock {SkIE}: A heterogeneous environment for {HPC} applications.
\newblock {\em Parallel Computing}, 25(13-14):1827--1852, 1999.

\bibitem{aspect-mpi-c++}
P.~V. Bangalore.
\newblock Generating parallel applications for distributed memory systems using
  aspects, components, and patterns.
\newblock In {\em The 6th AOSD Workshop on Aspects, Components and Patterns for
  Infrastructure Software (ACP4IS)}, Vancouver, BC, Canada, March 2006. ACM.

\bibitem{parUnifi}
J.~Barklund.
\newblock {\em Parallel Unification}.
\newblock PhD thesis, Uppsala University, 1989.

\bibitem{BenoitCGH05}
Anne Benoit, Murray Cole, Stephen Gilmore, and Jane Hillston.
\newblock Flexible skeletal programming with eskel.
\newblock In {\em Euro-Par}, pages 761--770, 2005.

\bibitem{CLOS}
Daniel~G. Bobrow, Richard~P. Gabriel, and Jon~L. White.
\newblock Clos in context: The shape of the design space.
\newblock In Andreas Paepcke, editor, {\em Object-oriented Programming}, pages
  29--61. MIT Press, Cambridge, MA, USA, 1993.

\bibitem{ac:superimpostion:99}
Jan Bosch.
\newblock Superimposition: a component adaptation technique.
\newblock {\em Information and Software Technology}, 41(5):257--273, 1999.

\bibitem{bromling:parco:2001}
S.~Bromling.
\newblock Generalising pattern-based parallel programming systems.
\newblock In G.~R. Joubert, A.~Murli, F.~J. Peters, and M.~Vanneschi, editors,
  {\em Parallel Computing: Advances and Current Issues. Proceedings of the
  International Conference ParCo2001}, pages 91--100. Imperial College Press,
  2002.

\bibitem{fractal:spec}
E.~Bruneton, T.~Coupaye, and J-B. Stefani.
\newblock {\em The Fractal Component Model, Technical Specification}.
\newblock ObjectWeb Consortium, 2003.

\bibitem{bruneton02asm}
Coupaye~T Bruneton~E, Lenglet~R.
\newblock Asm: a code manipulation tool to implement adaptable systems,
  grenoble, france.
\newblock Adaptable and Extensible Component Systems, Nov. 2002.

\bibitem{ejb3}
Bill Burke and Richard Monson-Haefel.
\newblock {\em Enterprise JavaBeans 3.0}.
\newblock O'Reilly, 5th edition, 2006.

\bibitem{caromel89wait}
D.~Caromel.
\newblock Service, asynchrony, and wait-by-necessity.
\newblock {\em Journal of Object-Oriented Programming}, Nov/Dec 1989.

\bibitem{caromel04asynchronous}
D.~Caromel, L.~Henrio, and B.~Serpette.
\newblock Asynchronous and deterministic objects, 2004.

\bibitem{caromel05theory}
Denis Caromel and Ludovic Henrio.
\newblock {\em A Theory of Distributed Object}.
\newblock Springer Verlag, 2005.

\bibitem{1066964}
Walter Cazzola, Antonio Cisternino, and Diego Colombo.
\newblock [a]c\#: C\# with a customizable code annotation mechanism.
\newblock In {\em SAC '05: Proceedings of the 2005 ACM symposium on Applied
  computing}, pages 1264--1268, New York, NY, USA, 2005. ACM.

\bibitem{92417}
Paolo Ciancarini.
\newblock Blackboard programming in shared prolog.
\newblock In {\em Selected papers of the second workshop on Languages and
  compilers for parallel computing}, pages 170--185, London, UK, UK, 1990.
  Pitman Publishing.

\bibitem{anacleto-australia}
S.~Ciarpaglini, M.~Danelutto, L.~Folchi, C.~Manconi, and S.~Pelagatti.
\newblock {ANACLETO:} a template-based {P3L} compiler.
\newblock In {\em Proceedings of the Parallel Computing Workshop (PCW'97)},
  1997.
\newblock Camberra, Australia.

\bibitem{5390}
Keith Clark and Steve Gregory.
\newblock Parlog: parallel programming in logic.
\newblock {\em ACM Trans. Program. Lang. Syst.}, 8(1):1--49, 1986.

\bibitem{eskel-site}
M.~Cole and A.~Benoit.
\newblock {\em The eSkel home page}, 2005.
\newblock \url{http://homepages.inf.ed.ac.uk/abenoit1/eSkel/}.

\bibitem{128874}
Murray Cole.
\newblock {\em Algorithmic skeletons: structured management of parallel
  computation}.
\newblock MIT Press, Cambridge, MA, USA, 1991.

\bibitem{cole:manifesto:02}
Murray Cole.
\newblock Bringing skeletons out of the closet: A pragmatic manifesto for
  skeletal parallel programming.
\newblock {\em Parallel Computing}, 30(3):389--406, 2004.

\bibitem{gcm:coregrid:07}
CoreGRID NoE deliverable series, Institute on Programming Model.
\newblock {\em Deliverable D.PM.04 {--} Basic Features of the Grid Component
  Model (assessed)}, February 2007.

\bibitem{10.1109/PDP.2007.20}
C.~A. Cunha and J.~L. Sobral.
\newblock An annotation-based framework for parallel computing.
\newblock In {\em Proceedings of the 15th International Euromicro Conference on
  Parallel, Distributed and Network-based Processing}, pages 113--120, Los
  Alamitos, CA, USA, 2007. IEEE.

\bibitem{genProgramming}
Krzysztof Czarnecki and Ulrich~W. Eisenecker.
\newblock {\em Generative Programming - Methods, Tools, and Applications}.
\newblock Addison--Wesley, June 2000.

\bibitem{mlws}
M.~Danelutto, R.~Di Cosmo, X.~Leroy, and S.~Pelagatti.
\newblock {Parallel Functional Programming with Skeletons: the OCAMLP3L
  experiment}.
\newblock In {\em ACM Sigplan Workshop on ML}, pages 31--39, 1998.

\bibitem{MDF:parco:99}
Marco Danelutto.
\newblock Dynamic run time support for skeletons.
\newblock In E.~H. D'Hollander, G.~R. Joubert, F.~J. Peters, and H.~J. Sips,
  editors, {\em Proceedings of the International PARCO 99: Parallel Computing},
  Parallel Computing Fundamentals \& Applications, pages 460--467. Imperial
  College Press, 1999.

\bibitem{Da01PPL}
Marco Danelutto.
\newblock Efficient support for skeletons on workstation clusters.
\newblock {\em Parallel Processing Letters}, 11(1):41--56, 2001.

\bibitem{muskel:qos:pdp:05}
Marco Danelutto.
\newblock {QoS} in parallel programming through application managers.
\newblock In {\em Proceedings of the 13th International Euromicro Conference on
  Parallel, Distributed and Network-based Processing}, pages 282--289, Lugano,
  Switzerland, February 2005. IEEE.

\bibitem{DaDa05parco}
Marco Danelutto and Patrizio Dazzi.
\newblock A java/jini framework supporting stream parallel computations.
\newblock In {\em Proceedings of the International PARCO 2005: Parallel
  Computing}, volume~33 of {\em John von Neumann Institute for Computing
  Series}, pages 681--688. Central Institute for Applied Mathematics,
  J{\"u}lich, Germany, September 2005.

\bibitem{muskaspects:cg_book:08}
Marco Danelutto and Patrizio Dazzi.
\newblock Workflows on top of a macro data flow interpreter exploiting aspects.
\newblock In Marco Danelutto, Paraskevi Frangopoulou, and Vladimir Getov,
  editors, {\em Making Grids Work}, CoreGRID. Springer, June 2008.

\bibitem{pal}
Marco Danelutto, Marcelo Pasin, Marco Vanneschi, Patrizio Dazzi, Luigi Presti,
  and Domenico Laforenza.
\newblock Pal: Exploiting java annotations for parallelism.
\newblock In Marian Bubak, Sergei Gorlatch, and Thierry Priol, editors, {\em
  Achievements in European Research on Grid Systems}, CoreGRID Series, pages
  83--96. Springer, Krakow, Poland, 2007.

\bibitem{stigliani:europar:00}
Marco Danelutto and Massimiliano Stigliani.
\newblock {SKElib}: parallel programming with skeletons in {C}.
\newblock In A.~Bode, T.~Ludwing, W.~Karl, and R.~Wism\"uller, editors, {\em
  Proc. of 6th Intl. Euro-Par 2000 Parallel Processing}, volume 1900 of {\em
  Lecture Notes in Computer Science}, pages 1175--1184, Munich, Germany, August
  2000. Springer Verlag.

\bibitem{653465}
Marco Danelutto and P.~Teti.
\newblock Lithium: A structured parallel programming environment in java.
\newblock In {\em ICCS '02: Proceedings of the International Conference on
  Computational Science-Part II}, pages 844--853, London, UK, 2002.
  Springer-Verlag.

\bibitem{darlington:parle:93}
J.~Darlington, A.~J. Field, P.G. Harrison, P.~H.~J. Kelly, D.~W.~N. Sharp,
  R.~L. While, and Q.~Wu.
\newblock Parallel programming using skeleton functions.
\newblock In {\em Proc. of Parallel Architectures and Langauges Europe
  (PARLE'93)}, volume 694 of {\em Lecture Notes in Computer Science}, pages
  146--160, Munich, Germany, June 1993. Springer Verlag.

\bibitem{alice}
J.~Darlington and M.~J. Reeve.
\newblock Alice and the parallel evaluation of logic programs.
\newblock In {\em Computer Architecture Symposium}, Stockholm, June 1983.

\bibitem{ac:safran:06}
Pierre-Charles David and Thomas Ledoux.
\newblock An aspect-oriented approach for developing self-adaptive fractal
  components.
\newblock In Welf L{\"o}we and Mario S{\"u}dholt, editors, {\em Proceedings of
  the 5th Intl Symposium Software on Composition}, volume 4089 of {\em Lecture
  Notes in Computer Science}, pages 82--97, Vienna, Austria, March 2006.
  Springer.

\bibitem{DenPerPriRib}
Alexandre Denis, Christian P\'erez, Thierry Priol, and Andr\'e Ribes.
\newblock Bringing high performance to the corba component model.
\newblock In {\em SIAM Conference on Parallel Processing for Scientific
  Computing}, February 2004.

\bibitem{aop1}
T.~Elrad, R.~E. Filman, and A.~Bader.
\newblock Aspect-oriented programming.
\newblock {\em Communications of the ACM}, 44(10), Oct. 2001.

\bibitem{DBLP:conf/ccgrid/FerreiraSP06}
Jo{\~a}o~Fernando Ferreira, Jo{\~a}o~Lu\'{\i}s Sobral, and Alberto~Jos{\'e}
  Proen\c{c}a.
\newblock Jaskel: A java skeleton-based framework for structured cluster and
  grid computing.
\newblock In {\em CCGRID}, pages 301--304. IEEE, 2006.

\bibitem{JBoss}
Marc Fleury.
\newblock {\em The Official JBoss Development and Administration Guide}.
\newblock Pearson Education, 2002.

\bibitem{163131}
Ivan Futo.
\newblock Prolog with communicating processes: from t-prolog to csr-prolog.
\newblock In {\em ICLP'93: Proceedings of the tenth international conference on
  logic programming on Logic programming}, pages 3--17, Cambridge, MA, USA,
  1993. MIT Press.

\bibitem{383872}
Vladimir Getov, Gregor von Laszewski, Michael Philippsen, and Ian Foster.
\newblock Multiparadigm communications in java for grid computing.
\newblock {\em Commun. ACM}, 44(10):118--125, 2001.

\bibitem{gorlatch:hoc:dagstuhl:05}
S.~Gorlatch and J.~D{\"u}nnweber.
\newblock From grid middleware to grid applications: Bridging the gap with
  {HOC}s.
\newblock In V.~Getov, D.~Laforenza, and A.~Reinefeld, editors, {\em Future
  Generation Grids}, CoreGRID series. Springer, November 2005.

\bibitem{wfls}
Survey on grid workflows, 2007.
\newblock http://wiki.cogkit.org/index.php/ Survey\_on\_Grid\_Workflows.

\bibitem{grimshaw93mentat}
Andrew~S. Grimshaw.
\newblock The mentat computation model data-driven support for object-oriented
  parallel processing.
\newblock Technical report, Dept. Comp. Science, Univ. Virginia, 28 1993.

\bibitem{38811}
Andrew~S. Grimshaw and Jane W.~S. Liu.
\newblock Mentat: An object-oriented macro data flow system.
\newblock In {\em OOPSLA '87: Conference proceedings on Object-oriented
  programming systems, languages and applications}, pages 35--47, New York, NY,
  USA, 1987. ACM.

\bibitem{ngg3:06}
Next Generation GRIDs~Expert Group.
\newblock {\em NGG3, Future for European Grids: GRIDs and Service Oriented
  Knowledge Utilities. Vision and Research Directions 2010 and Beyond}.
\newblock Next Generation GRIDs Expert Group, January 2006.

\bibitem{bruno04}
Bruno Harbulot and John~R. Gurd.
\newblock Using aspectj to separate concerns in parallel scientific java code.
\newblock In {\em Proceedings of the 3rd International Conference on
  Aspect-Oriented Software Development}, pages 122--131, Lancaster, UK, March
  2004.

\bibitem{Spring}
Rod Johnson, Juergen Hoeller, Alef Arendsen, Thomas Risberg, and Dmitriy
  Kopylenko.
\newblock {\em Professional Java Development with the Spring Framework}.
\newblock Wrox Press Ltd., Birmingham, UK, UK, 2005.

\bibitem{1013209}
Wesley~M. Johnston, J.~R.~Paul Hanna, and Richard~J. Millar.
\newblock Advances in dataflow programming languages.
\newblock {\em ACM Computing Surveys}, 36(1):1--34, 2004.

\bibitem{PaulKelly}
Paul~H. Kelly.
\newblock {\em Functional Programming for Loosely-Coupled Multiprocessors}.
\newblock MIT Press, Cambridge, MA, USA, 1989.

\bibitem{grads:overview}
K.~Kennedy, M.~Mazina, J.~Mellor-Crummey, K.~Cooper, L.~Torczon, F.~Berman,
  A.~Chien, H.~Dail, O.~Sievert, D.~Angulo, I.~Foster, D.~Gannon, L.~Johnsson,
  C.~Kesselman, R.~Aydt, D.~Reed, J.~Dongarra, S.~Vadhiyar, and R.~Wolski.
\newblock Toward a framework for preparing and executing adaptive {Grid}
  programs.
\newblock In {\em Proceedings of the NSF Next Generation Systems Program
  Workshop (IPDPS 2002)}, 2002.

\bibitem{AC:vision:2003}
J.~O. Kephart and D.~M. Chess.
\newblock The vision of autonomic computing.
\newblock {\em IEEE Computer}, 36(1):41--50, 2003.

\bibitem{aspectjnew}
Gregor Kiczales, Erik Hilsdale, Jim Hugunin, Mik Kersten, Jeffrey Palm, and
  William Griswold.
\newblock Getting started with aspectj.
\newblock {\em Communications of the ACM}, 44(10):59--65, 2001.

\bibitem{AOP}
Gregor Kiczales, John Lamping, Anurag Mendhekar, Chris Maeda, Cristina Lopes,
  Jean-Marc Loingtier, and John Irwin.
\newblock Aspect-oriented programming.
\newblock In {\em Proceedings of the European Conference on Object-Oriented
  Programming}, volume 1241, page 220–242, 1997.

\bibitem{krall}
A.~Krall.
\newblock Eficientjavavm just-in-timecompilation.
\newblock In Jean-Luc Gaudiot, editor, {\em Proceedings of the International
  Conference on Parallel Architectures and Compilation Techniques}, pages
  205--212, Paris, 1998.

\bibitem{kuchen-optim}
H.~Kuchen.
\newblock Optimizing sequences of skeleton calls.
\newblock In D.~Batory, C.~Consel, C.~Lengauer, and M.~Odersky, editors, {\em
  Domain-Specific Program Generation}, number 3016 in Lecture Notes in Computer
  Science, pages 254--273. Springer Verlag, 2004.

\bibitem{kuchen:europar:2002}
Herbert Kuchen.
\newblock A skeleton library.
\newblock In B.~Monien and R.~Feldman, editors, {\em Proceedings of the 8th
  International Euro-Par Conference}, volume 2400 of {\em Lecture Notes in
  Computer Science}, pages 620--629, Paderborn, Germany, August 2002. Springer.

\bibitem{muesli-home}
Herbert Kuchen.
\newblock The muesli home page, 2006.
\newblock \url{http://www.wi.uni-muenster.de/PI/forschung/Skeletons/}.

\bibitem{156619}
Vipin Kumar, Ananth Grama, Anshul Gupta, and George Karypis.
\newblock {\em Introduction to parallel computing: design and analysis of
  algorithms}.
\newblock Benjamin-Cummings Publishing Co., Inc., Redwood City, CA, USA, 1994.

\bibitem{JavaLang}
T.~Lindholm and F.~Yellin.
\newblock {\em The Java Virtual Machine Specification}.
\newblock Addison-Wesley, 1999.

\bibitem{shaeffer-europar00}
S.~McDonald, D.~Szafron, J.~Schaeffer, and S.~Bromling.
\newblock Generating parallel program frameworks from parallel design patterns.
\newblock In A.~Bode, T.~Ludwing, W.~Karl, and R.~Wism\"uller, editors, {\em
  Proceedings of the 6th International Euro-Par Conference}, LNCS, No. 1900,
  pages 95--105. Springer Verlag, August/September 2000.

\bibitem{ac:automate:06}
Manish Parashar, Hua Liu, Zhen Li, Vincent Matossian, Cristina Schmidt,
  Guangsen Zhang, and Salim Hariri.
\newblock {AutoMate}: Enabling autonomic applications on the {Grid}.
\newblock {\em Cluster Computing}, 9(2):161--174, 2006.

\bibitem{libro-susanna}
S.~Pelagatti.
\newblock {\em Structured Development of Parallel Programs}.
\newblock Taylor and Francis, 1998.

\bibitem{12074}
Luis~Moniz Pereira, Luis Monteiro, Jose Cunha, and Joaquim~N Aparicio.
\newblock Delta prolog: a distributed backtracking extension with events.
\newblock In {\em Proceedings on Third international conference on logic
  programming}, pages 69--83, New York, NY, USA, 1986. Springer Verlag.

\bibitem{kuchen-farm}
M.~Poldner and H.~Kuchen.
\newblock Scalable farms.
\newblock In {\em Proceedings of the Parallel Computing (ParCo)}, 2005.
\newblock Malaga (post-proceedings, to appear, available at
  \texttt{http://danae.uni-muenster.de/ lehre/ kuchen/ papersI.html}.

\bibitem{fraclet}
R.~Rouvoy, N.~Pessemier, R.~Pawlak, and P.~Merle.
\newblock Using attribute-oriented programming to leverage fractal-based
  developments.
\newblock In {\em Proceedings of the 5th Fractal workshop at ECOOP 2006}, July
  2006.

\bibitem{AttributeOP2}
Romain Rouvoy and Philippe Merle.
\newblock Leveraging component-oriented programming with attribute-oriented
  programming.
\newblock In {\em Proceedings of the 11th International ECOOP Workshop on
  Component-Oriented Programming}, volume 2006-11, Nantes, France, July 2006.
  Karlsruhe University.

\bibitem{ac:proxy:04}
S.~M. Sadjadi and P.~K. McKinley.
\newblock Transparent self-optimization in existing {CORBA} applications.
\newblock In {\em Proceedings of the 1st International Conference on Autonomic
  Computing (ICAC'04)}, pages 88--95, Washington, DC, USA, 2004. IEEE.

\bibitem{serot02}
J.~Serot.
\newblock Tagged-token data-flow for skeletons.
\newblock {\em Parallel Processing Letters}, 11(4), 2001.
\newblock 2001.

\bibitem{772854}
Jocelyn Serot and Dominique Ginhac.
\newblock Skeletons for parallel image processing: an overview of the skipper
  project.
\newblock {\em Parallel Computing}, 28(12):1685--1708, 2002.

\bibitem{39085}
E.~Shapiro, editor.
\newblock {\em Concurrent prolog: collected papers}.
\newblock MIT Press, Cambridge, MA, USA, 1987.

\bibitem{silc98asynchrony}
J.~Silc, B.~Robic, and T.~Ungerer.
\newblock Asynchrony in parallel computing: From dataflow to multithreading,
  1998.

\bibitem{ILParallelism}
Dezso Sima, Terence Fountain, and Peter Karsuk.
\newblock {\em Advanced Computer Architectures: A Design Space Approach}.
\newblock Addison-Wesley, 1997.

\bibitem{LISP}
B.C. Smith and J.~des Rivieres.
\newblock {\em Interim 3-LISP Reference Manual}.
\newblock XEROX, Palo Alto, June 1984.

\bibitem{SobralIpdps2006}
J.~Sobral.
\newblock Incrementally developing parallel applications with aspectj.
\newblock In {\em Proceedings of the 20th IEEE International Parallel and
  Distributed Processing Symposium}. IEEE Press, 4 2006.

\bibitem{Sobral:aosd-acp4is06}
Joao~L. Sobral, Miguel~P. Monteiro, and Carlos~A. Cunha.
\newblock Aspect-oriented support for modular parallel computing.
\newblock In {\em Proceedings of the 5th Workshop on Aspects, Components, and
  Patterns for Infrastructure Software}, pages 37--41, Bonn, Germany, 2006.
  Published as University of Virginia Computer Science Technical Report
  \mbox{CS--2006--01}.

\bibitem{proactive}
OASIS team.
\newblock Proactive home page, 2006.
\newblock \texttt{http://www-sop.inria.fr/oasis/proactive/}.

\bibitem{tesiteti}
P.~Teti.
\newblock Lithium: a java skeleton environment \textit{in italian}.
\newblock Master's thesis, Dept. Computer Science, University of Pisa, October
  2001.

\bibitem{javaVMSpec}
Frank~Yellin Tim~Lindholm.
\newblock {\em The Java Virtual Machine Specification}.
\newblock Sun Microsystems Press, second edition edition, 2004.

\bibitem{ac:wrapping:01}
E.~Truyen, B.~J{\o }rgensen, W.~Joosen, and P.~Verbaeten.
\newblock On interaction refinement in middleware.
\newblock In J.~Bosch, C.~Szyperski, and W.~Weck, editors, {\em Proceedings of
  the 5th International Workshop on Component-Oriented Programming}, pages
  56--62, 2001.

\bibitem{Ued86}
Kazunori Ueda.
\newblock {\em Guarded Horn clauses}.
\newblock D.eng. thesis, University of Tokyo, Tokyo, Japan, March 1986.

\bibitem{van:assist:02}
Marco Vanneschi.
\newblock The programming model of {ASSIST}, an environment for parallel and
  distributed portable applications.
\newblock {\em Parallel Computing}, 28(12):1709--1732, December 2002.

\bibitem{AttributeOP1}
Hiroshi Wada and Junichi Suzuki.
\newblock Modeling turnpike frontend system: A model-driven development
  framework leveraging uml metamodeling and attribute-oriented programming.
\newblock In {\em MoDELS}, pages 584--600, 2005.

\bibitem{612574}
Paul~G. Whiting and Robert S.~V. Pascoe.
\newblock A history of data-flow languages.
\newblock {\em IEEE Annals of the History of Computing}, 16(4):38--59, 1994.

\end{thebibliography}
\fussy
%bibliografie/bibbe,bibliografie/biblioAspectIW07,bibliografie/biblioJJPF,bibliografie/biblioMuskelPAPP,bibliografie/biblioMuskelSTPE,bibliografie/biblioWRASQ,bibliografie/biblioPAL,bibliografie/biblioPDP08,bibliografie/biblioProposal}
%}

\insertblankpage
\makecopyright
\insertblankpage

\end{document}